\documentclass[11pt]{article}
\usepackage{amsmath}
\usepackage{amsbsy}
\usepackage{amscd}
\usepackage{amsopn}
\usepackage{amstext}
\usepackage{amsxtra}
\usepackage{graphicx}
\usepackage{epsfig}
\usepackage{authblk}
\usepackage{cite}
\usepackage{url}

\usepackage{color} 

\begin{document}
\title{A Simple Generative Model of Collective Online Behaviour}
\author[1]{\rm James P. Gleeson}
\author[2]{\rm Davide Cellai}
\author[3]{\rm Jukka-Pekka Onnela}
\author[4]{\rm Mason A. Porter}
\author[5]{\rm Felix Reed-Tsochas}
\affil[1,2]{MACSI, Department of Mathematics \& Statistics, University of Limerick, Ireland}
\affil[3]{Department of Biostatistics, Harvard School of Public Health, Boston, USA}
\affil[4]{Oxford Centre for Industrial and Applied Mathematics, Mathematical Institute, University of Oxford, Oxford OX2 6GG, UK; CABDyN Complexity Centre, University of Oxford, Oxford, OX1 1HP, UK}
\affil[5]{CABDyN Complexity Centre, Sa\"{i}d Business School, University of Oxford, OX1 1HP, UK; Institute for New Economic Thinking at the Oxford Martin School, University of Oxford, Oxford OX1 3BD, UK}
\date{28 May 2014}

\maketitle





\textbf{Human activities increasingly
take place in online environments, providing novel opportunities for relating individual
behaviours to population-level outcomes.
In this paper, we introduce a simple generative
model for the collective behaviour of millions of social networking site users who
are deciding between different software applications. Our model incorporates two
distinct components: one is associated with recent decisions of users, and the other
reflects the cumulative popularity of each application. Importantly, although various
combinations of the two mechanisms yield long-time behaviour that is consistent
with data, { the only models that reproduce the observed temporal dynamics are those that strongly emphasize the recent popularity of applications
over their cumulative popularity.
This demonstrates---even when using purely observational data without experimental design---that temporal data-driven modelling can effectively distinguish between competing microscopic mechanisms, allowing us to uncover new aspects of collective online behaviour.}}\\





SIGNIFICANCE STATEMENT: One of the most common strategies in studying complex systems is to
investigate and interpret whether any ``hidden order'' is present by fitting observed statistical regularities via data analysis and then reproducing such regularities with long-time or equilibrium dynamics from some
generative model.  Unfortunately, many different models can
possess indistinguishable long-time dynamics, so the above recipe
is often insufficient to discern the relative quality of competing
models.  In this paper, we use the example of collective online behaviour to
illustrate that, by contrast, time-dependent modeling \emph{can} be very
effective at disentangling competing generative models of a complex system.\\


 The recent availability of data sets that capture the behaviour of individuals participating in online social systems has helped drive the emerging field of computational social science \cite{comp-ss}, as large-scale empirical data sets enable the development of detailed computational models of individual and collective behaviour \cite{Aral12,fowler-face,Gonzalez11}.
Choices of which movies to watch, which mobile applications (``apps'') to download, or which messages to retweet are influenced by the opinions of our friends, neighbours, and colleagues \cite{Bentleybook}. Given the difficulty in distinguishing between potential explanations of observed behaviour at the individual level \cite{shalizi},
 it is useful to examine population-level models and attempt to reproduce empirically-observed popularity distributions using the simplest possible assumptions about individual behaviour. Such generative models have arisen in a wide range of disciplines---including economics \cite{Simon55,deVanybook}, evolutionary biology \cite{Yule25,Ewens04}, and physics \cite{Redner98}. { When studying generative models, the microscopic dynamics are known exactly, so it is possible to explore the population-level mechanisms that emerge in a controlled manner. This contrasts with  studies driven by empirical data, in which confounding effects can always be present \cite{shalizi}. The value of explanations based on mechanisms has long been appreciated in sociology \cite{Hedstrombook,Granovetter78,Schellingbook}, and they have recently received increased attention due to the availability of extensive data from online social networks \cite{Onnela10,Romero11,Bakshy11,Lerman12}}.

One well-studied rule for choosing between multiple options is cumulative advantage (a.k.a. preferential attachment), in which popular options are more likely to be selected than unpopular ones. This leads to a ``rich-get-richer'' agglomeration of popularity \cite{Yule25,Simon55,Price76,Barabasi99,Simkin11}.
Bentley et al.~\cite{Bentley04,Bentley11,Bentleybook} proposed an alternative model, in which members of a population randomly copy the choices made by other members in the recent past.
As a result, products whose popularity levels have recently grown the fastest are the most likely to be selected (whether or not they are the most popular overall).
{ In the present paper, we
 show that models of app-installation decisions that are biased heavily towards recent 
popularity rather than cumulative
popularity provide the best fit to empirical data on the installation of Facebook apps.} We use the model to identify the timescales over which the influence of Facebook users upon each others' choices is strongest, and we argue that the interaction between these timescales and the diurnal variation in Facebook activity yields many of the observed features of the popularity distribution of apps.  More generally, we illustrate how to incorporate temporal dynamics in modelling and data analysis to differentiate between competing models that produce the same long-time (i.e., after transients have died out) behaviour.


We use the Facebook apps data set that was first reported in Ref.~\cite{Onnela10}. This data includes the records, for every hour from 25 June 2007 to 14 August 2007, of the number of times that every Facebook app (of the $N = 2705$ total available during this period) was installed.
At the time, Facebook users had two streams of information about apps: a \emph{cumulative information} stream gave an ``all-time best-seller'' list, in which all apps were ranked by their cumulative popularity (i.e., the total number of installations to date), and a \emph{recent activity information} stream consisted of updates provided by Facebook on the recent app installation activity by a user's friends.  Users could also visit the profiles of their friends to see which applications a friend had installed.

The data thus consists of $N$ time series $n_i(t)$, where the \emph{popularity} $n_i(t)$ of app $i$ at time $t$ is the total number of users who have installed
app $i$ by hour $t$ of the study period. The discrete time index $t$ counts hours from the start of the study period ($t=0$) to the end ($t=t_\text{max}\equiv 1209$). The distribution of $n_i$ values is heavy-tailed [see Fig.~S1 of the Supporting Information Appendix (SI)], so the popularities $n_i(t)$ of the apps cover a very wide range of scales. Facebook apps first became available on 24 May 2007, corresponding to $t\approx -720$ in our notation. By time $t=0$, when the data collection began, 980 apps had already launched { (with unknown launch times)}; the remaining apps in our data set were launched during the study period.
Among the latter, we pay particular attention to those for which we have at least $t_\text{LES}\equiv 650$ hours (i.e., more than half of the data-collection window) of data. We call these apps the \emph{Launched-Early-in-Study (LES)} apps. Denoting by $t_i$ the launch time of app $i$, the 921 LES apps $i$ are those that  satisfy $t_i>0$ and $t_i<t_\text{max}-t_\text{LES}=559$.
{ We set $t_i=0$ for apps that were launched prior to the study period.}

To measure the change in app popularity during hour $t$, we define the \emph{increment} in popularity of app $i$ at time $t$ as $f_i(t)=n_i(t)-n_i(t-1)$ (with $f_i(t)=0$ for $t \le t_i$) \cite{Onnela10}.  The total app installation activity of users during hour $t$ is then
\begin{equation}
	F(t) = \sum_{i=1}^{N} f_i(t)\,. \label{1}
\end{equation}
We show in  SI1 that $F(t)$ has large diurnal fluctuations superimposed {on a linear-in-time aggregate growth.}

{ We define the \emph{age-shifted popularity} $\tilde n_i(a) = n_i(t_i+a)$ and \emph{age-shifted increment} $\tilde f_i(a) = f_i(t_i+a)$ of app $i$ at age $a$
to enable comparison of apps when they are the same age (i.e., at the same number of hours after their launch).} An examination of the trajectories of the  largest LES apps
 reveals that their popularity grows exponentially for some time before reaching a steady-growth regime in which $\tilde n_i(a)$ increases approximately linearly with age. The corresponding age-shifted increment functions $\tilde f_i(a)$ reach a ``plateau" at large $a$, though they have a superimposed 24-hour oscillation (see Figs.~S3 and S4 in the SI Appendix).
In order to study the entire set of LES apps, we scale the increment $\tilde f_i$ of app $i$ by its temporal average $\tilde \mu_i = \left(\sum_{a=1}^{t_\text{LES}}\tilde f_i(a)\right)/t_\text{LES}$ over the first $t_\text{LES}=650$ observations for each app.
This weights very popular apps and other (less popular) apps in a similar manner \cite{Szabo10}. For a given set $\cal I$ of LES apps,
 we define the \emph{mean scaled age-shifted growth rate}
\begin{equation}
	r(a) = \left\langle \frac{\tilde f_i(a)}{\tilde \mu_i}\right\rangle_{\cal I}\,, \label{2}
\end{equation}
where $\langle \cdot \rangle$ denotes an ensemble average over all apps in the set $\cal I$.

\begin{figure}
\centering
\epsfig{figure=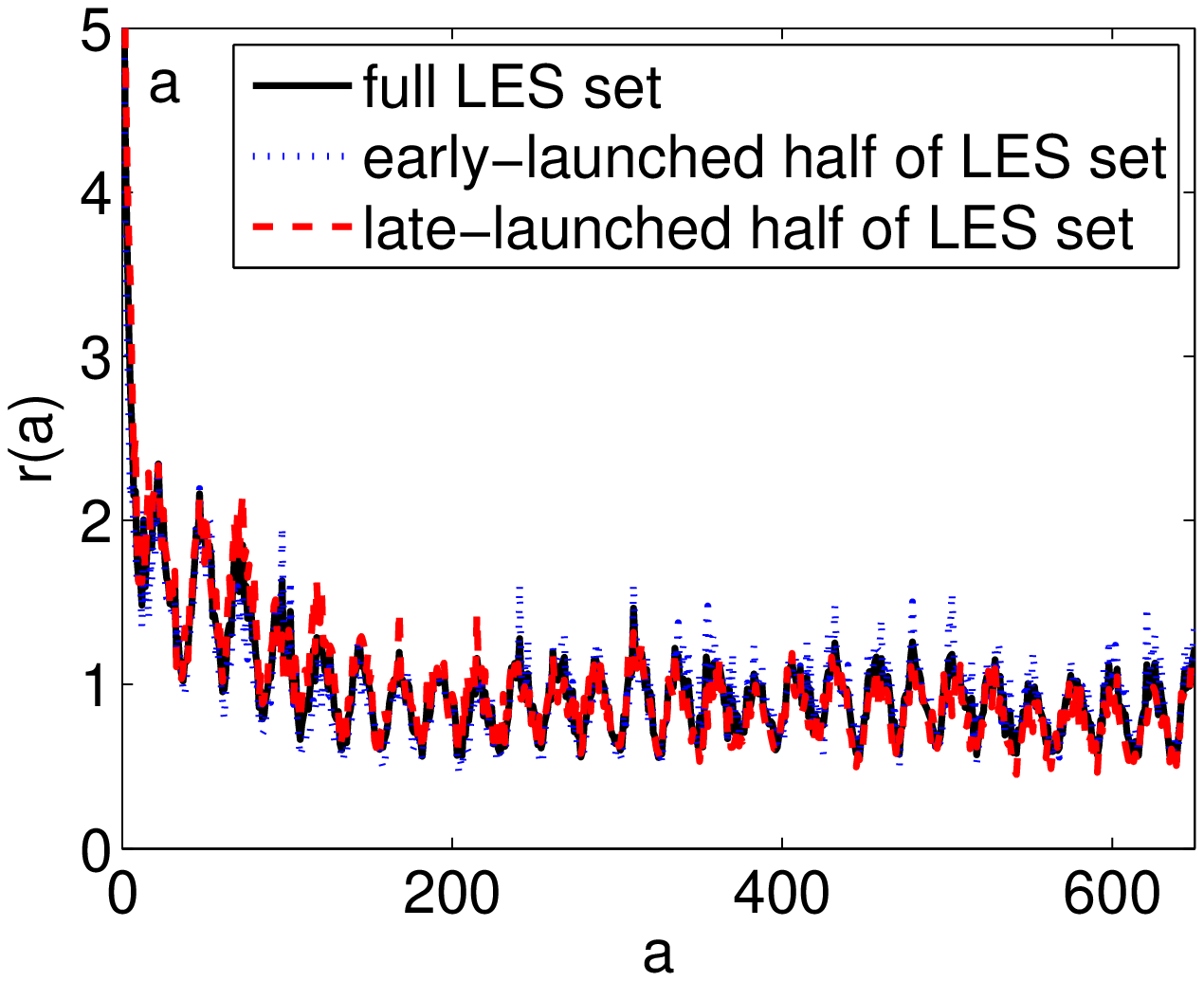,width=5.25cm} 
\epsfig{figure=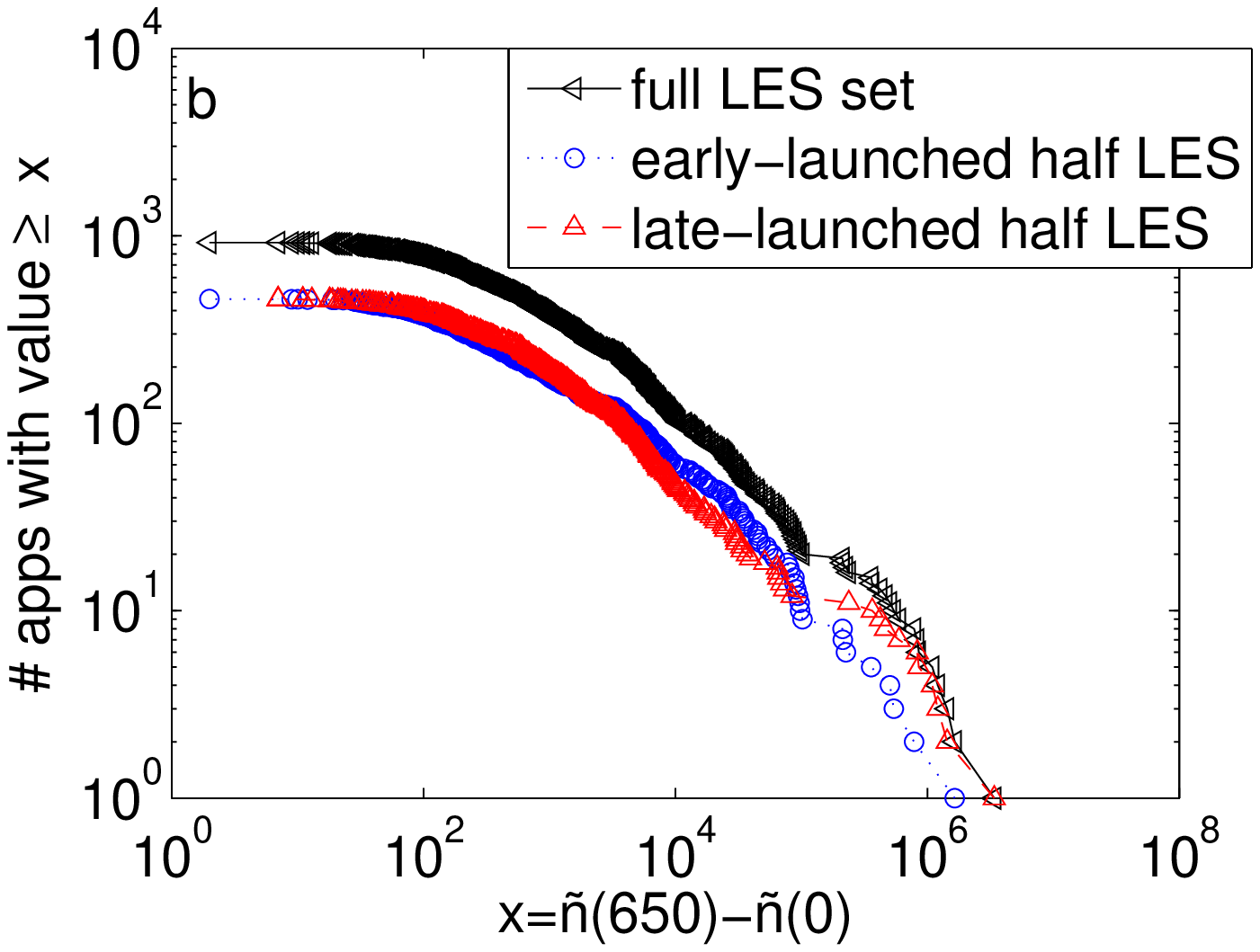,width=5.25cm} 
\epsfig{figure=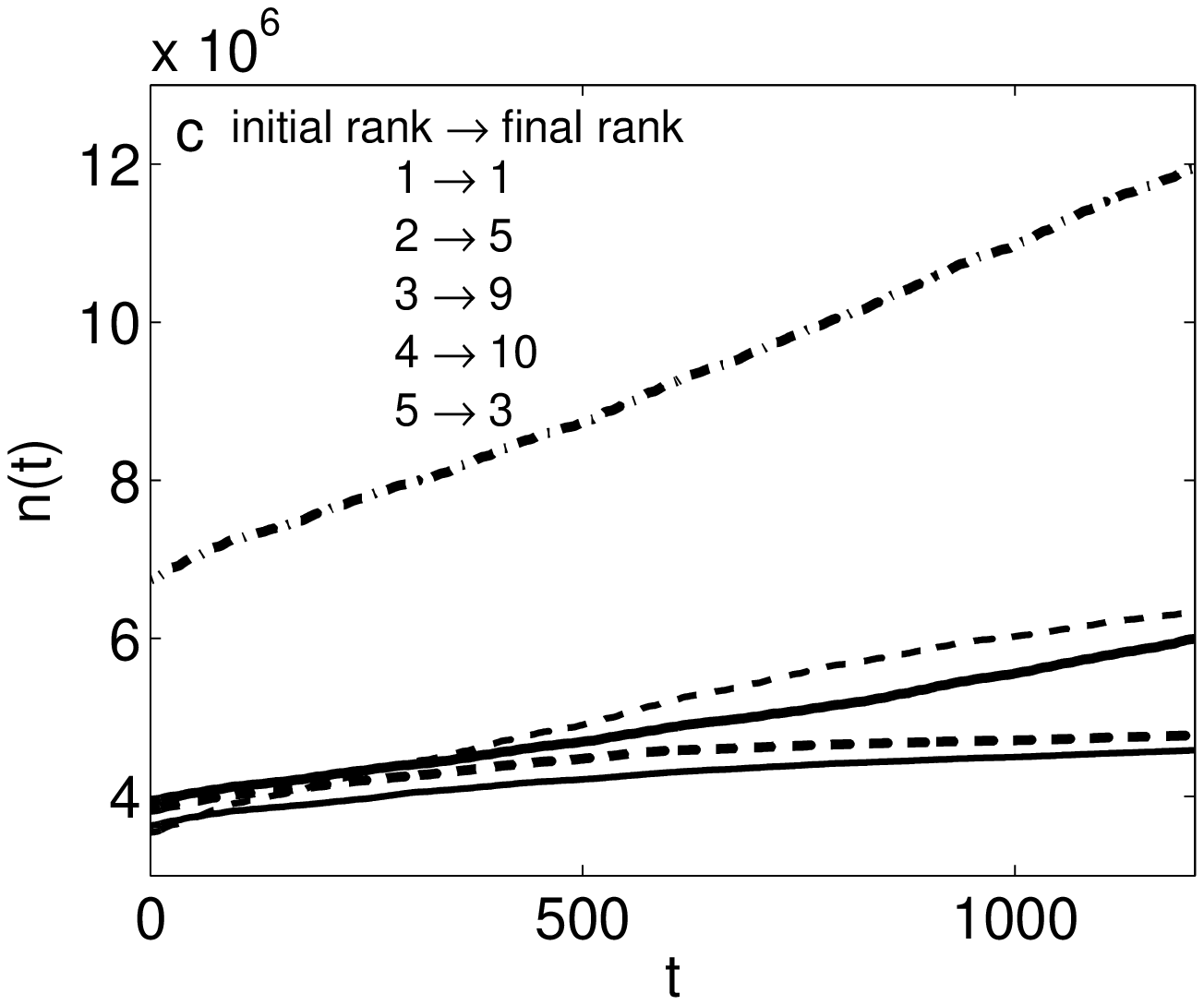,width=5.25cm} 
\epsfig{figure=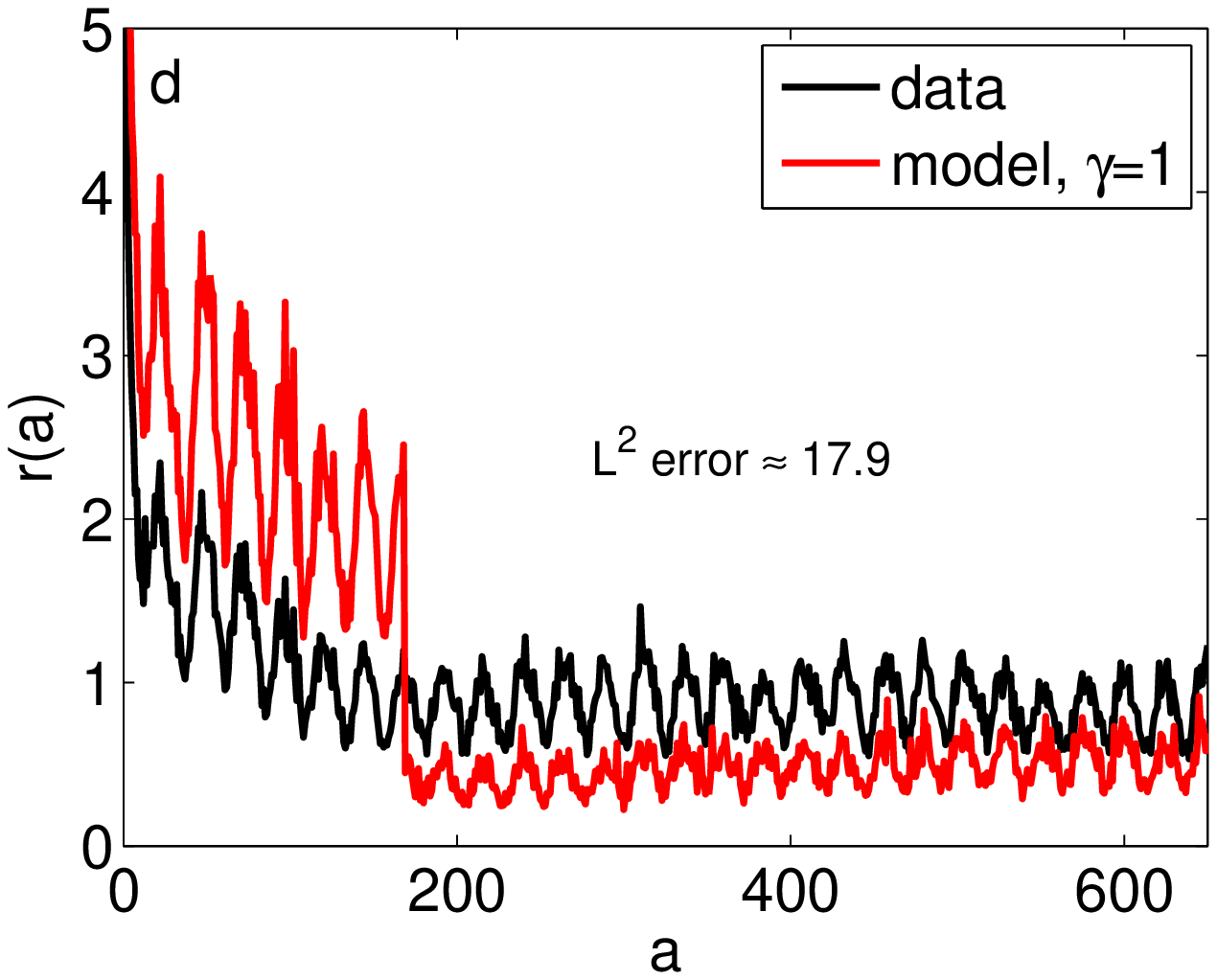,width=5.25cm}
\epsfig{figure=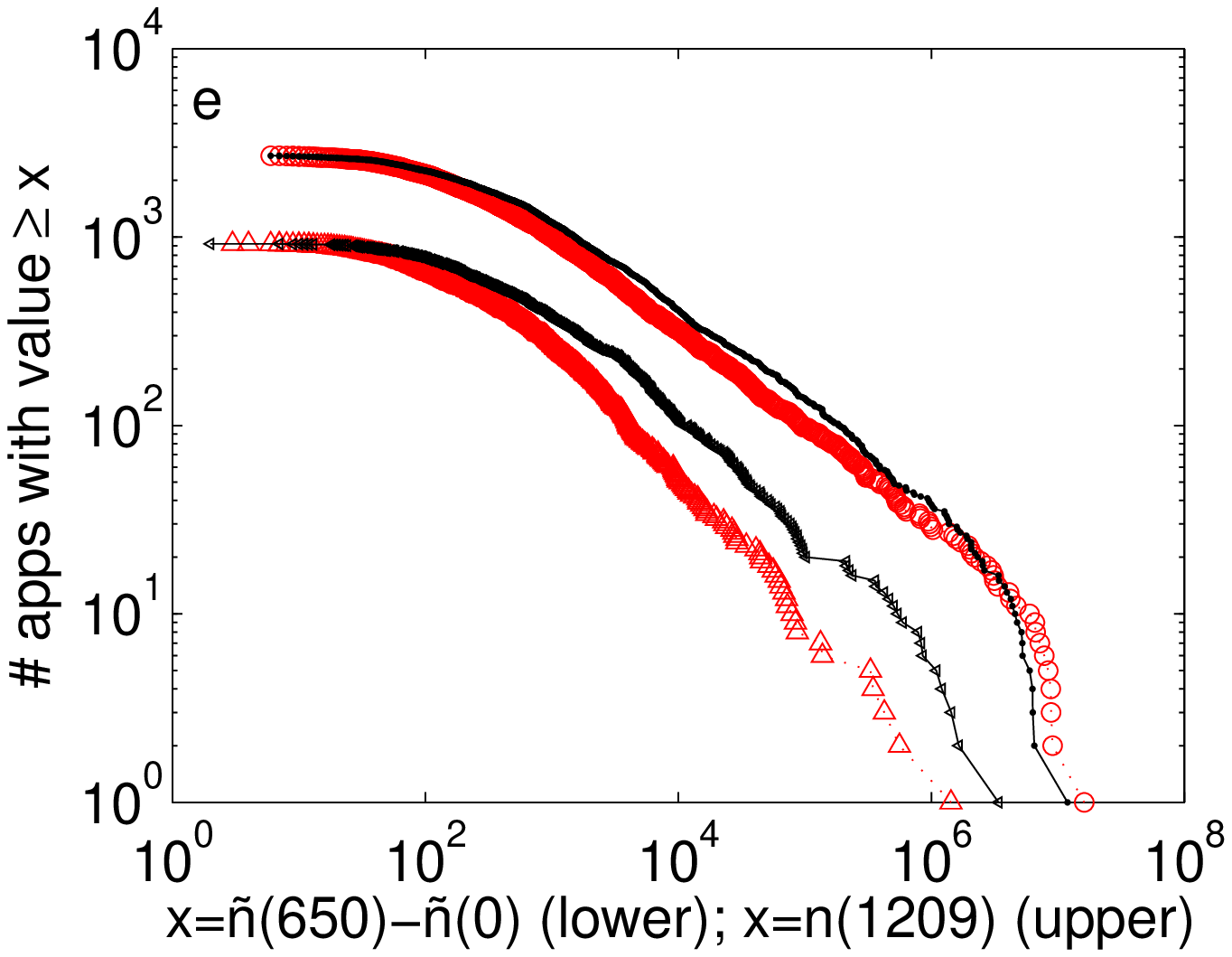,width=5.25cm}
\epsfig{figure=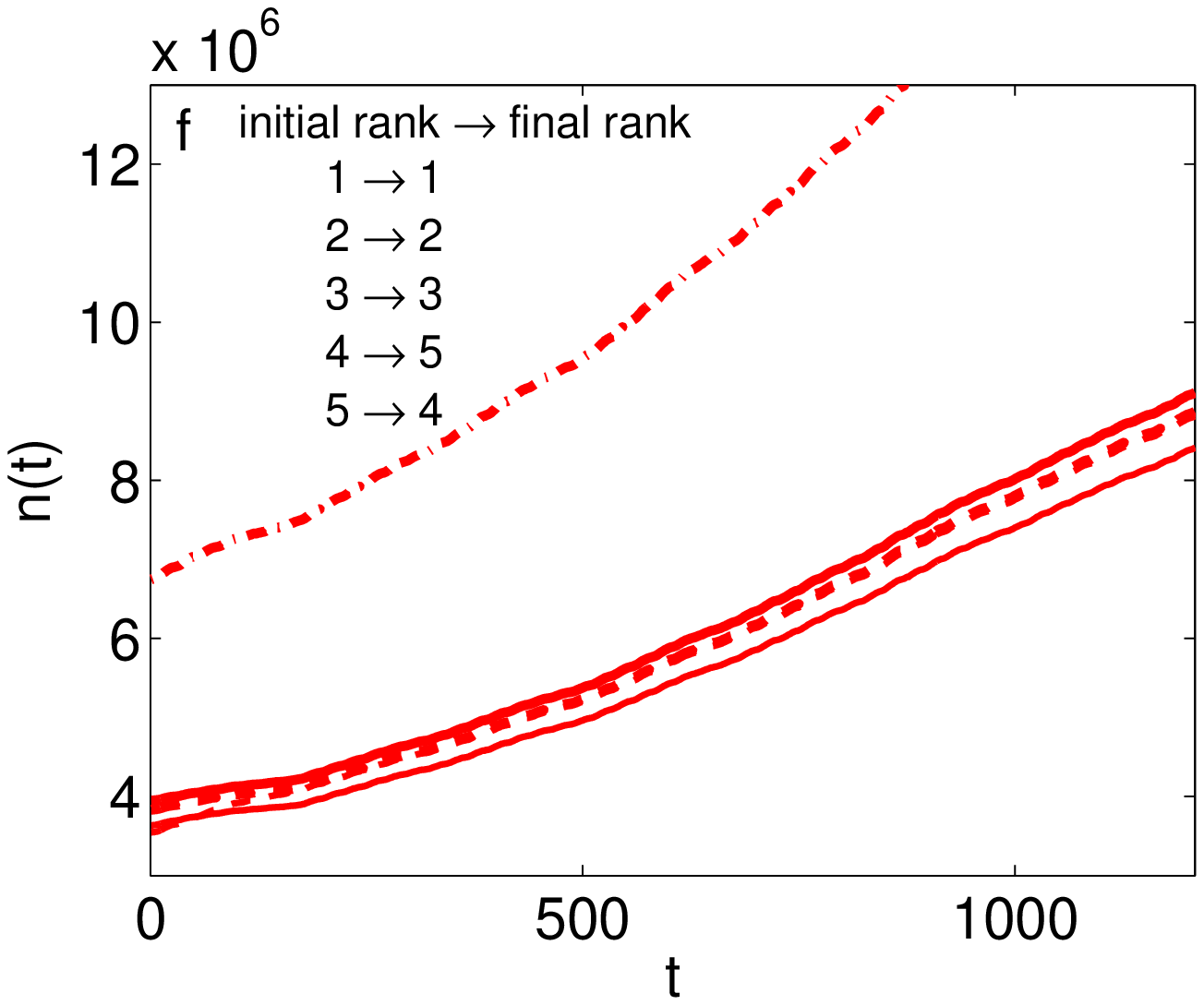,width=5.25cm} 
\epsfig{figure=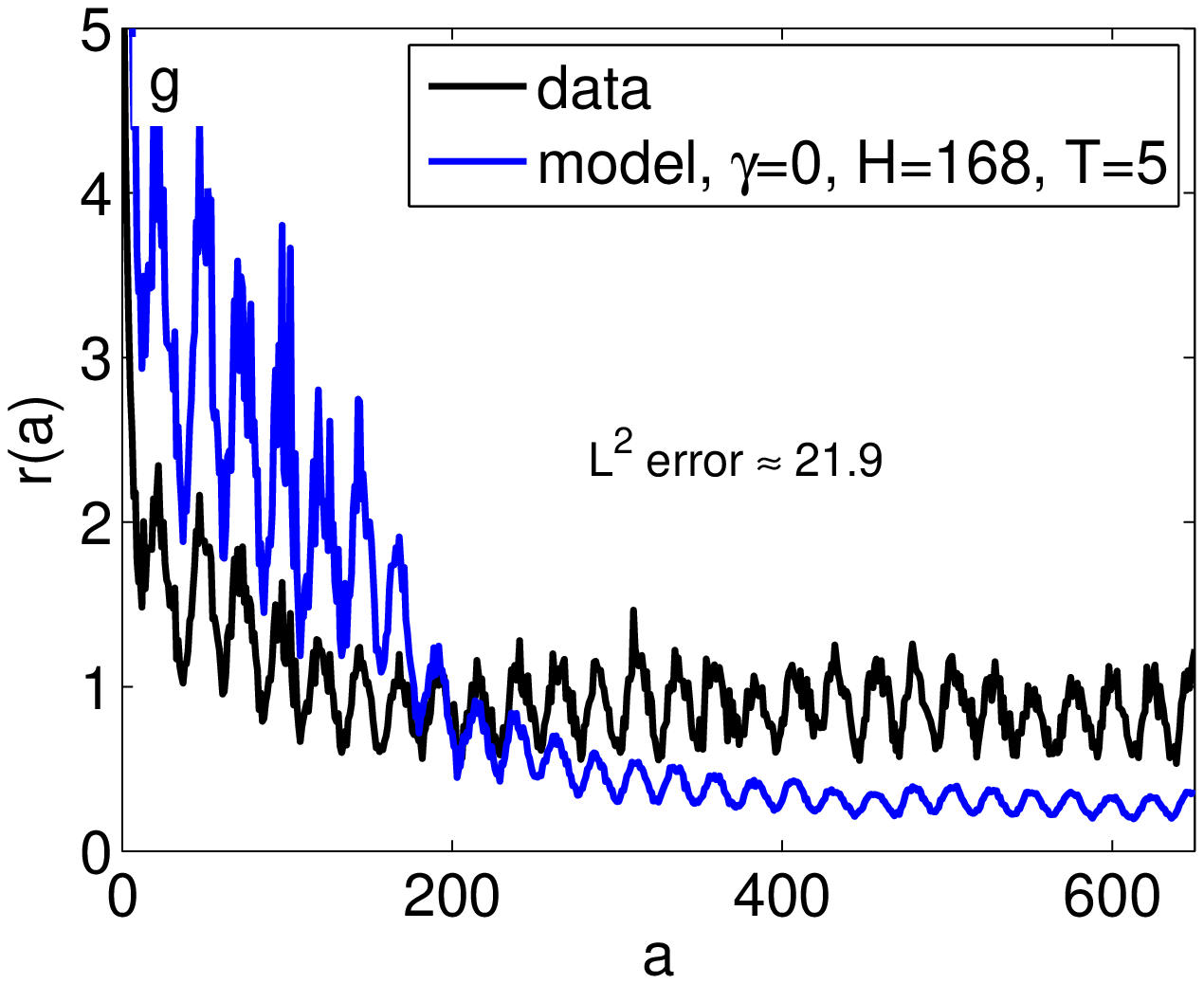,width=5.25cm}
\epsfig{figure=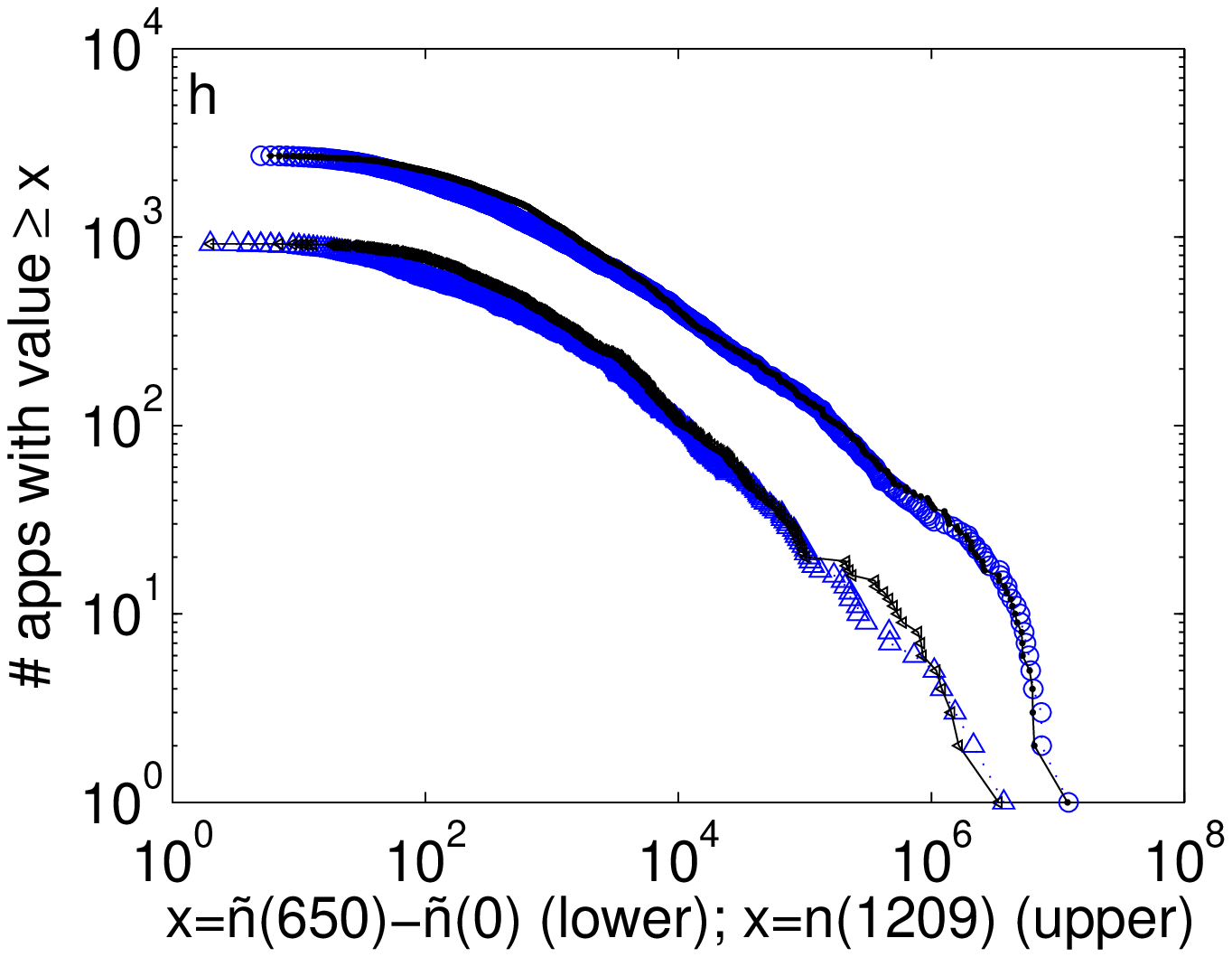,width=5.25cm}
\epsfig{figure=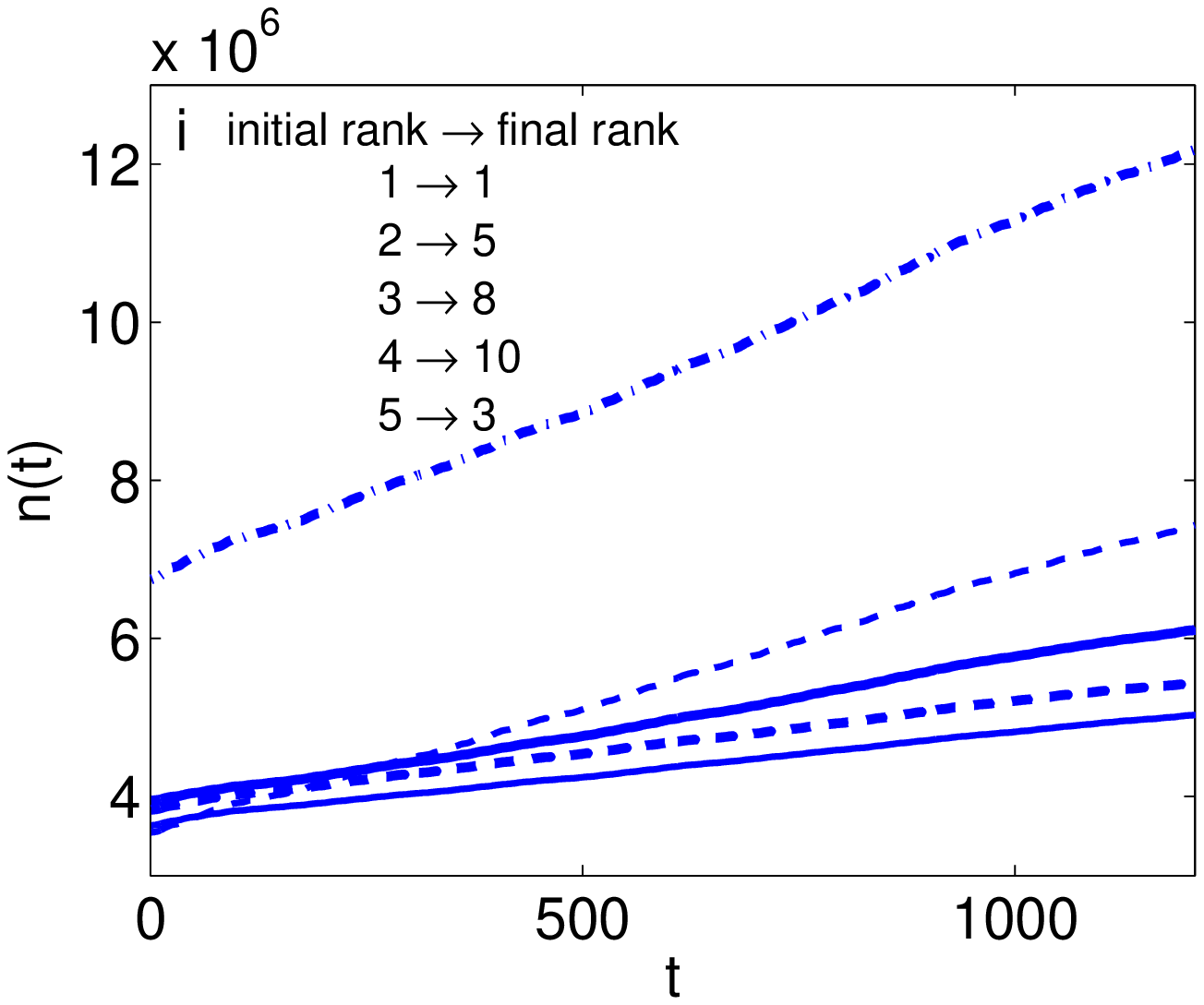,width=5.25cm} 
\epsfig{figure=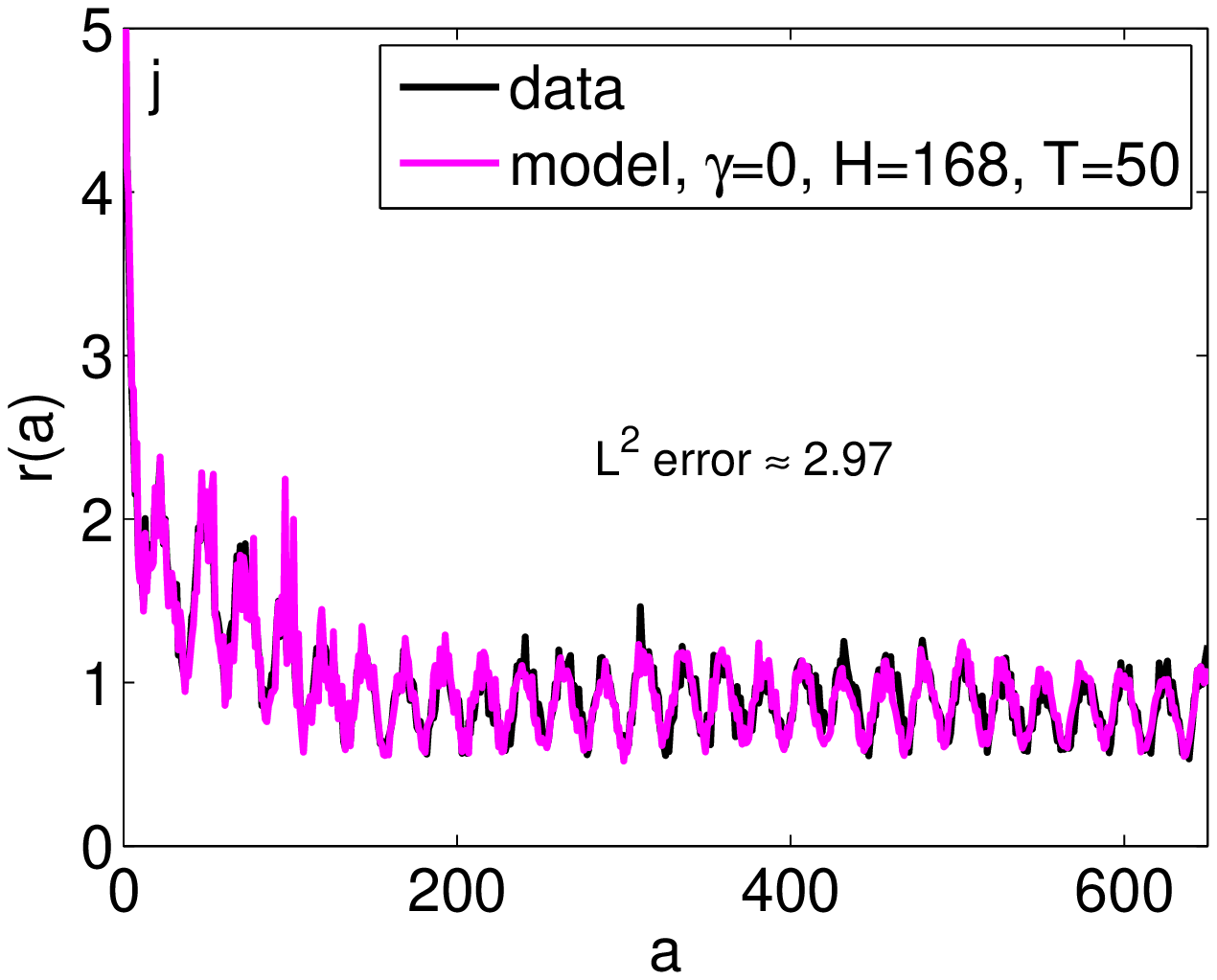,width=5.25cm}
\epsfig{figure=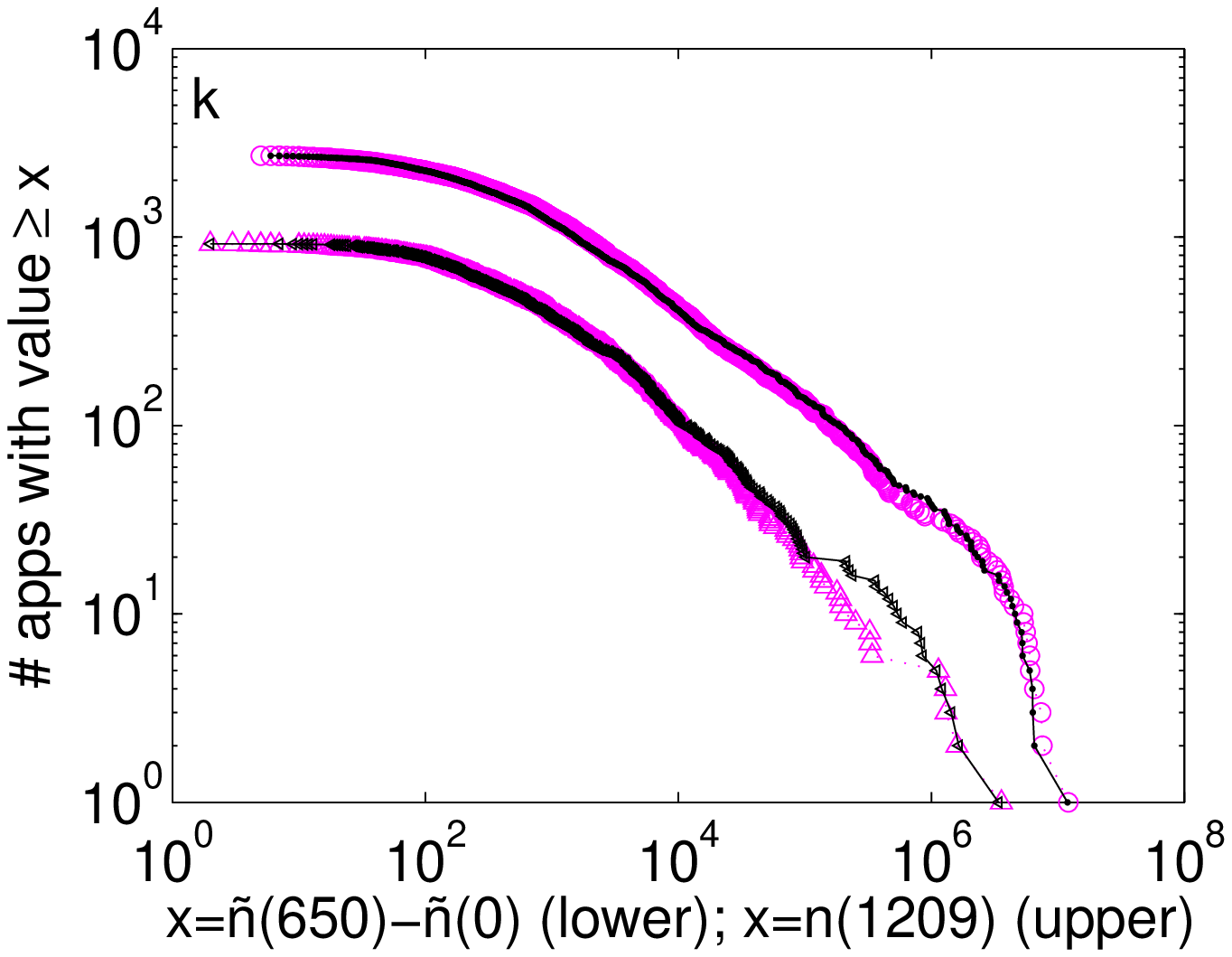,width=5.25cm}
\epsfig{figure=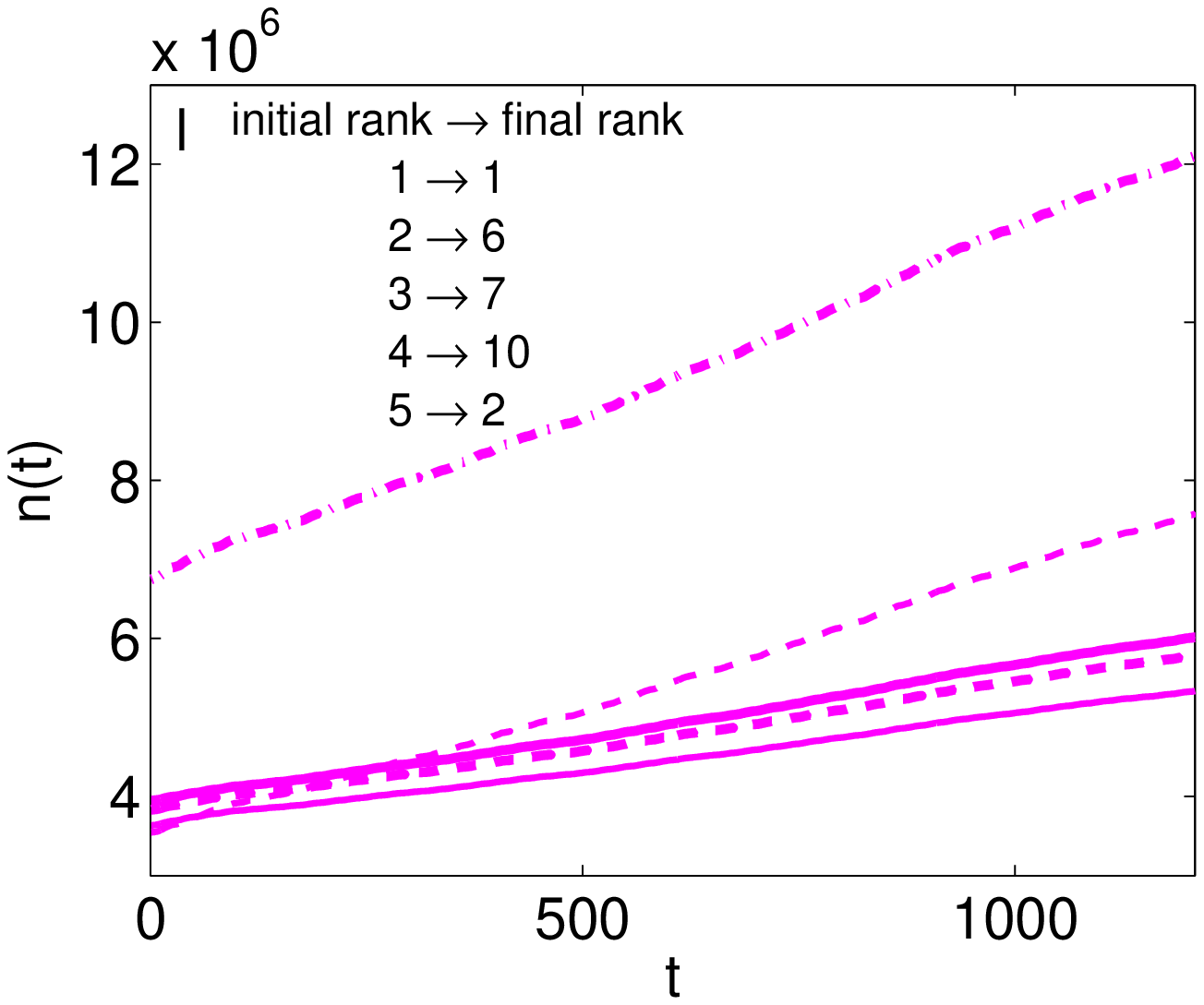,width=5.25cm} 
\caption{{ (Left column) Mean scaled age-shifted growth rate $r(a)$ , (center column) distributions of app popularity, and (right column) popularity over time for the top-5 apps, showing turnover.  {(a, b)} Behaviour of the entire LES set of applications and its two subsets (which are described in the text); (c) trajectories of the top-5 apps in the data set (ordered by popularity at $t=0$; note apps that were not in the $t=0$ top-5 are not shown here but can be seen in Fig.~S7).
{(d,e,f)} Cumulative-information model ($\gamma=1$), for which (e) shows  popularity distributions at $t = t_\text{max}$ (upper symbols) and for LES app growth to age $a = t_\text{LES}$ (lower symbols); empirical data is in black. {(g,h,i)} Recent-activity model with short memory ($\gamma=0$, $H=168$, $T=5$). {(j,k,l)} Recent-activity model with long-memory ($\gamma=0$, $H=168$, $T=50$).
}}
\label{fig2}
\end{figure}

The mean scaled age-shifted growth rate
reveals several interesting features (see Fig.~\ref{fig2}a). First, at large ages (e.g., $a \ge 150$ hours), the function $r(a)$ has
24-hour oscillations superimposed on a nearly constant curve. The behaviour of $r(a)$ is very different for smaller ages;
we dub this the \emph{novelty regime}, as it represents the (approximately one-week) time period that immediately follows the launch of apps. The $r(a)$ curve for the entire LES set is similar to those found by splitting the LES set into two disjoint subsets based on ordered launch times---the 460 applications with earlier launch times ($t_i \le 260$; early-launch) and the 461 applications with later launch times ($t_i \ge 261$; late-launch).
The small difference between the $r(a)$ curves for these cases
gives an estimate of the inherent variability within the data and sets a natural target for how well stochastic simulations can fit
the data{. We find similar results for other subsets of the same size (see SI3).}


To directly measure the growth of new apps in their first $t_\text{LES}$ hours, we show the distribution of $\tilde n_i(t_\text{LES})-\tilde n_i(0)$ for the entire LES set in Fig.~\ref{fig2}b.  We also show the corresponding distributions for the two LES subsets (early and late launch).
The similarity of distributions for early-born apps and late-born apps implies that { the launch time}, at least in the period that we examined, does not have a strong effect on the growth of new apps. This contrasts with Yule-Simon models of popularity \cite{Simon55,Cattuto07,Simkin11} and related preferential-attachment models used to model citations \cite{Redner98}. { In these models, early-born apps  have more time to accumulate popularity and hence  exhibit a different aging behaviour to later-born apps \cite{Simkin07}.}


{ In Fig.~\ref{fig2}c, we examine changes in the rank order of the top-5 list of apps by plotting the trajectories of the largest apps (ranked by their popularity at time $t=0$) over the duration of the study (and see Fig.~S7 for plots of top-10 lists). Reproducing realistic levels of turnover in such lists is a challenging test for models of popularity dynamics \cite{Bentley11,Evans11}.
}


The popularity dynamics for the novelty regime 
seem to be app-specific (see Figs.~\ref{fig2}a and S4), but a simple model can satisfactorily describe the post-novelty regime. We introduce a general stochastic simulation framework with a \emph{history-window parameter} $H$  and consider an app to be within its \emph{history window} for the first $H$ hours that data on the app is available.
 The history window of LES apps extends from their launch time to $H$ hours later; for non-LES apps, we define the history window to be the first $H$ hours ($t=0$ to $t=H$) of the study.
We conduct stochastic simulations by modelling
$F(t)$ computational ``agents'' in time step $t$, each of whom installs one app at that time step. { We take the values of $F(t)$ from the data [see Eq.(\ref{1})].} Note that our simulated agents do not correspond directly to Facebook users, as we do not have data at the level of individual users. In reality, a Facebook user can, for example, install several different apps during an hour; in our simulations, however, such actions would be modelled by the choices of several
agents.

We simulate the choices of the agents as follows. First, for any app $i$ that is in its history window at time $t$, we copy the increment $f_i(t)$ directly from the data. This determines the choices of $F_H(t)$ of the agents, where $F_H(t)$ is the number of installations of all apps that are within their history window at time $t$.
Each of the remaining $F(t)-F_H(t)$ agents then installs any one of the apps that are not in their history window. An installation probability $p_i(t)$ is allocated based on model-specific rules (see below), and the $F(t)-F_H(t)$ agents each independently choose app $i$ with probability $p_i(t)$. These rules ensure that the total number of installations in each hour exactly matches the data and that the history window of each app is reproduced exactly.

\begin{figure}
\centering
\epsfig{figure=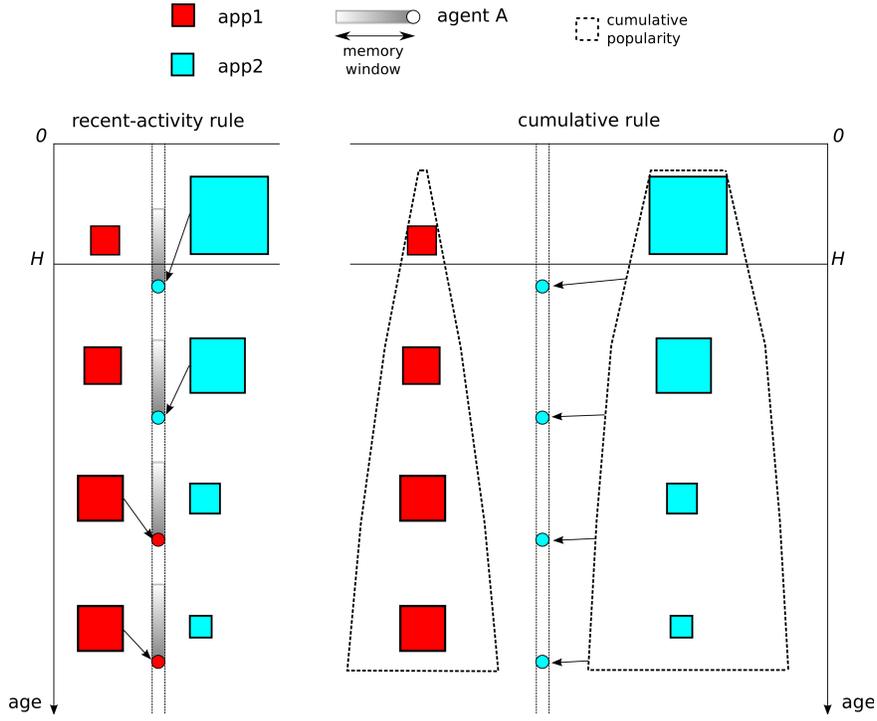,width=11.5cm}
\caption{{ Schematic of the model. The squares indicate the number of installations at time $t$ of two example apps; their size represents the number of installations of an app in a particular hour. Circles represent agents,
 and arrows indicate the adoption of an app.
 In the history window (ages $0$ to $H$), we copy the installation history directly from the data. Outside of the history windows, we simulate the actions of $F(t)$
agents by assigning probabilistic rules for how they choose which app to install.  An agent who uses (left) the \emph{recent-activity rule} at a given time
copies the choice of an agent who acted in the recent past, so apps that were recently more popular are more likely to be chosen.
By contrast, an agent who uses (right) the \emph{cumulative rule} at a given time installs the app with the larger number of accumulated installations. We represent this cumulative popularity using the dashed contour, which increases in width with time as more installations occur.}
}\label{cartoon}
\end{figure}

We investigate several possible choices for $p_i(t)$ by comparing the results of simulations with the characteristics of the data highlighted in Figs.~\ref{fig2}a,b,c. The history-window parameter $H$ plays an important role in capturing the app-specific novelty regime. However, if $H$ is very large, then most of the simulation is copied directly from the data and the decision probability $p_i(t)$ becomes irrelevant.  It is therefore desirable to find models that fit the data well while keeping $H$ as small as possible. Motivated by the information available to Facebook users during the data collection period, we propose a model based on a combination of a \emph{cumulative rule} $p_i^c(t)$ and a \emph{recent activity rule} $p_i^r(t)${. See the schematic in Fig.~\ref{cartoon}.}

An agent who uses the \emph{cumulative rule} at time $t$ chooses app $i$ with a probability proportional to its cumulative popularity $n_i(t-1)$, yielding
\begin{equation}
	p_i^c(t) = K \, n_i(t-1) \label{3}\,,
\end{equation}
where the constant $K$ is determined by the normalization $\sum_i p_i^c(t)=1$. In contrast, an agent who  follows the \emph{recent-activity rule} at time $t$ copies the installation choice of an agent who acted in an earlier time step, with some memory weighting (see Eq.~(\ref{4}) below).
Consequently, apps that were recently installed by many agents (i.e., apps with large $f_i(\tau)$ values for $\tau \approx  t$) are more likely to be installed at time step $t$ even if these apps are not yet globally popular (i.e., $n_i(t-1)$ can be small).
 In reality, the information available to Facebook users on the recent popularity of apps was limited to  observations of the installation activity of their network neighbours. As we lack any information on the real network topology, we make the simplest
  possible assumption: that the network is sufficiently well-connected (see  \cite{Traud12} for a study of Facebook networks from 2005) to enable all agents in the model to have information on the aggregate (system-wide) installation activity.
When applying the recent-activity rule, an agent chooses app $i$ with a probability proportional to the recent level of that app's installation activity:
\begin{equation}
	p_i^r(t) = L \,\sum_{\tau=0}^{t-1} W(t-\tau) f_i(\tau)\,, \label{4}
\end{equation}
where $L$ is determined by the normalization
$\sum_i p_i^r(t) =1 $. The
\emph{memory function} $W(\tau)$ determines the weight assigned to activity from $\tau$ hours ago and thereby incorporates human-activity timescales \cite{bursty}.
In the SI Appendix, we consider several examples of plausible memory functions and also examine the possibility of  heterogeneous app fitnesses.

If our data set included the early growth of every app, {then} a constant weighting function $W(t)\equiv 1$ would reduce $p_i^r$ to $p_i^c$. However, because of our finite data window, many apps have large values of $n_i(0)$, so we cannot capture the cumulative rule by using a suitable weighting function in the recent-activity rule. Instead, we introduce a tunable parameter $\gamma \in [0,1]$
so that
 the population-level installation probability $p_i$ used in the simulation is
a weighted sum,
\begin{equation}\label{weused}
 	p_i(t) = \gamma \,p_i^c(t) + (1-\gamma)\, p_i^r(t)\,,
\end{equation}	
that interpolates between the extremes of $\gamma=0$ (recent-activity rule) and $\gamma=1$ (cumulative rule). { The model ignores externalities between apps, an assumption  that is supported by the results of \cite{Onnela10}.
}

To explore our model, we start by considering the case $\gamma=1$, in which agents consider only cumulative information.  In Figs.~\ref{fig2}d,e,f, we compare the results of stochastic simulations with the data (see Figs.~\ref{fig2}a,b,c) using a history window of $H=168$ hours (i.e., 1 week). Clearly, the cumulative
model does not match the data well. Although the app popularity distributions at $t=t_\text{max}$ are reasonably similar
(see Fig.~\ref{fig2}e), the largest popularities are overpredicted by the model. By contrast,
the popularity of the LES apps---which include many of the less popular apps---is underpredicted. 
In particular, their mean scaled age-shifted growth rate has a lower long-term mean than that of the data  (see Fig.~\ref{fig2}d). { Recall from Eq.~(\ref{2}) that each app's increments are scaled by their temporal average $\tilde{\mu}_i$ before ensemble averaging to calculate $r(a)$. As a result, any error in predicting the value of $\tilde{\mu}_i$ has an effect on the entire $r(a)$ curve. This explains why, for example, the values of $r(a)$ for $a<H$ are overpredicted in Fig.~\ref{fig2}d, despite the fact that the increments in this regime are copied from the data. The corresponding temporal averages are too low, so the scaled increment values are too high. In Figure~\ref{fig2}f, we illustrate that the ordering among the top-5 apps does not change in time for this model, so it does not produce realistic levels of app-popularity turnover (see Figs.~\ref{fig2}c and S7). In SI6 and SI7, we demonstrate that several alternative models based on cumulative information also match the data poorly.
}

\begin{figure}
\centering
\epsfig{figure=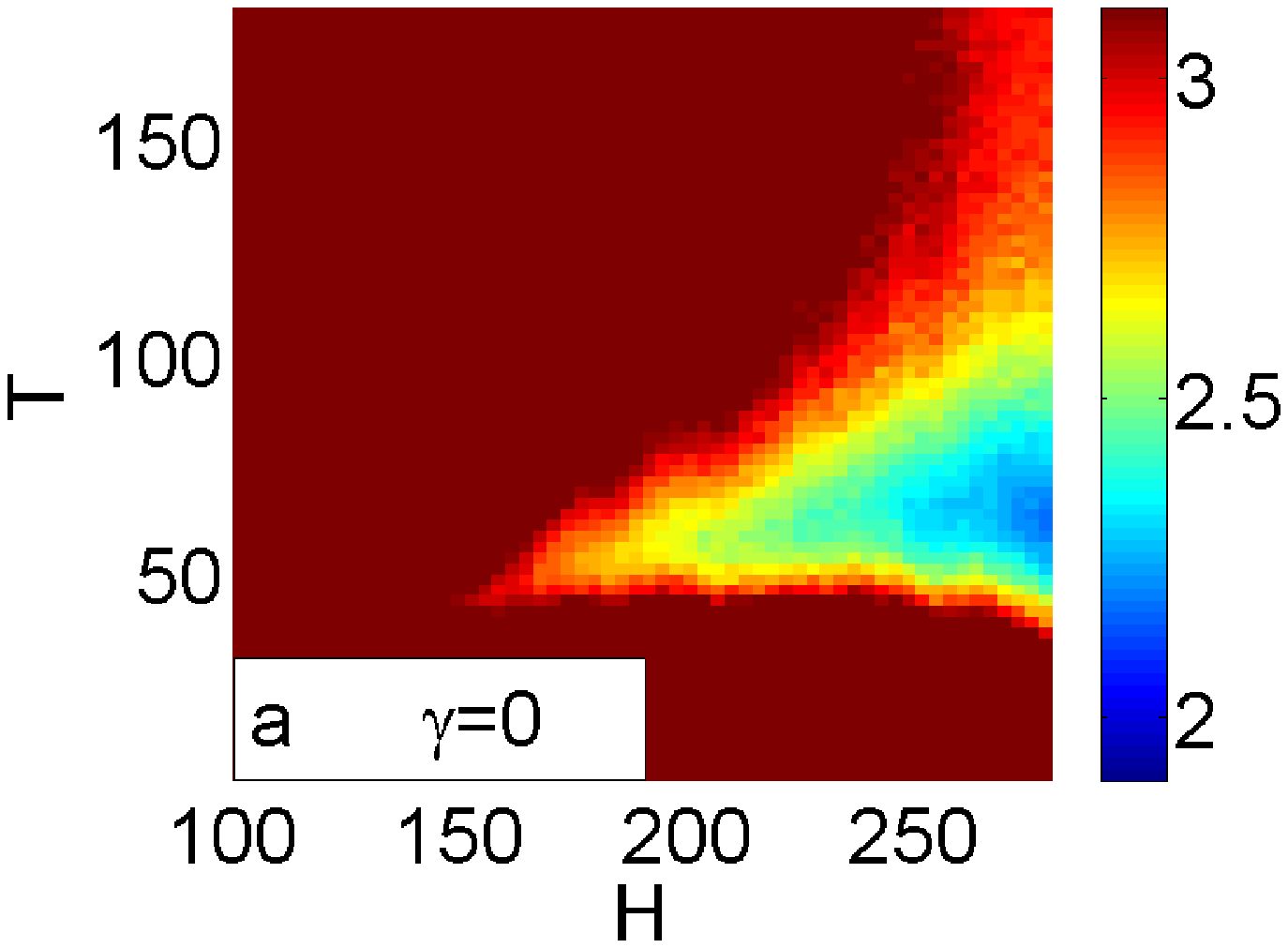,width=5.2cm}
\epsfig{figure=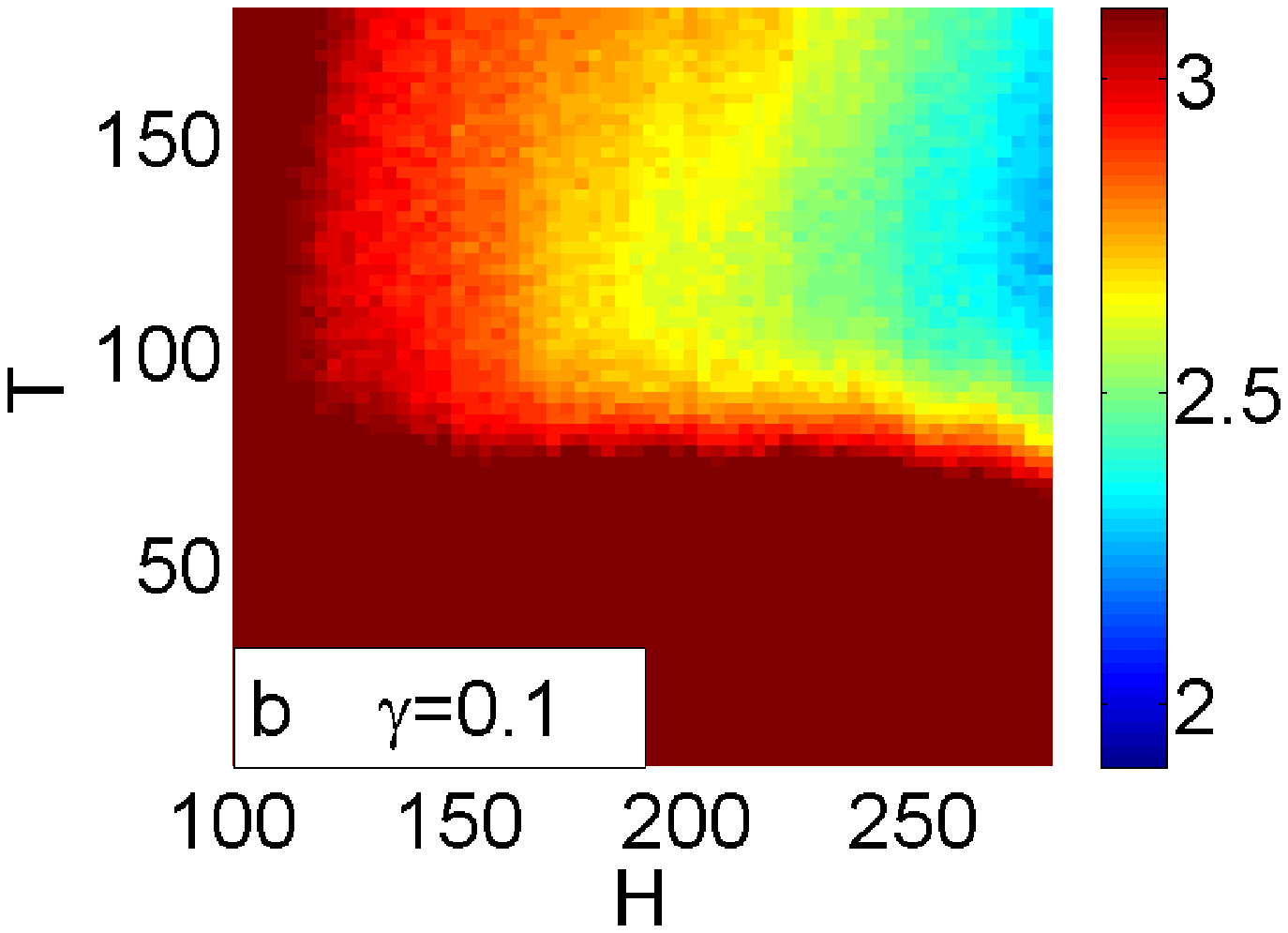,width=5.2cm}
\epsfig{figure=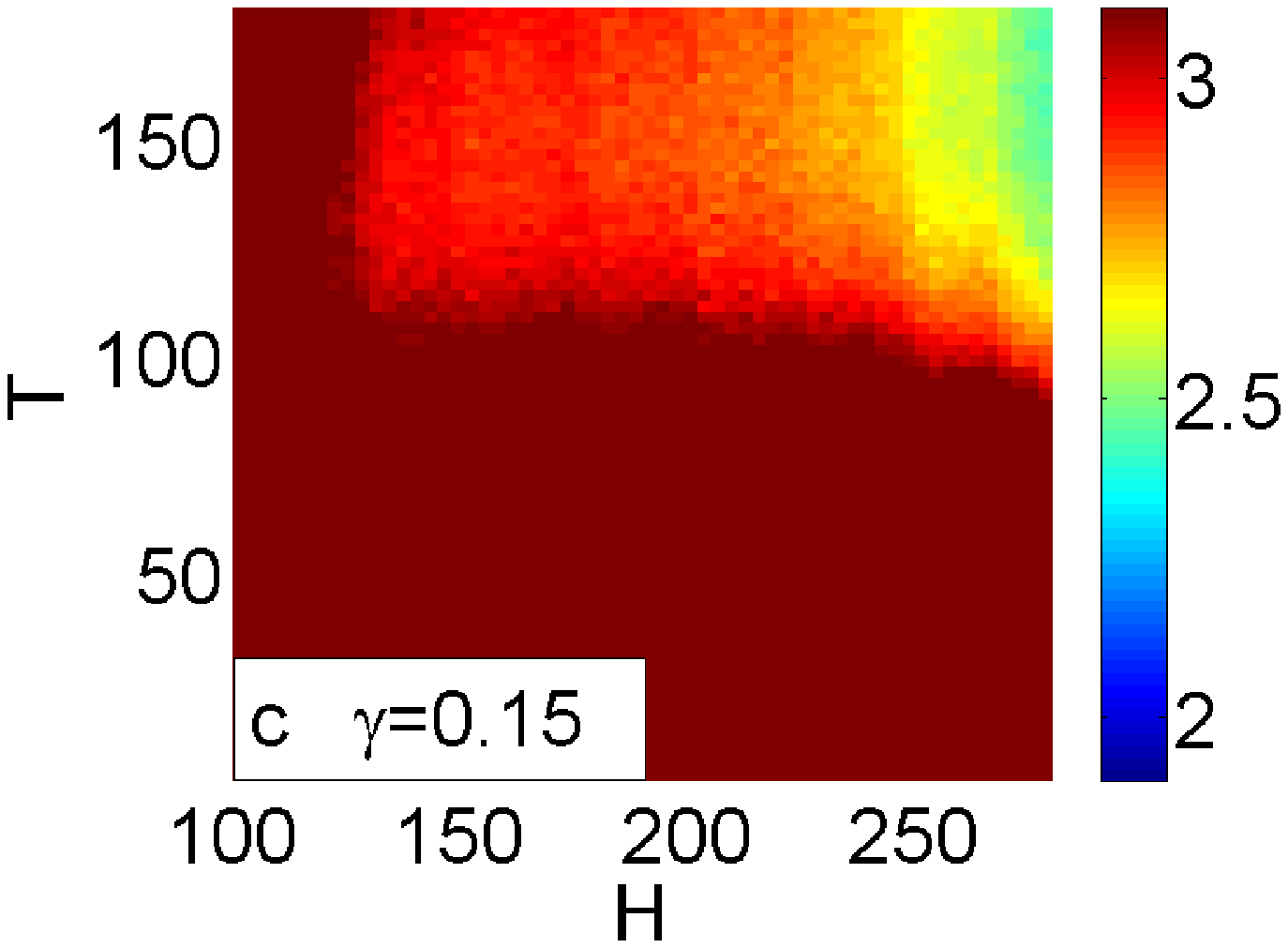,width=5.2cm}
\caption{Parameter planes showing the $L^2$ error (see SI3) for the $r(a)$ curve for the recent-activity-dominated model described in the text.  The parameter $H$ is the length of the history window, and $T$ is the mean of the exponential response-time distribution.  For each point in the plane, we average values of the $L^2$ error over 24 realizations. We show all values above 3.11 as dark red.
} \label{fig5}
\end{figure}

We next consider the case in which $\gamma$ is small, so recent information dominates
\cite{Bentleybook,Bentley11}. In Fig.~\ref{fig5}, we show results for stochastic simulations using an exponential response-time distribution $P(t)=\frac{1}{T}e^{-t/T}$  to determine the weights $W(t)$ assigned to activity from $t$ hours earlier for varying history-window lengths $H$ and response-time parameters $T$. The colours in the $(H,T)$ parameter plane represent the $L^2$ error, which is given by the $L^2$ norm of the difference between the simulated $r(a)$ curve and the $r(a)$ curve from the data.
 A value of 3.11 is representative of inherent fluctuations in the data (see SI3), and
 the bright colours in Fig.~\ref{fig5} represent parameter values for which the difference between the model's mean growth rate and the empirically observed growth rate is less than the magnitude of fluctuations present in the data. Observe that the model requires a history window of approximately 1 week (i.e., $H\approx 168$ hours) to match the data. As $\gamma$ increases, cumulative information is weighted more heavily, and the region of ``good-fit'' parameters moves towards larger $T$ and larger $H$ (see SI3). As noted previously, large-$H$ models trivially provide good fits (because they mostly copy directly from data), but the $\gamma=0$ case provides a good fit to the data even with a relatively short history window $H$.

In Figs.~\ref{fig2}g,h,i, we compare model results with data for parameter values $H=168$, $T=5$, and $\gamma=0$ (i.e., the ``recent-activity, short-memory'' case). This reproduces the app popularity distributions of the data rather well, but the mean scaled age-shifted growth rates are markedly different. In contrast, Figs.~\ref{fig2}j,k,l compare model results with data for parameter values $H=168$, $T=50$, and $\gamma=0$ (i.e., the ``recent-activity, long-memory'' case).
These parameters are just inside the good-fit region of Fig.~\ref{fig5}a, so the $r(a)$ curve in Fig.~\ref{fig2}j matches the data well. Moreover, the popularity distributions at $t=t_\text{max}$ and at age $t_\text{LES}$ (see Fig.~\ref{fig2}k) are both reasonably matched by the model{, which also allows realistic turnover in the top-10 list (see Figs.~\ref{fig2}l and S7).} These considerations highlight the importance of using temporal data to develop and fit models of complex systems. Distributions at single times can be insensitive to model differences, and the $r(a)$ curves are crucial for distinguishing between competing models. In SI4, we  show that the recent-activity ($\gamma=0$) case still gives good fits to the data if the exponential response-time distribution is replaced by a lognormal, gamma, or uniform distribution.

Another noteworthy feature of the recent-activity case is its ability to produce heavy-tailed popularity distributions in stochastic simulations even if no history is copied from the data ($H=0$). Even if all apps initially have the same number of installations,  random fluctuations lead to some apps becoming more popular than others, and the aggregate popularity distribution becomes heavy-tailed \cite{Bentley04,Ewens04,Evans07,Bentley11}. In SI5, we show that this situation is described by a near-critical branching process, for which power-law popularity distributions are expected \cite{Harrisbook,Zapperi95,Adami02,Goh03,Gleeson13}




{ Our model suggests that app adoption among Facebook users was guided more by recent popularity of apps (as reflected in installations by friends within 2 days) than by cumulative popularity.} The fact that the model is a near-critical branching process might help to explain the prevalence of heavy-tailed popularity distributions that have been
observed in information cascades on social networks, such as the spreading of retweets on Twitter \cite{Gonzalez11,Bakshy11,Lerman12} or news stories on Digg \cite{verSteeg11}. { The branching-process analysis is also applicable to the random-copying models of Bentley et al.~\cite{Bentley04,Bentley11,Bentleybook}. Although most random-copying models consider only short (e.g., single time-step) memory~\cite{Bentley04,Bentleybook}, the simulation study of Ref.~\cite{Bentley11} includes a uniform response-time distribution and demonstrates the role of memory effects in generating turnover.
As shown in Fig.~\ref{1} and detailed in SI7, generating realistic turnover of rank order in the top-10 apps is a significant challenge for all models based on cumulative information, even those that include a time-dependent decay of novelty \cite{Wu07,Wang13}. In SI9, we show that our model can also explain the results 
of the fluctuation-scaling analysis of the Facebook apps data in Ref.~\cite{Onnela10} that highlighted the existence of distinct scaling regimes (depending on app popularity).
}

Our approach also highlights the need to address temporal dynamics when modelling complex social systems.
Online experiments have been used successfully in computational social science \cite{comp-ss}, but it is challenging to run experiments in online environments that people actually use (as opposed to creating new online environments with potentially distinct behaviours). If longitudinal data is available, as in the present case, it is possible to evaluate a model's fit based not only on long-time behaviour but also on dynamical behaviour. Given that several models successfully produce similar long-time behaviour, the investigation of temporal dynamics is critical for distinguishing between competing models. As more observational data with high temporal resolution from online social networks becomes available, we believe that this modelling strategy, which leverages temporal dynamics, will become increasingly essential.

\subsection*{Acknowledgements} We thank Andrea Baronchelli, Ken Duffy, James Fennell, James Fowler, Sandra Gonz\'alez-Bail\'on, Stephen Kinsella, Jack McCarthy, Yamir Moreno, Peter Mucha, Puck Rombach, and Frank Schweitzer for helpful discussions. We thank the SFI/HEA Irish Centre for High-End Computing (ICHEC) for the provision of computational facilities. We acknowledge funding from Science Foundation Ireland grant 11/PI/1026 (JPG, DC), the FET-Proactive project PLEXMATH FP7-ICT-2011-8 grant \#317614 (JPG, DC, MAP), the FET-Open project FOC-II FP7-ICT-2007-8-0  grant \#255987 (FRT), the John Fell Fund from University of Oxford (MAP), and DeGruttola NIAID R01AI051164 (JPO).

\clearpage

\section*{SUPPLEMENTARY INFORMATION}

\renewcommand{\thefigure}{S\arabic{figure}}
\setcounter{figure}{0}
\renewcommand{\thetable}{S\arabic{table}}
\setcounter{table}{0}
\renewcommand{\theequation}{S\arabic{equation}}
\setcounter{equation}{0}


\section*{SI1: Data Cleaning and Aggregate Installation Activity}

The data was downloaded from Facebook for all existing 2720 applications (``apps") between 25 June 2007 (shortly after applications were introduced) and 14 August 2007 \cite{Onnela10}.  The data consists of time series $n_i(t)$, where $i \in \{1,2,\ldots,2720\}$, discrete time is indexed by the (real-time) hour $t \in \{0,1,2, \ldots, 1209\}$, and $n_i(t)$ corresponds to the aggregate number of users who have application $i$ installed at time $t$. Data for 15 applications was corrupted, so we omitted these from our investigation and examined a total of $N = 2705$ applications. This data covers 100\% of the population of 50 million potential
app adopters and about 99\% (2705 of 2720) of all applications that could be adopted. This thereby gives an almost complete view of system-wide adoptions during the time period of the data collection. We define the \emph{launch time} $t_i$ of app $i$ as the smallest value of $t$ for which $n_i(t)>0$, and we define the \emph{increment} in hour $t$ for app $i$ to be $f_i(t) = n_i(t) - n_i(t-1)$.

\begin{figure}
\centering
\epsfig{figure=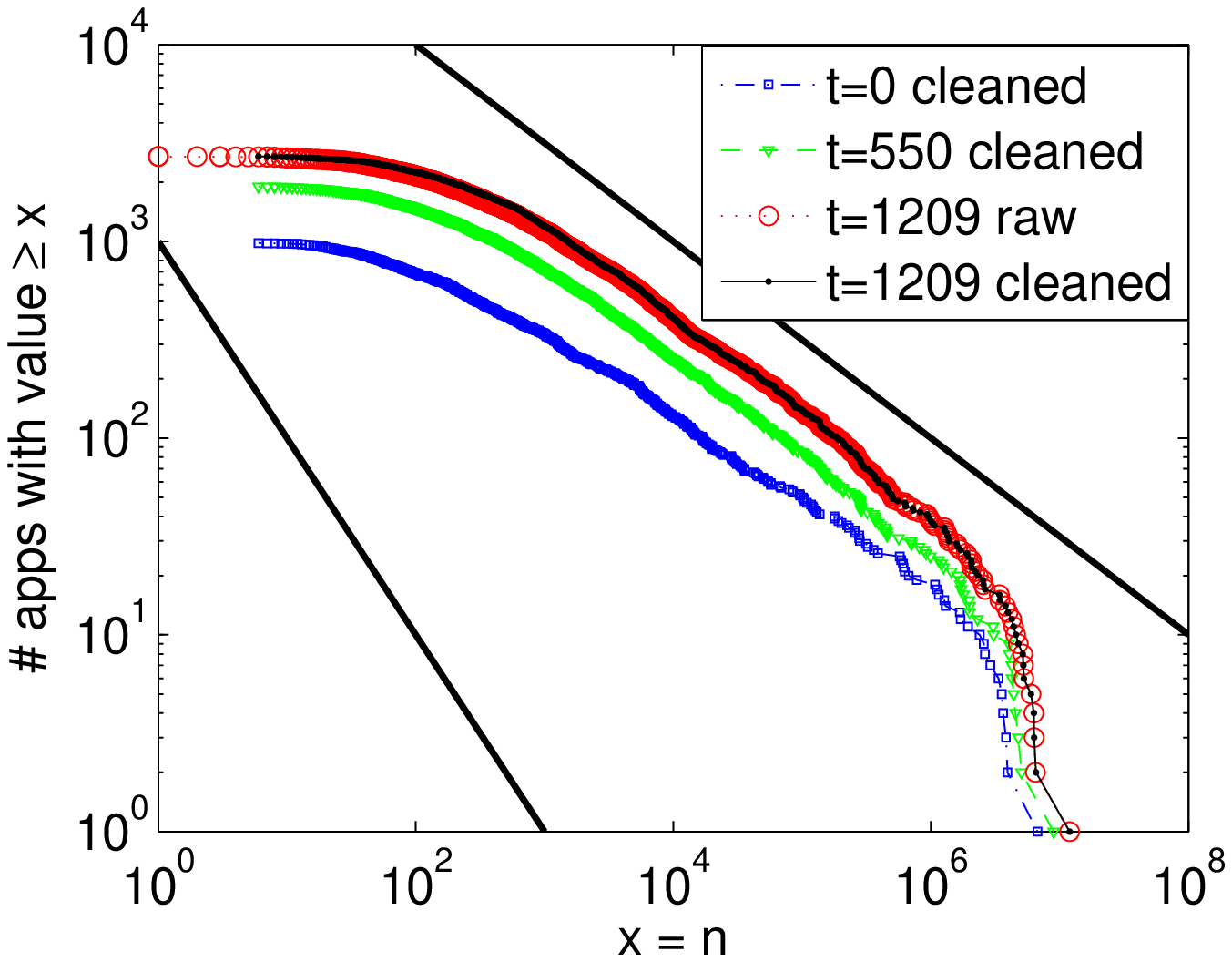,width=8cm}
\caption{Distribution functions showing the number of applications (``apps") with popularity greater than or equal to $n$ at times $t=0$, $t=550$, and $t=t_\text{max}$. Red circles are from the raw data (i.e., prior to the cleaning process that we describe in Section SI1). The straight lines indicate the scalings corresponding to probability distribution functions with scalings $P(n)\sim n^{-\alpha}$ with $\alpha=3/2$ (upper line) and $\alpha=2$ (lower line).
}
\label{fig1}
\end{figure}

The data-cleaning process involves removing any undefined values within the data and imputing replacement values. For each app $i$, if $f_i(t)$ is undefined for $t>t_i$, then we copy the most recent well-defined increment value for app $i$ into $f_i(t)$. A second cleaning step entails removing negative values of $f_i(t)$. Such values correspond to the (rare) cases in which deinstallations exceeded installations of an app in a given hour. We do this by setting any instances in the data with $f_i(t)<0$ to $f_i(t)=0$. The effects of the data cleaning are small in the context of the aggregate statistical characteristics of the data.
In Fig.~\ref{fig1},  the distribution function of the popularity at $t=t_\text{max}$ for the cleaned data is shown in black.  We show the corresponding function that uses the raw (pre-cleaning) $n_i(t)$ time series as red circles. The two distributions are almost indistinguishable, except for the smallest (i.e., least popular) apps, indicating that the cleaned data is very similar to the original data.

\begin{figure}
\centering
\epsfig{figure=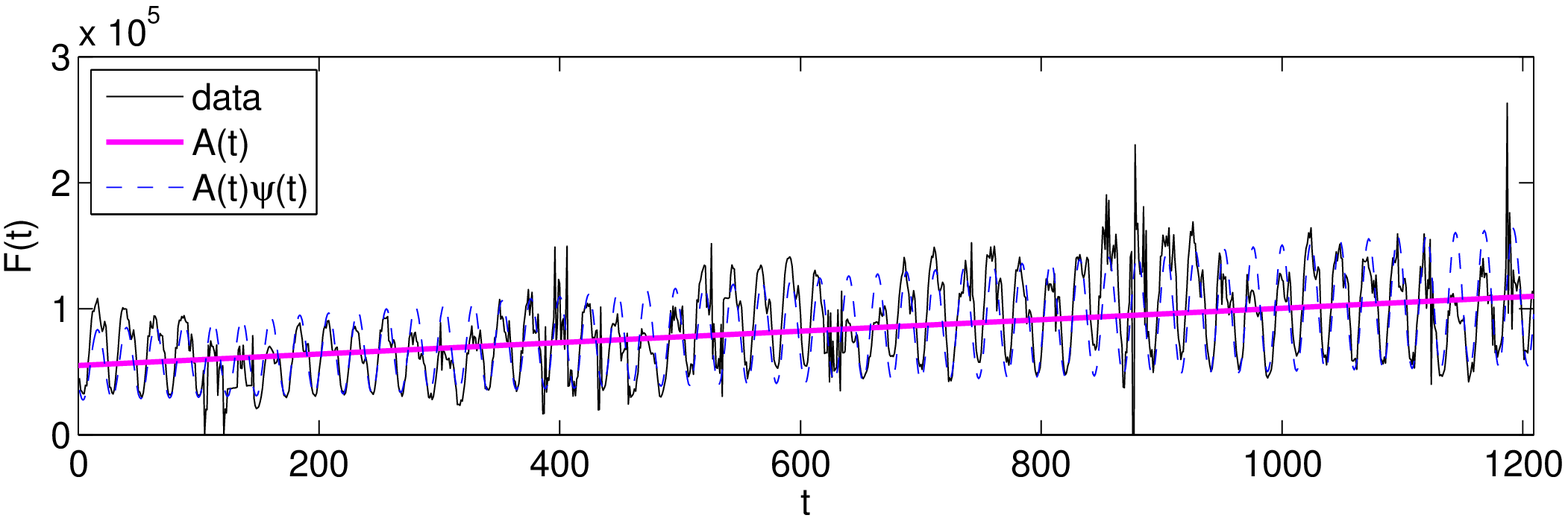,width=15cm}
\caption{Total installation activity in hour $t$, as defined in Eq.~(1) in the main text. We describe the linear growth function $A(t)$ and the 24-hour oscillation function $\psi(t)$ in Section SI1.
} \label{figSI1}
\end{figure}

Figure~\ref{fig1} shows that the popularities $n_i(t)$ of the apps cover a range of scales from very small to extremely popular and that the distribution of $n_i$ values is heavy-tailed. In Fig.~\ref{figSI1}, we show the total app installation activity $F(t)$, which is defined by Eq.~(1) of the main text, of Facebook users during hour $t$. This function exhibits slow growth and 24-hour oscillations. We highlight these features by also plotting a linear growth function $A(t) =  c_1 +c_2 t$ and (as a guide to the eye) a growing oscillation $A(t)\psi(t)$, where $\psi(t)= 1+0.5 \cos(2\pi(t+8)/24)$ gives the oscillatory part of the function. Least-squares fitting gives $c_1 \approx 5.5\times 10^4$ and $c_2 \approx 49$.  Thus, by $t=t_\text{max}$ (i.e., the end of the data-collection period), the mean hourly installation rate is approximately twice as large as it was at $t=0$.



\section*{SI2: Top Ten Launched-Early-in-Study (LES) Apps }

\begin{figure}
\centering
\epsfig{figure=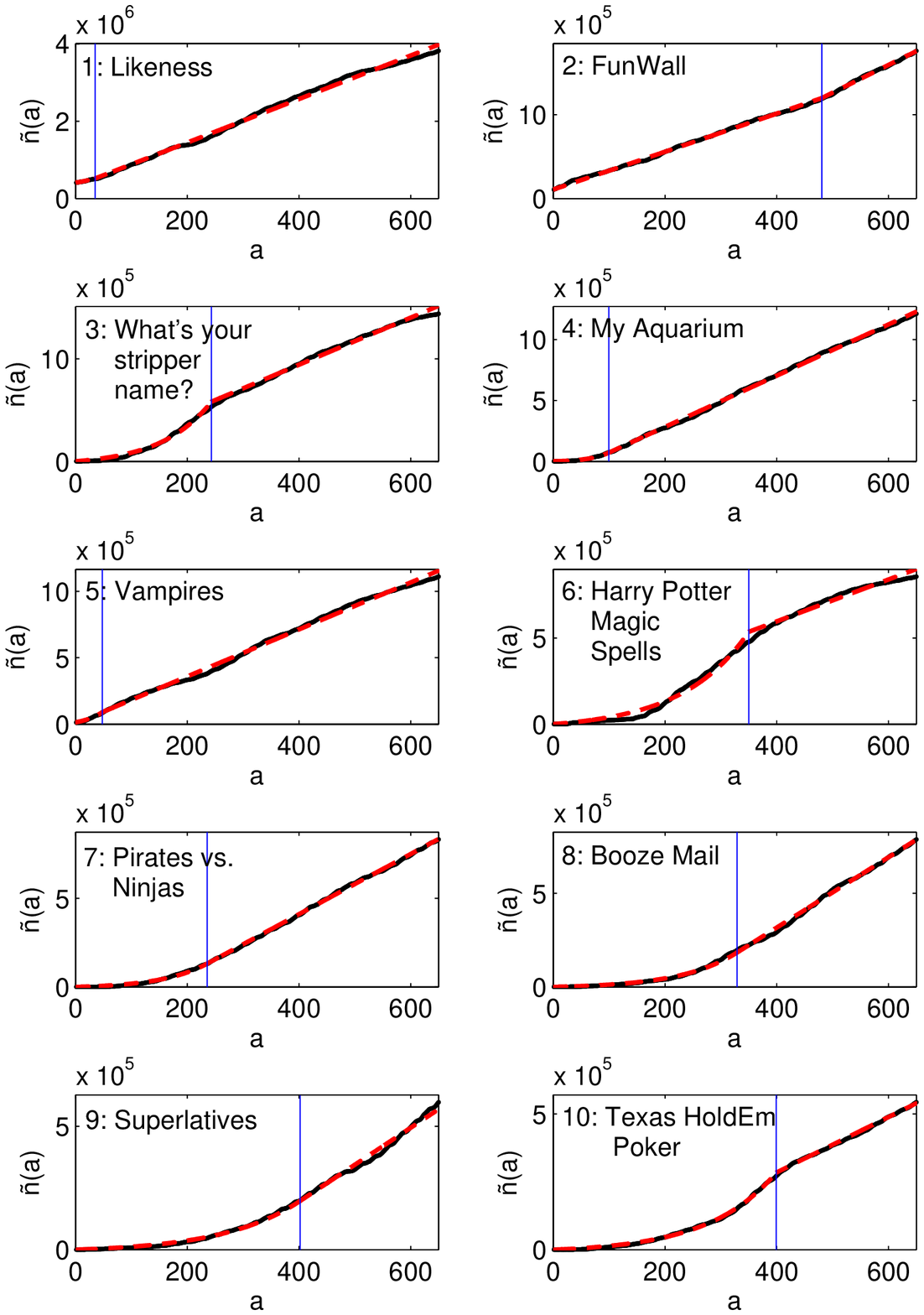,width=15cm}
\vspace{-1cm}
\caption{Growth trajectories of the 10 most popular LES apps, which we order according to their popularity when their age is $t_\text{LES}$ hours. The data values are in black, and red dashed curves show the fitting function $m(a)+\tilde n_i(0)$ described in Eq.~(\ref{fitting}) with parameter values from Table~\ref{tabSI1}{. The vertical lines mark the age $\theta$ at which the fitting function in Eq.~(\ref{fitting}) changes from an exponential to a linear function.}
} \label{figSI2}
\end{figure}

In Fig.~\ref{figSI2}, we show the ten most popular Launched-Early-in-Study (LES) apps.  We order them by $\tilde n_i(t_\text{LES})$, which denotes the number of installations by age $t_\text{LES}\equiv 650$.  To highlight common features of app growth, we use the heuristic fitting function
\begin{align}\label{fitting}
	m(a) = \left\{ \begin{array}{cc}
	C(e^{D a}-1)\,, &  \text{ if } a\le \theta,\\
	C(e^{D \theta}-1)+ E (a-\theta)\,,  &  \text{ if } a > \theta\,,
\end{array} \right.
\end{align}
where the parameters $C$, $D$, $E$, and $\theta$ are determined by least-squares fitting of $m(a)$ to $\tilde n_i(a)-\tilde n_i(0)$ for each app $i$.  We give the values of these parameters in Table~\ref{tabSI1}. The parameter values that we obtain are sensitive to the initial guesses that are used in the fitting routine, but it is nevertheless clear that most apps exhibit exponential growth in a \emph{novelty regime} (i.e., when age $a < \theta$) followed by linear growth at later ages (i.e., $a>\theta$).\footnote{The notable exception among the top 10 in terms of fitting quality is Harry Potter Magic Spells (the 6th most popular app).}

\begin{figure}
\centering
\epsfig{figure=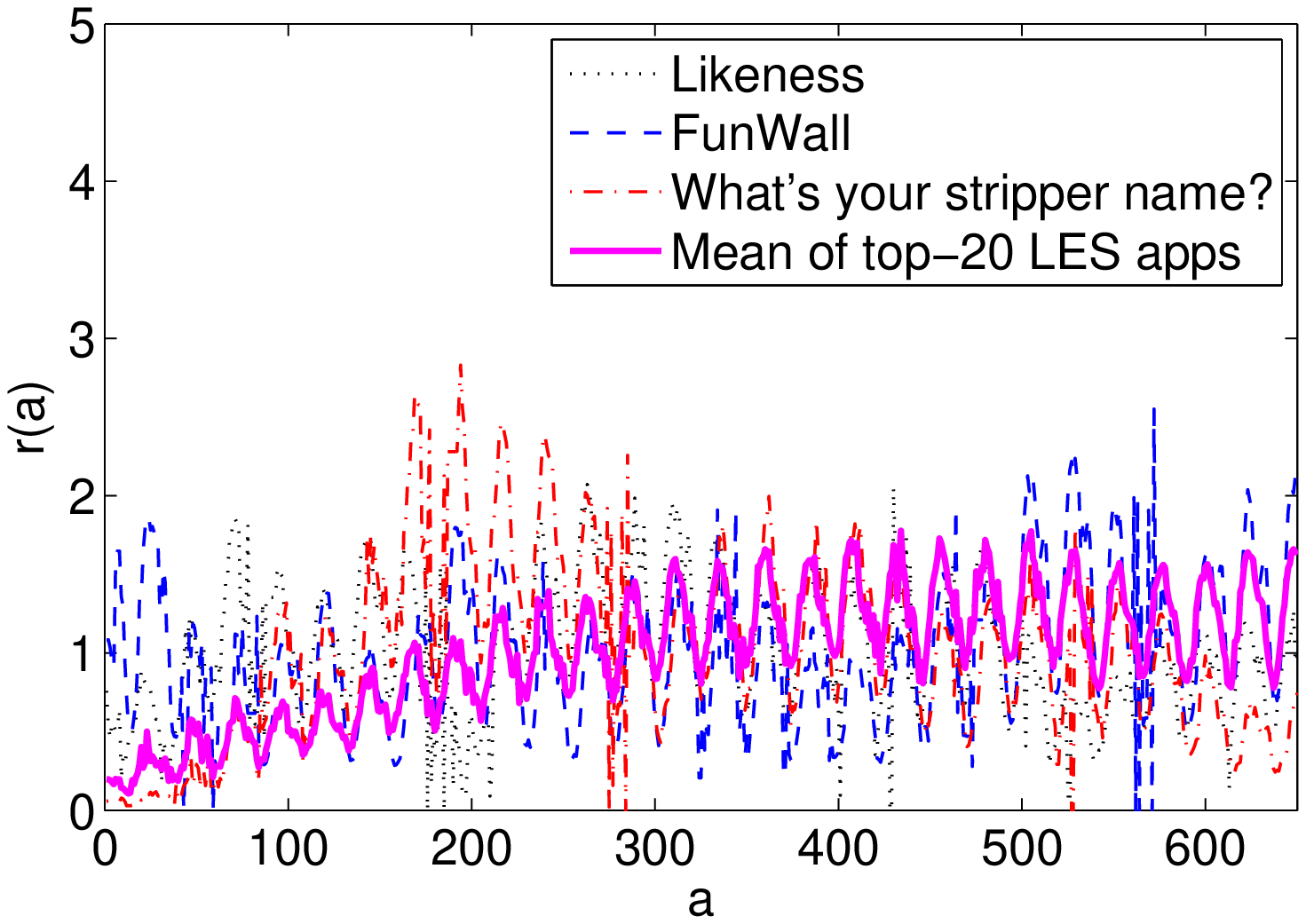,width=11cm}
\caption{Scaled age-shifted growth rate functions $\tilde f_i(a)/\tilde \mu_i$ for the three most popular LES apps and the mean scaled growth rate of the 20 most popular LES apps.
} \label{figSI3}
\end{figure}

\begin{table}
\begin{center}
\begin{tabular}{|c|c|c|c|c|c|}
\hline
Rank & Name & $C$ & $D$ & $E$ & $\theta$\\
\hline \hline
1 & Likeness & $1.59\times 10^5$ & $1.48\times 10^{-2}$& $5.63\times 10^{3}$& $35$ \\ \hline
2 & FunWall & $4.25\times 10^7$ & $5.26\times 10^{-5}$& $3.31\times 10^{3}$& $481$ \\ \hline
3 & What's your stripper name? & $4.10\times 10^4$ & $1.12\times 10^{-2}$& $2.30\times 10^{3}$& $243$ \\ \hline
4 & My Aquarium & $4.44\times 10^3$ & $2.85\times 10^{-2}$& $2.10\times 10^{3}$& $99$ \\ \hline
5 & Vampires & $3.14\times 10^4$ & $2.58\times 10^{-2}$& $1.77\times 10^{3}$& $48$ \\ \hline
6 & Harry Potter Magic Spells & $3.10\times 10^4$ & $8.30\times 10^{-3}$& $1.22\times 10^{3}$& $350$  \\ \hline
7 & Pirates vs.~Ninjas & $7.65\times 10^3$ & $1.23\times 10^{-2}$& $1.70\times 10^{3}$& $236$ \\ \hline
8 & Booze Mail & $7.22\times 10^3$ & $9.95\times 10^{-3}$& $1.88\times 10^{3}$& $329$ \\ \hline
9 & Superlatives & $1.14\times 10^4$ & $7.24\times 10^{-3}$& $1.50\times 10^{3}$& $402$ \\ \hline
10 & Texas HoldEm Poker & $1.10\times 10^4$ & $8.24\times 10^{-3}$& $1.02\times 10^{3}$& $399$\\
 \hline
\end{tabular}
\end{center}
\caption{Parameter values for the fitting functions $m(a)$ used in Fig.~\ref{figSI2}.}
\label{tabSI1}
\end{table}

In Fig.~\ref{figSI3}, we show the scaled age-shifted growth rates $\tilde f_i(a)/\tilde\mu_i$ for the three most popular LES apps and the mean scaled age-shifted growth rate $r(a)$ (as defined in Eq.~(2) of the main text) for the set of top-20 LES apps. At large values of $a$, the function $r(a)$ is qualitatively similar to that of the full LES set in Fig.~1a in the main text, as it exhibits a ``quasi-stationary" (i.e., constant plus 24-hour oscillations) behaviour. However, the small-$a$ novelty regime is different in the two cases; this reflects differences in early-stage growth patterns. In particular, the most popular apps exhibit steadily growing popularity during the novelty regime. This is consistent with the exponential growth in Fig.~\ref{figSI2}, but it contrasts with the decrease in novelty experienced by the majority of apps in their early stages (and reflected in the $r(a)$ curve in Fig.~1a of the main text).




\section*{SI3: Further Information on Figure 1 of the Main Text}
\subsection*{SI3.1 Discussion of the $L^2$ Error in the Mean Scaled Age-Shifted Growth Rate}

In Fig.~1a of the main text, we saw that the mean scaled age-shifted growth rate $r(a)$ for the entire LES set is similar to the corresponding $r(a)$ curves that we obtained by splitting the LES set into two disjoint subsets: the early-launch subset and the late-launch subset. To quantify the level of inherent diversity within the data, we calculate the $L^2$ norm of the difference between the $r(a)$ curves and call this the \emph{$L^2$ error} of the partition:
\begin{equation}
	E_{L^2} = \sqrt{ \sum_{a=1}^{t_\text{LES}}\left(r_\text{LES}(a)-r_\text{subset}(a)\right)^2 }\,.
\end{equation}
For the aforementioned subsets, we find that the $L^2$ error is less than 3.11, and we take this value to represent a natural target for how well stochastic simulations can be fit to the data.
In Fig.~3 of the main text, we showed all $L^2$ error values above 3.11 as dark red, and we concentrated on the light-coloured regions of the $(H,T)$ parameter plane, as these constitute the locations where high-quality fits are possible.

\begin{figure}
\centering
\epsfig{figure=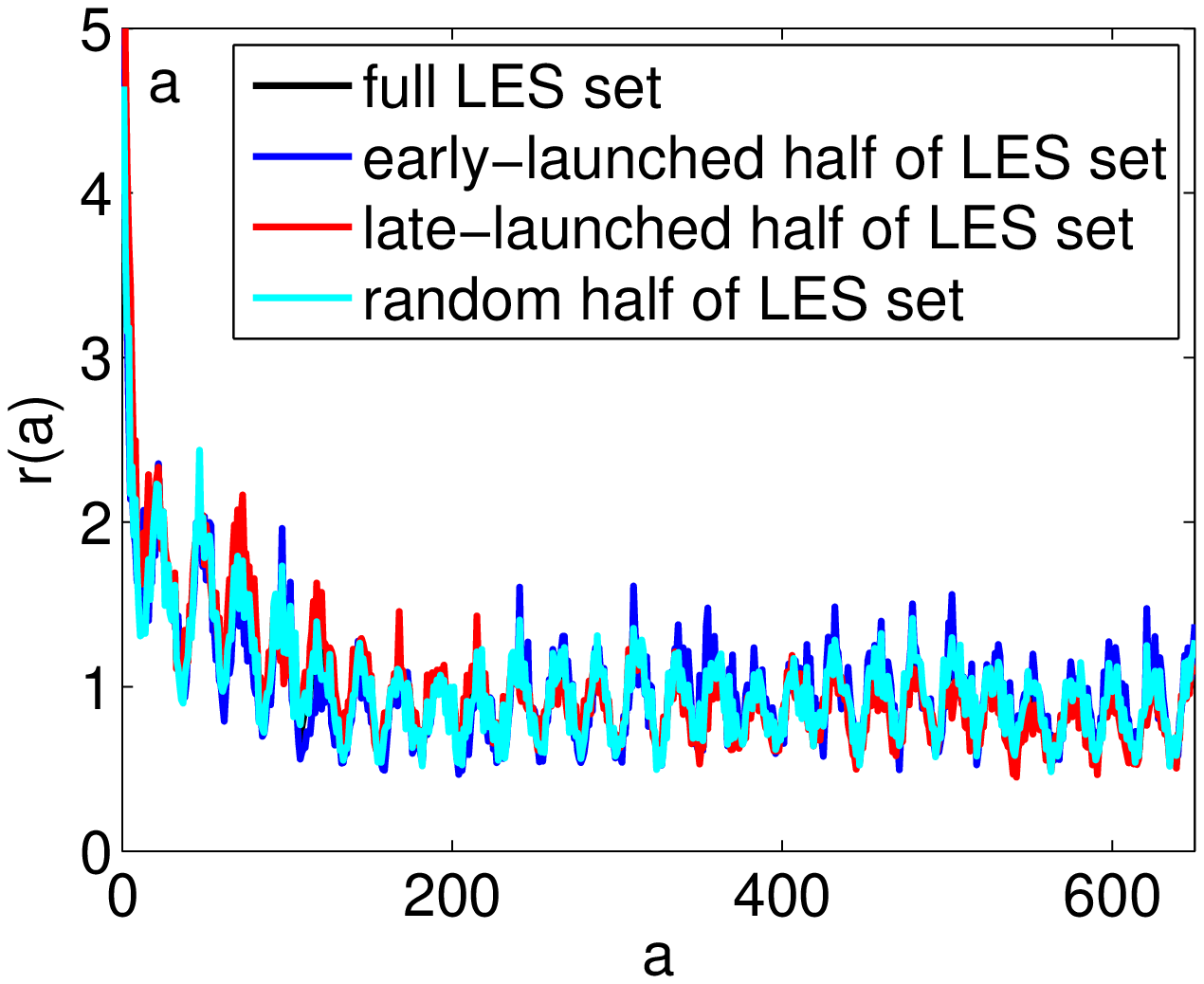,width=8cm}
\epsfig{figure=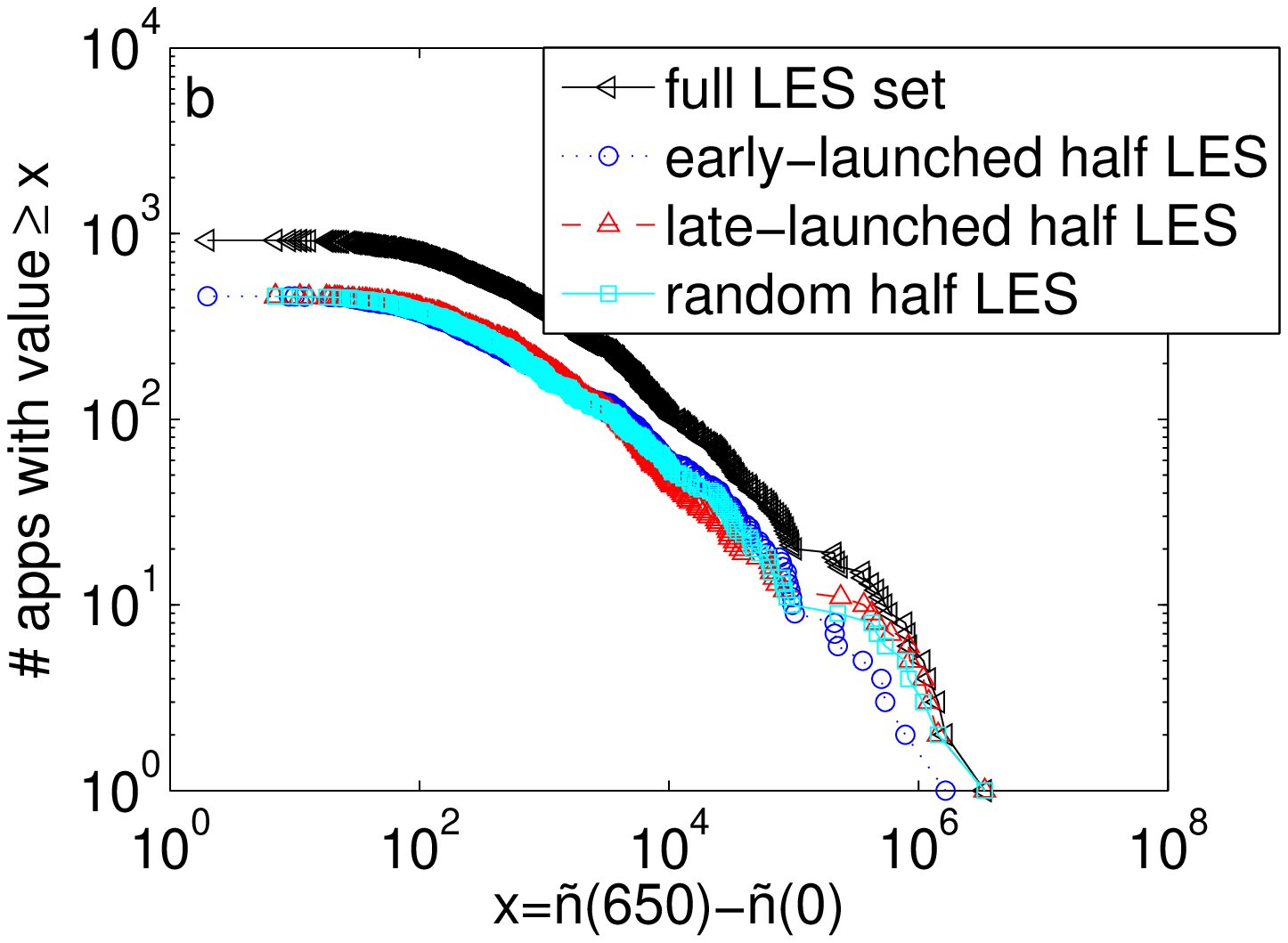,width=8cm}
\caption{{ As in Figs.~1a,b of the main text, but including results for a randomly-chosen subset that includes 460 of the LES apps.  (We chose the subset uniformly at random and without replacement.)
}
}
\label{paperfig2plus}
\end{figure}

{ We obtain similar results for any partition into two disjoint subsets of the same sizes as above. In Fig.~\ref{paperfig2plus}, we again show the results of Figs.~1a,b of the main text, but we now also include curves for which we only use a subset (chosen uniformly at random and without replacement) that includes 460 of the LES apps.
 The randomly-chosen subset has very similar characteristics to the early-launch and late-launch subsets. Using 5000 realization of randomly drawn subsets of the same size, the mean $L^2$ error is $1.99$ (with a standard deviation of $0.13$.}


\begin{figure}
\centering
\epsfig{figure=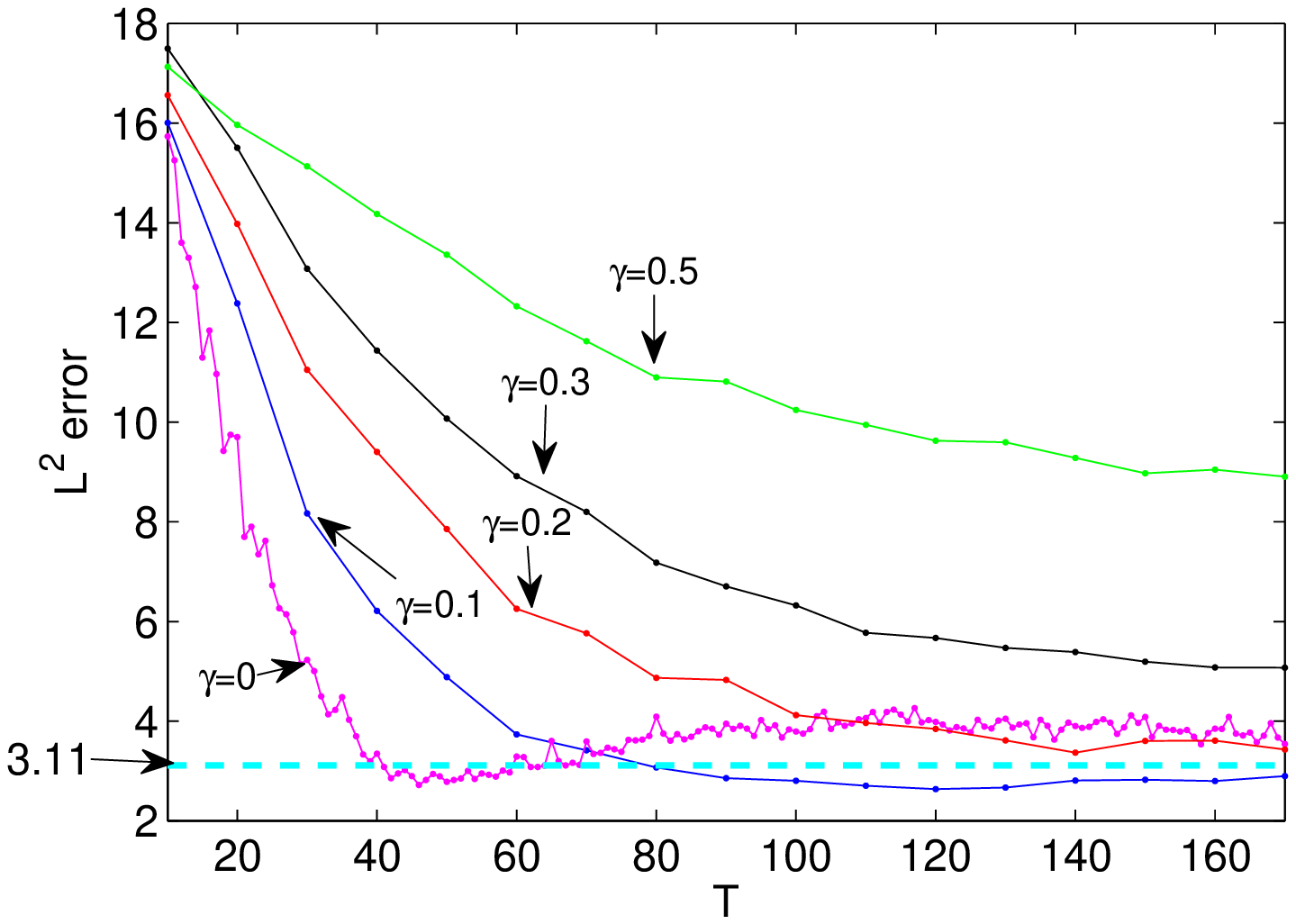,width=10cm}
\caption{The $L^2$ error $E_{L^2}$ in the mean scaled age-shifted growth rate $r(a)$ as a function of the memory time $T$ for exponential memory function $W(\tau)$, history window $H = 168$, and several values of the parameter $\gamma$. Each point is the result of a single realization of a stochastic simulation of our model.
}
\label{figSI10}
\end{figure}

In Fig.~\ref{figSI10}, we show the $L^2$ error $E_{L^2}$ as a function of the memory time $T$ for exponential memory function $W(\tau)$ for a fixed history window of length $H = 168$ and several values of the parameter $\gamma$ (see Fig.~3 of the main text).  The dashed line indicates the threshold for the ``good-fit" regime of $E_{L^2} \leq 3.11$.  The error tends to increase with increasing $\gamma$ and is unacceptably high for all values of $T$ for $\gamma > 0.2$. { It is interesting to note that the good-fit regime moves towards larger $T$ values as $\gamma$ increases. This seems to be a characteristic feature of the model---it appears also in Figs.~\ref{figSI6} and \ref{figSI7}---but we do not, as yet, have an explanation for it.}
{
\subsection*{SI3.2: Turnover in the Top-10}

The right column of Fig.~1 of the main text shows the popularity of those apps that are in the top-5 list at $t=0$. Figure~\ref{figtop10} shows more detail for each of the four cases (data plus three models) corresponding to Fig.~1c,f,i,l. In each panel of Fig.~\ref{figtop10}, the apps in the $t=0$ top-5 are shown with solid lines, while dashed lines show the popularity of those apps that make up the remainder of the top-10 list at $t=t_\text{max}$. As in Fig.~1, the change in rankings (turnover) is given in the legend of each panel, but here all apps in the top-10 are shown.
\begin{figure}
\centering
\epsfig{figure=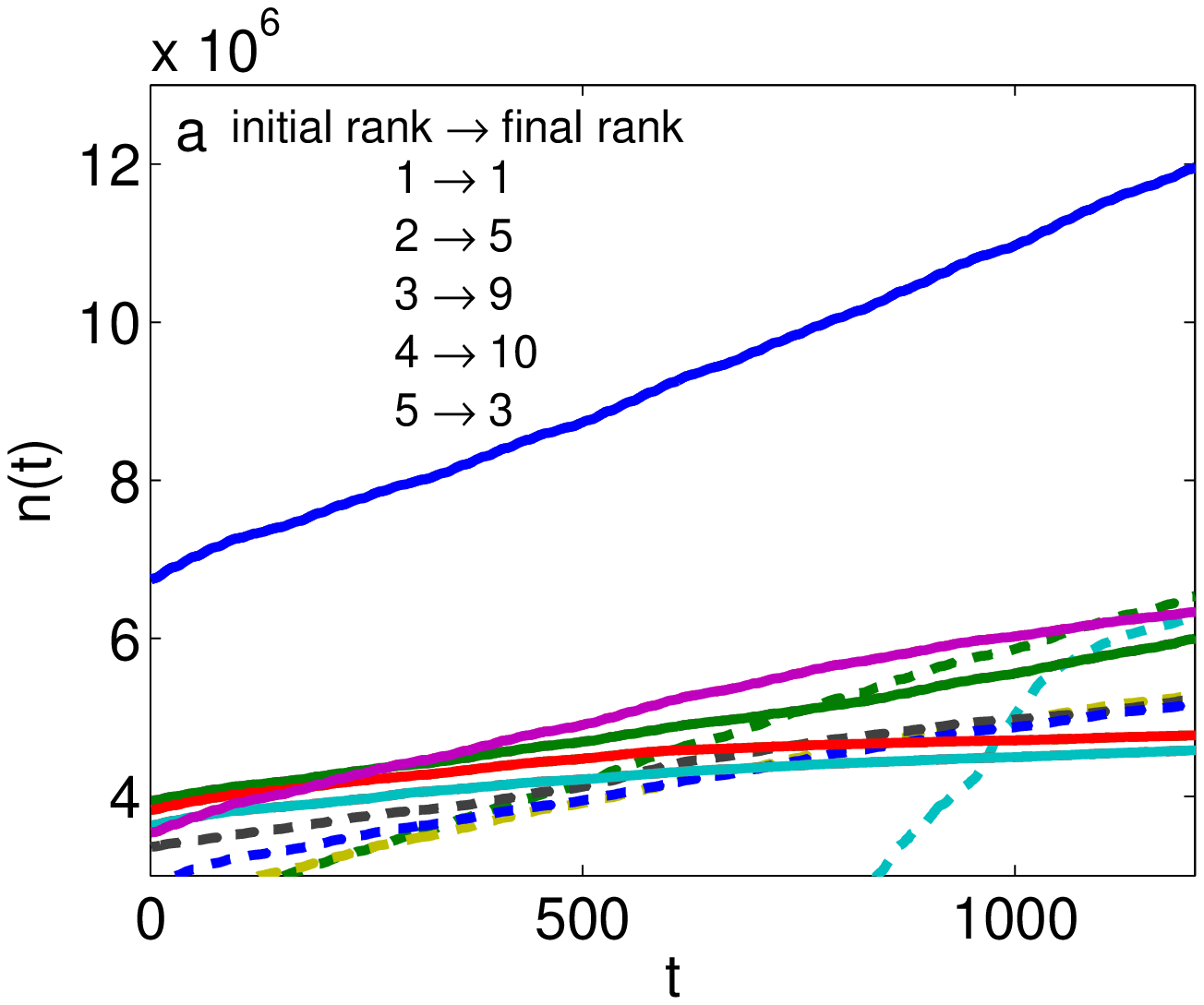,width=7cm}
\epsfig{figure=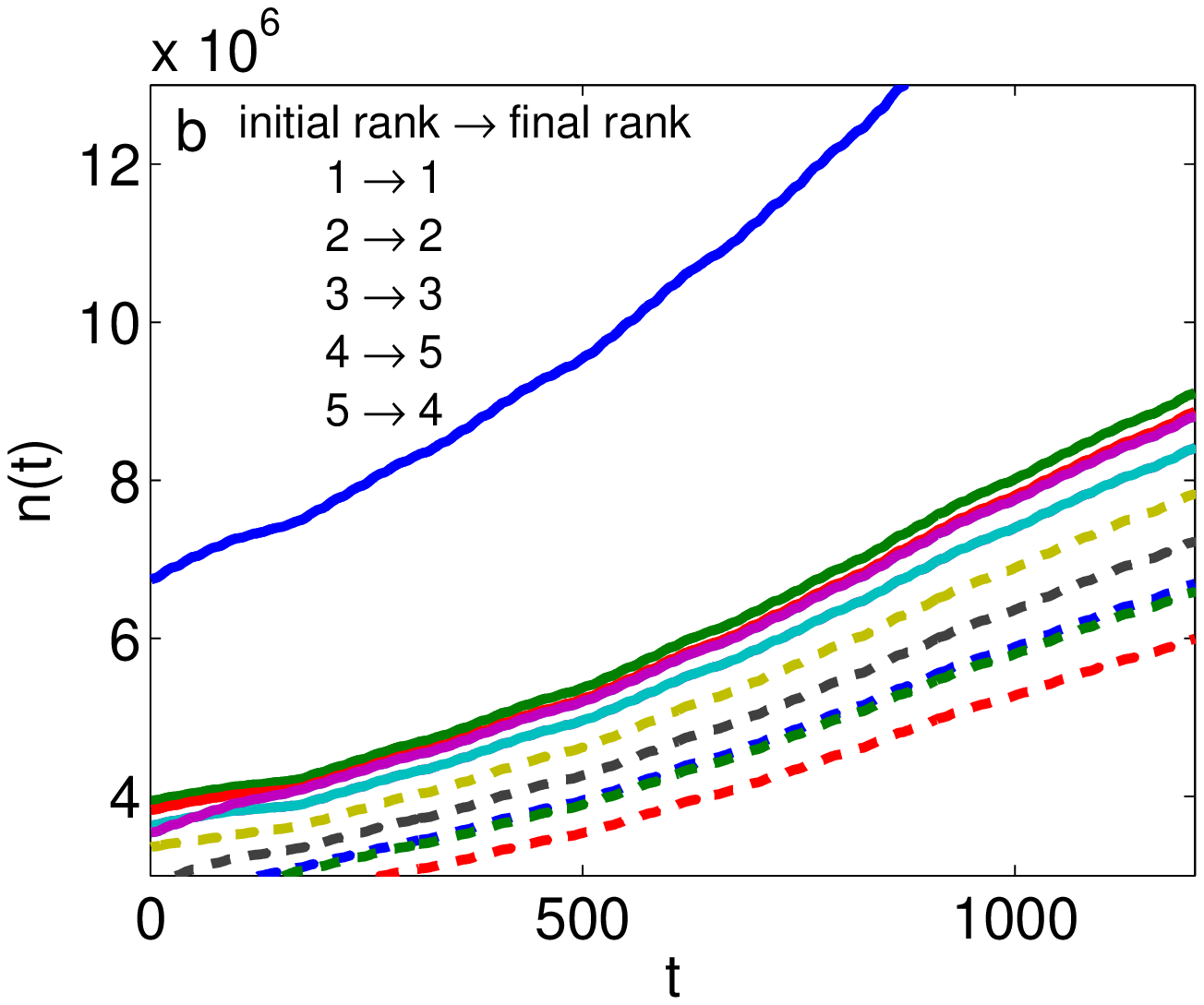,width=7cm}
\epsfig{figure=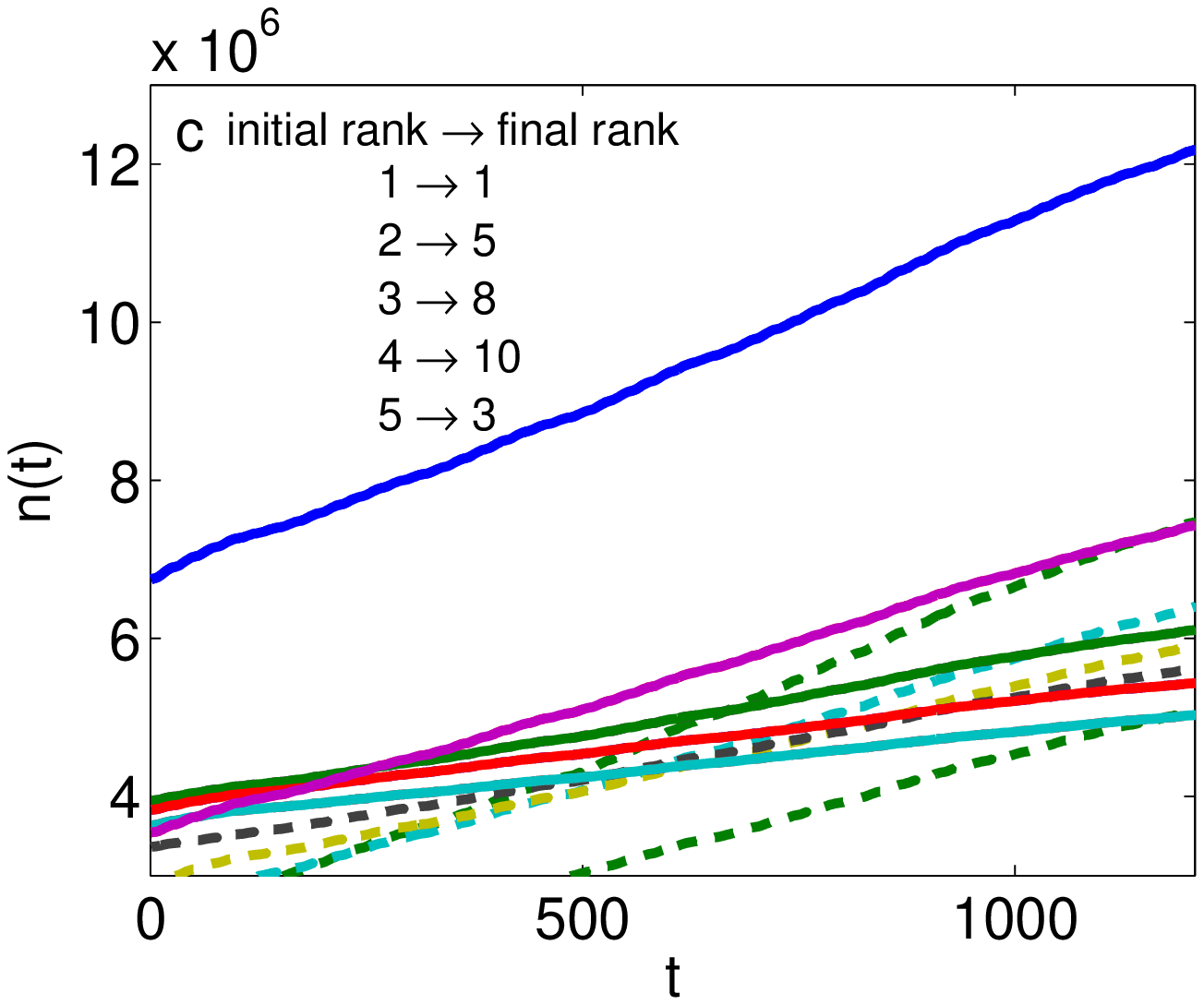,width=7cm}
\epsfig{figure=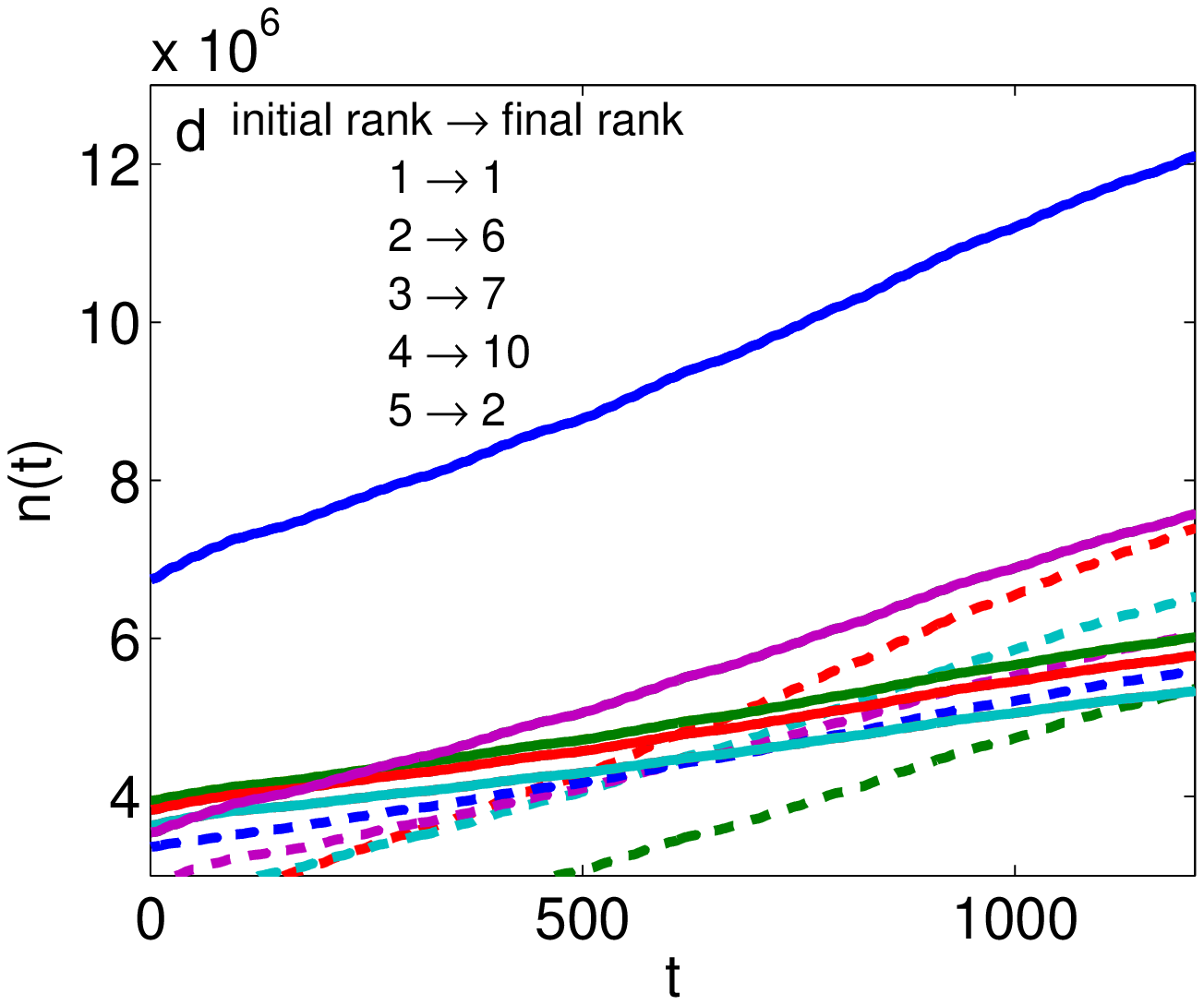,width=7cm}
\caption{{Growth trajectories of the 10 most popular apps in (a) data (as Fig.~1c), (b) simulation with cumulative-information model ($\gamma=1$, as Fig.~1f), (c) simulation with recent-activity model with short memory (as Fig.~1i), (d) simulation with recent-activity model with long memory (as Fig.~1l). The solid curves show the popularities of the top-5 apps from $t=0$; the dashed curves show the popularities of the remainder of the top-10 apps from $t=t_\text{max}$.
}}
\label{figtop10}
\end{figure}

}


\section*{SI4: Response Functions Generating Memory Weighting}

If one assumes that the total installation activity $F(t)$ is constant in time, then the memory function $W(\tau)$ introduced in Eq.~(4) of the main text is
proportional to the probability that an agent copies the installation choice of an agent from $\tau$ hours in the past (see SI5 for details). Consequently, we consider weighting functions that are related to previous empirical studies of the distribution of response times for e-mails \cite{Iribarren09,Iribarren11,Vazquez07}.  Consider an update message that informs a Facebook user---which we model as a single computational agent---that a friend has installed a certain app, which can then lead to the user subsequently installing the app. Let $\tau^\prime$ denote the time between receiving the update message and installing the app, and let $P(\tau^\prime)$ denote the probability distribution function (PDF) of these ``response times'' across the user population. We coarse-grain to the one-hour temporal resolution of the data by setting $W(\tau) = \int_{\tau-1}^\tau P(\tau^\prime) \,d\tau^\prime$ (for $\tau \in \{1,2,\ldots\}$), with an initial condition of $W(0)=0$.

In the main text, we showed an example in which $P(t)$ is an exponential distribution. We now consider alternative assumptions on the underlying response-time distribution $P(t)$ and show results corresponding to Fig.~3 of the main text for the $L^2$ error in the mean scaled age-shifted growth rate.   We find similar results for lognormal, gamma, and uniform distributions. In all of these cases, we obtain good results with a history window parameter of $H \approx 168$ hours (i.e., 1 week). Interestingly, when $H = 168$, the results for all distributions are very similar to those shown in Figs.~1j,k,l of the main text if the characteristic response-time $\left\langle \tau \right\rangle = \sum_{\tau=1}^{168} \tau \, W(\tau)/\sum_{\tau=1}^{168} W(\tau)$ is  about $45$ hours (i.e., approximately 2 days\footnote{The value of  $\left\langle \tau \right\rangle$ is similar to the mean response time if most of the probability mass lies in the range $\tau<H$. The cutoff at $\tau=H$ reflects the fact that apps at early stages in their simulated growth possess a window of only approximately $H$ hours of previous-installation history to drive their temporal evolution.}).

\begin{figure}
\centering
\epsfig{figure=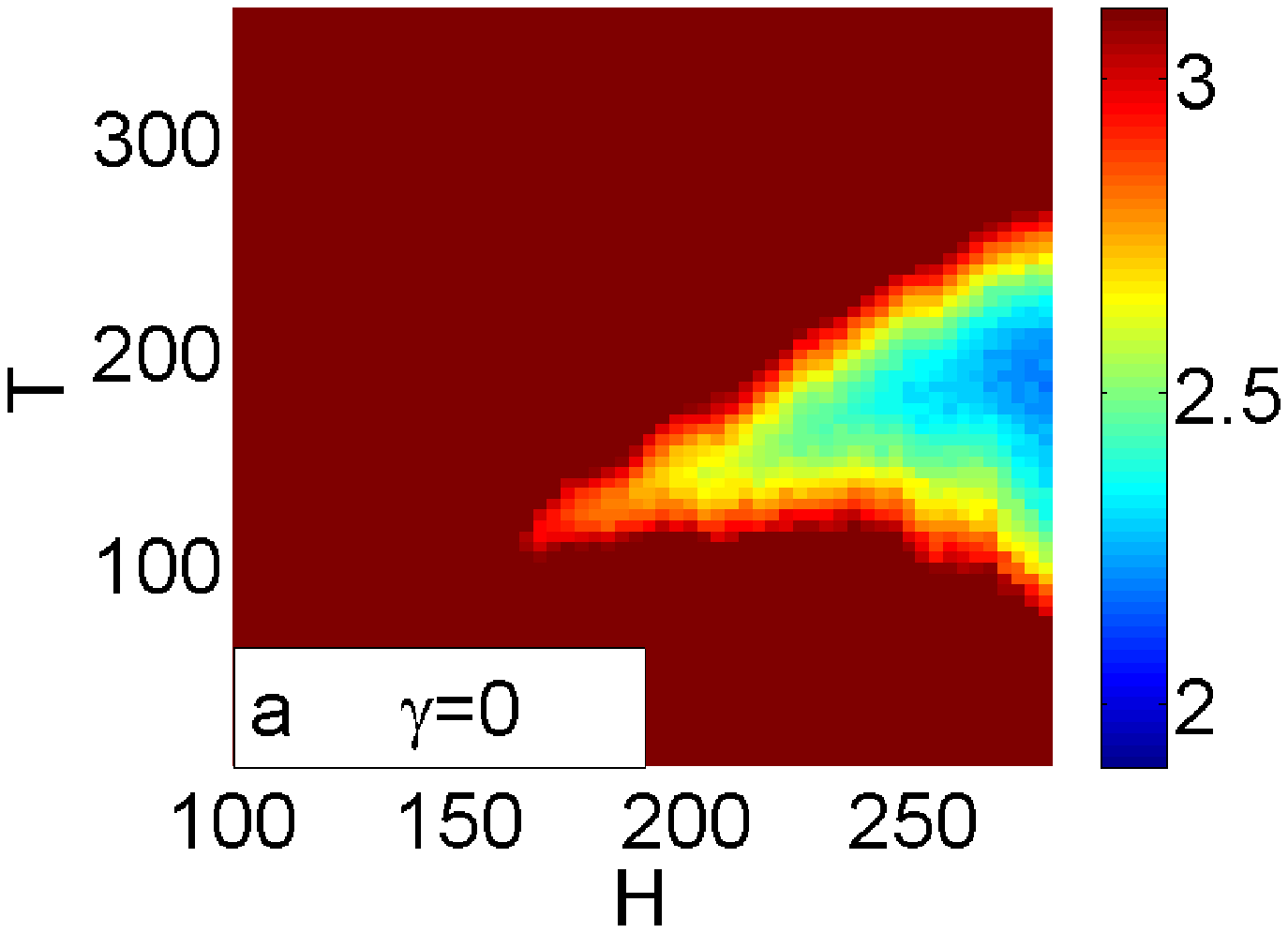,width=5.2cm}
\epsfig{figure=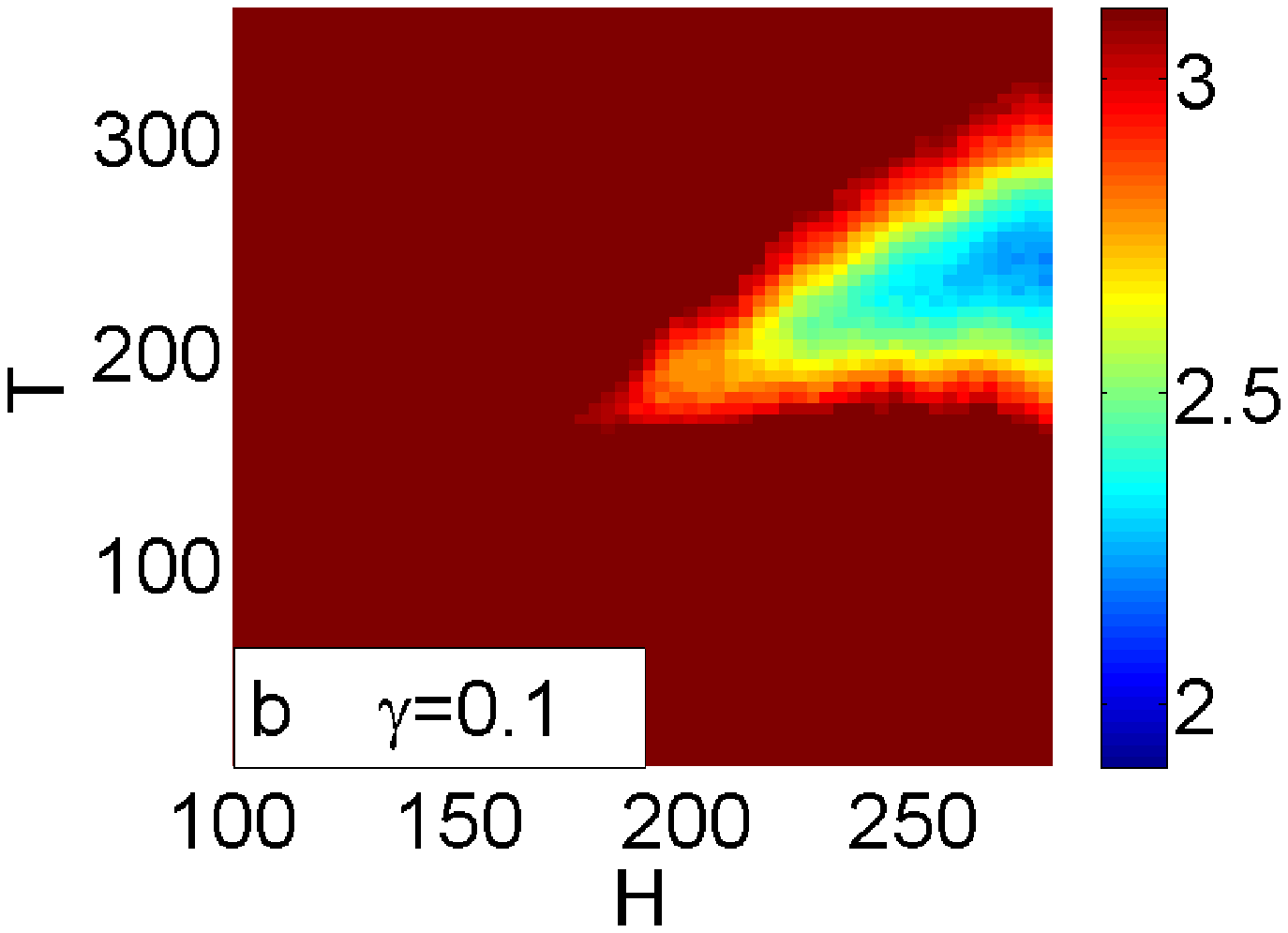,width=5.2cm}
\epsfig{figure=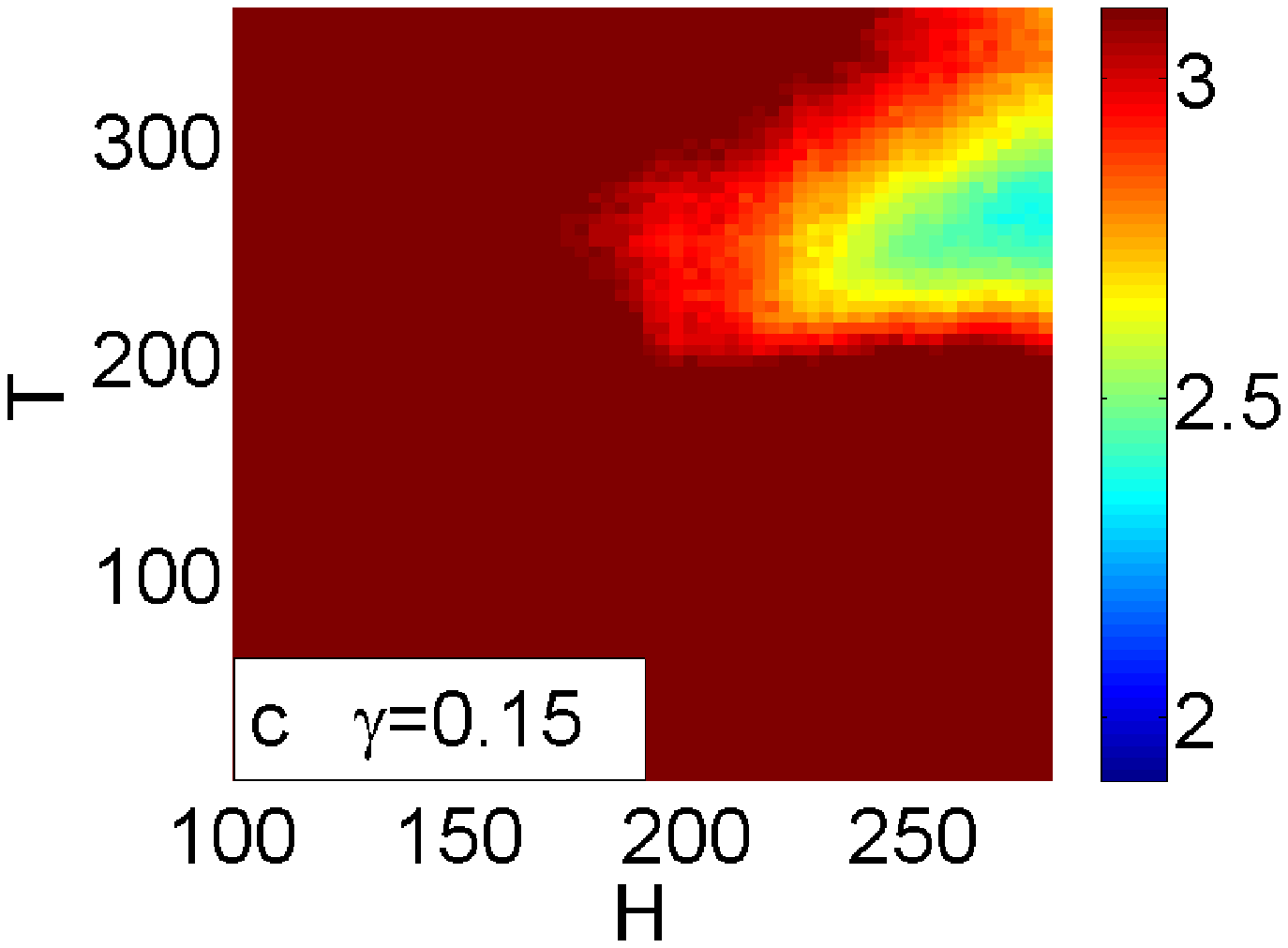,width=5.2cm}
\caption{As in Fig.~3 of the main text, but now we use a memory function $W$ that is generated from the uniform distribution (\ref{Punif}).
} \label{figSI6}
\end{figure}

In Fig.~\ref{figSI6}, we show results for the uniform distribution given by
\begin{equation}
	P(t) = \left\{ \begin{array}{cc}
	\frac{1}{T}\,, &  \text{ if } t\le T\,,\\
	0\,,  &  \text{ if } t > T\,,
	\end{array} \right.\label{Punif}
\end{equation}
where $T$ is the cutoff time.  (The mean response time is $T/2$.)  As with Fig.~3 in the main text, we show results in the $(H,T)$ parameter plane to highlight the roles of both the history window $H$ and the memory cutoff $T$. The three panels illustrate the effects of using increasing amounts of cumulative information (i.e., progressively larger values of $\gamma$) in the installation probability $p_i$. Moving from left to right, the weighting of cumulative information increases from $0$ to $0.1$ and $0.15$. As this weight increases, observe that the ``good-fit'' region of parameters moves to higher values of $H$ and $T$. This supports our conclusion in the main text that the recent-activity case $\gamma=0$ is ``optimal" in the sense of requiring only a relatively small history window size $H$ to fit the data. Similar conclusions were also reached in Ref.~\cite{Zeng13}.  We have also confirmed that the other main results for the exponential distribution (e.g., the ones depicted in Fig.~1 of the main text) are closely reproduced using the uniform distribution (where we set $T\approx100$ so that the mean response times are equal in the two cases).

\begin{figure}
\centering
\epsfig{figure=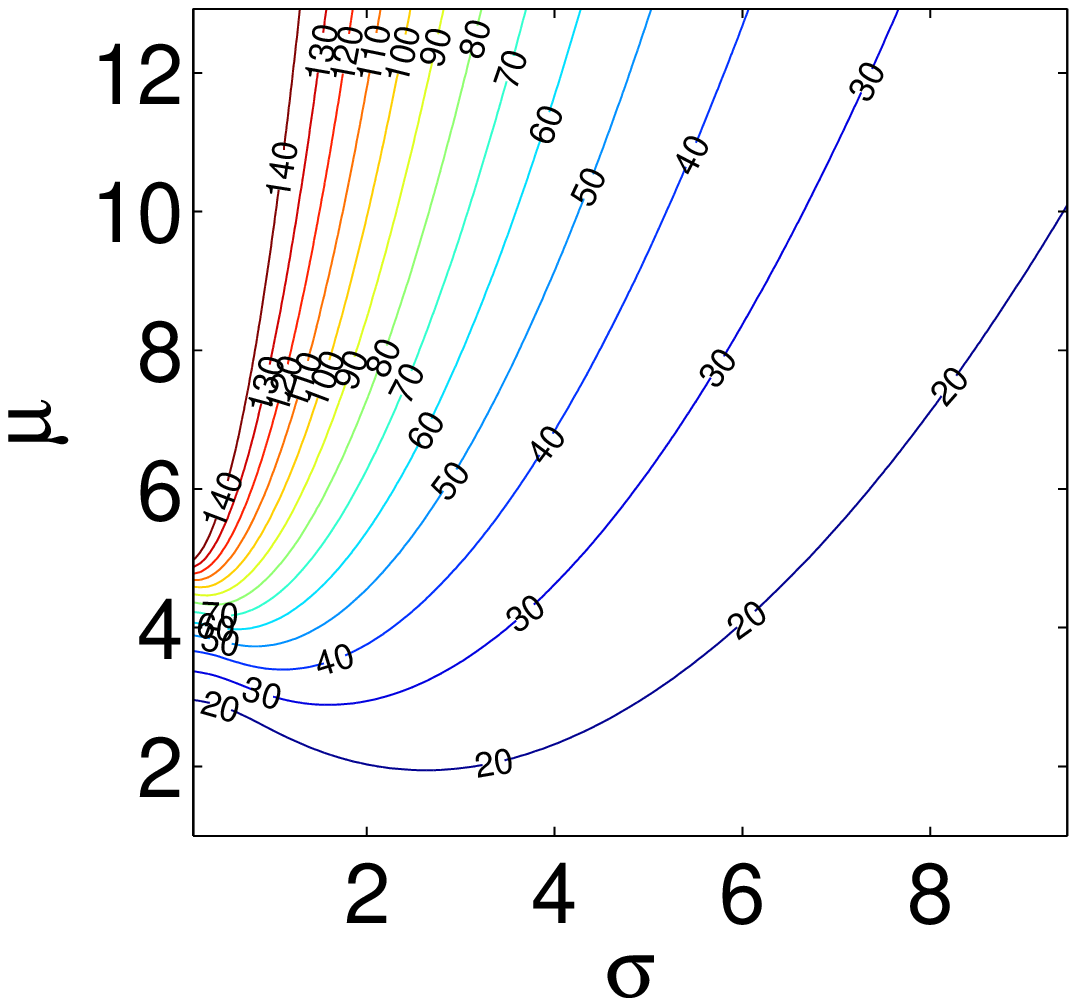,height=4.2cm}
\epsfig{figure=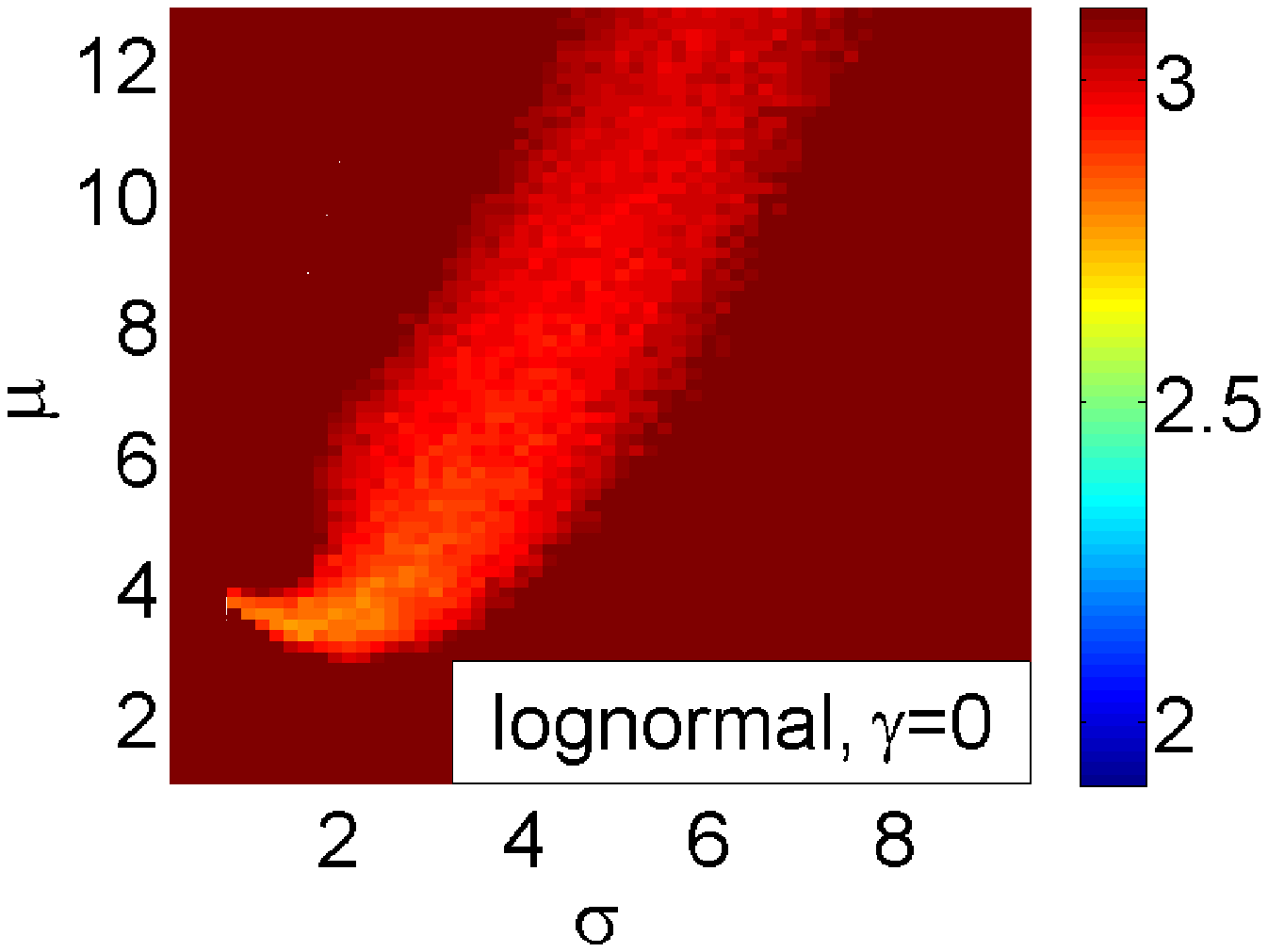,height=4.2cm}
\epsfig{figure=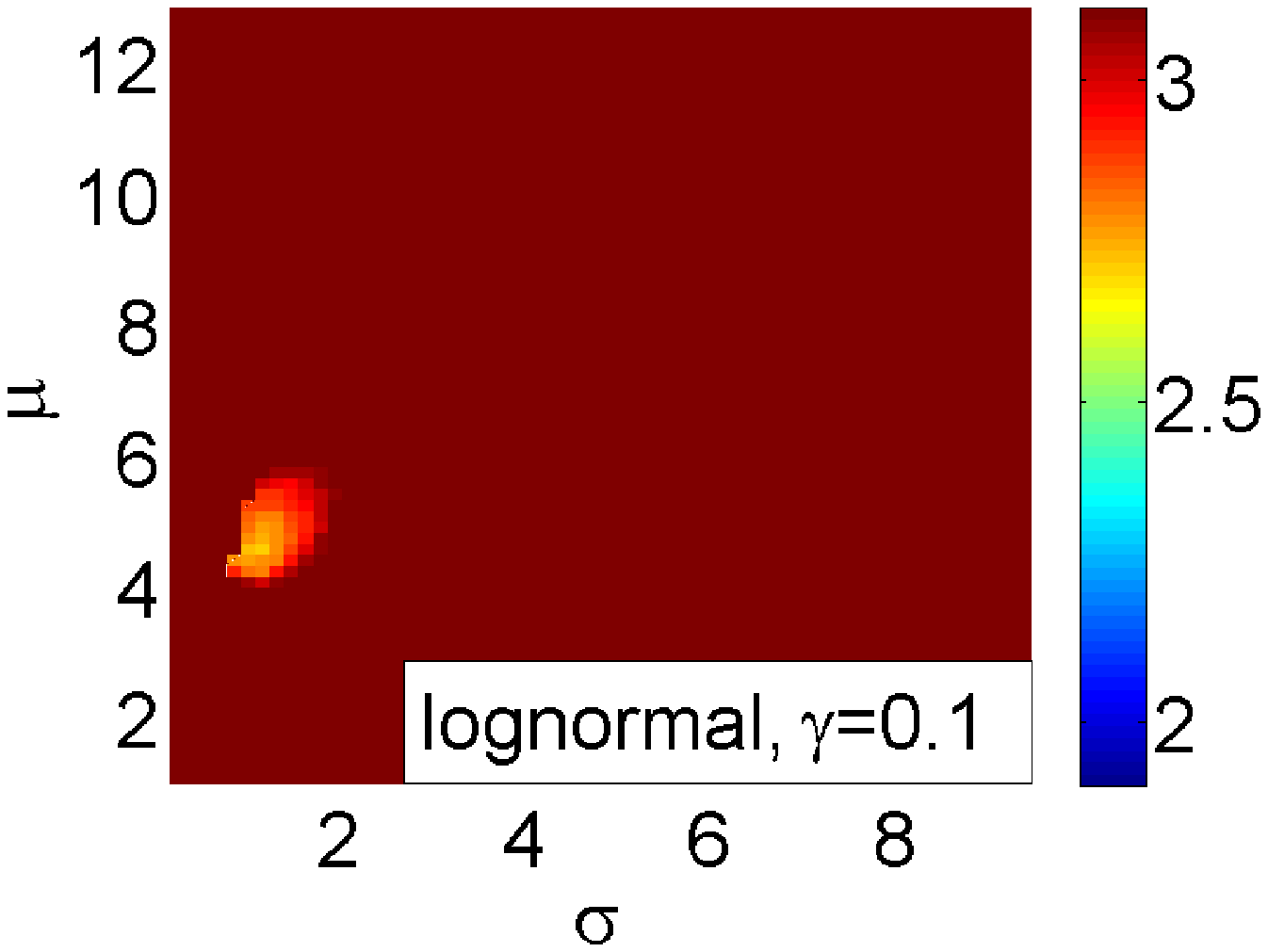,height=4.2cm}
\epsfig{figure=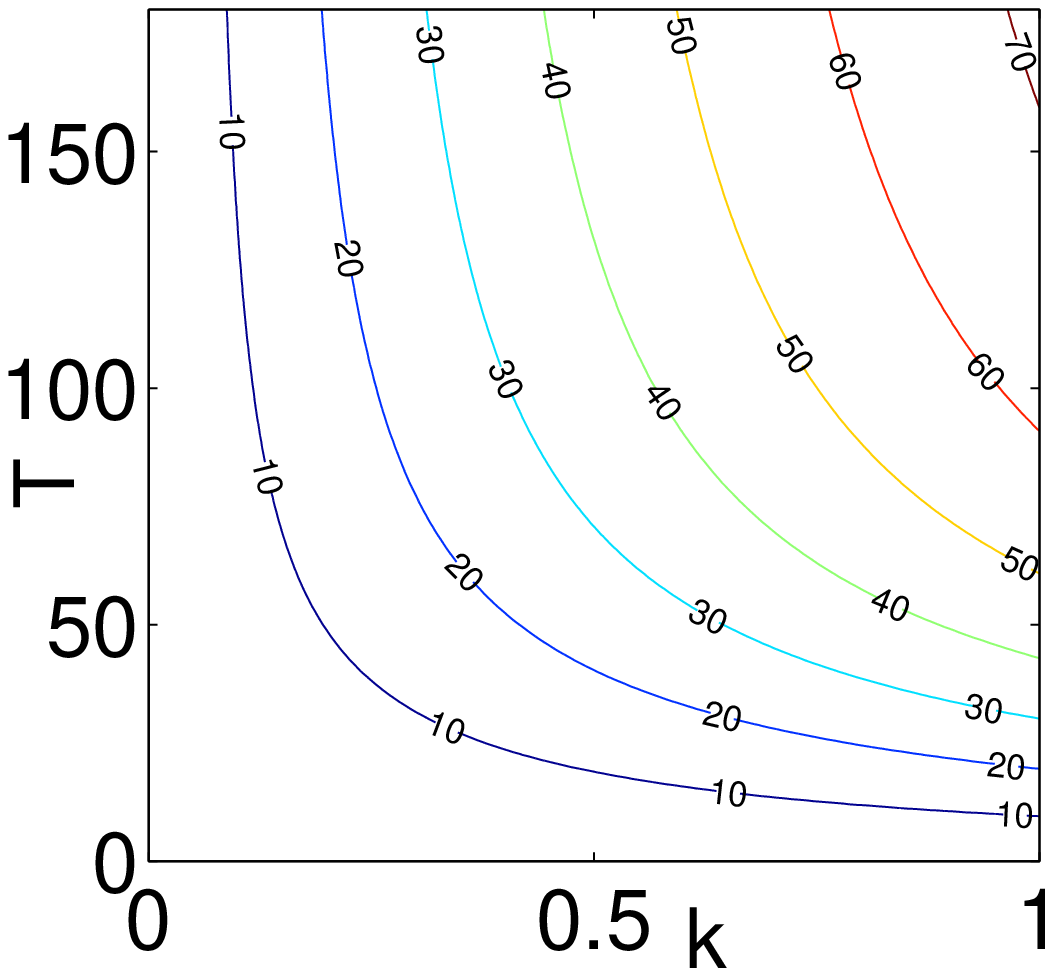,height=4.2cm}
\epsfig{figure=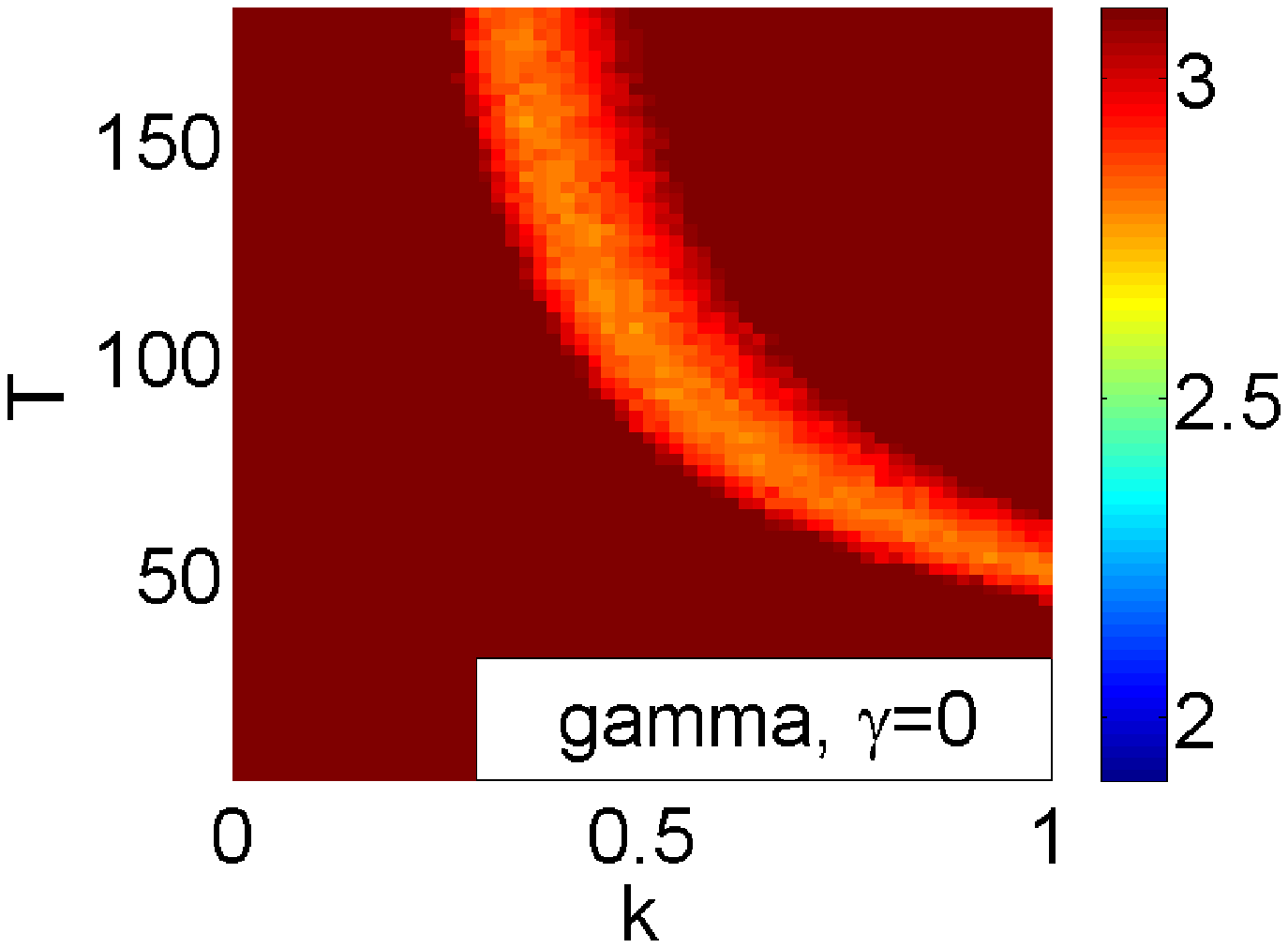,height=4.2cm}
\epsfig{figure=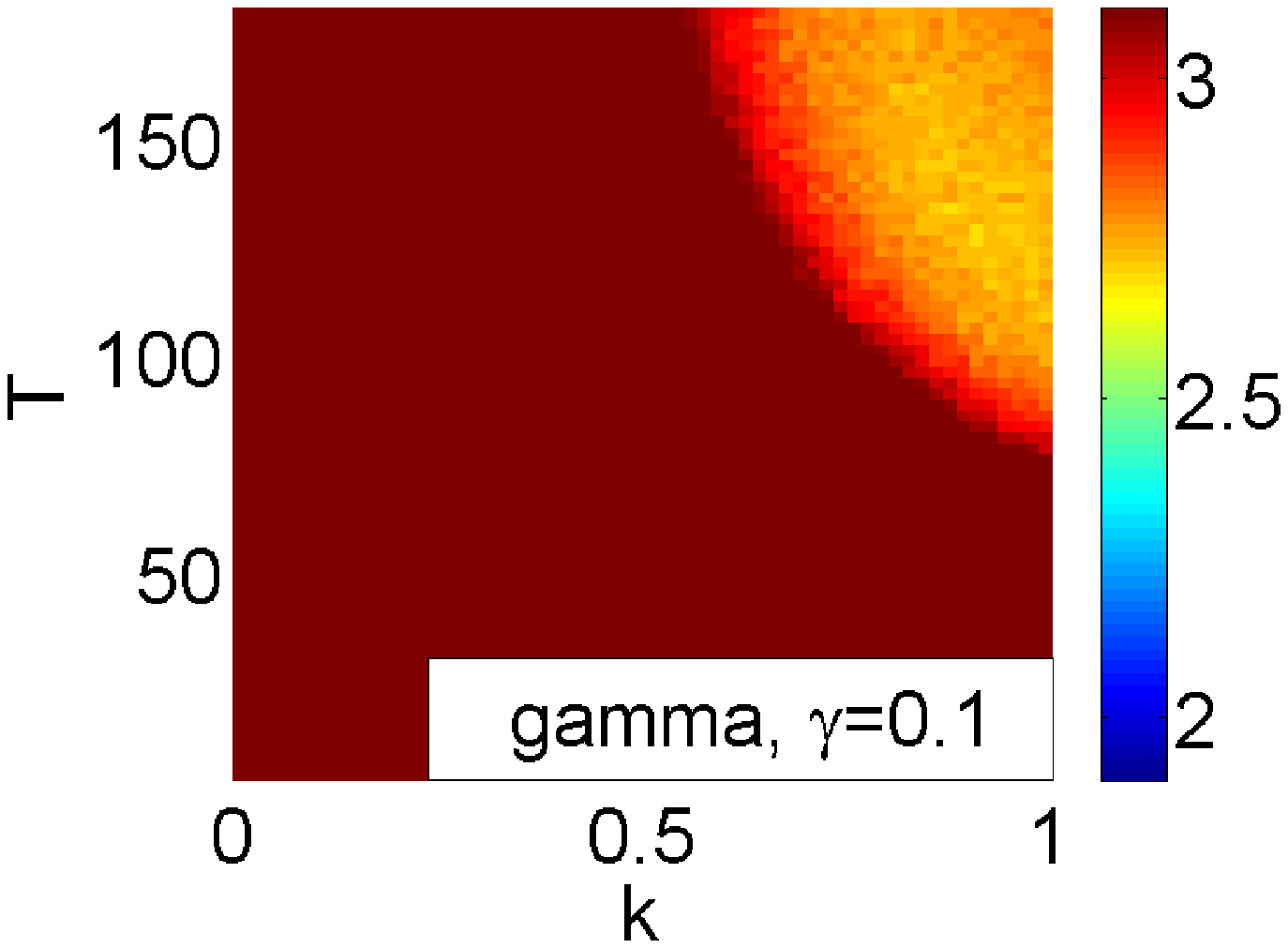,height=4.2cm}
\caption{As in Fig.~3 of the main text, but now we use a (top row) lognormal distribution and (bottom row) gamma distribution of response times. The left panel in each row shows the contours of the cutoff mean response time $\left\langle \tau\right\rangle$, which we defined in Eq.~(\ref{meanT}).} \label{figSI7}
\end{figure}

In Fig.~\ref{figSI7}, we show results in which the response-time distribution $P(t)$ is given by (top row) lognormal and (bottom row) gamma distributions. These distributions both have two parameters, so we fix the history window size $H$ to be 168 hours (i.e., 1 week) and consider the effect of the parameters that define the distributions. The lognormal distribution with parameters $\mu$ and $\sigma$ is
\begin{equation*}
	P(t) = \frac{1}{t \sqrt{2 \pi \sigma^2}} \exp\left\{{-\frac{(\ln t-\mu)^2}{2 \sigma^2}}\right\}\,,
\end{equation*}
and the gamma distribution with parameters $k$ and $T$ is
\begin{equation*}
	P(t) = \frac{1}{\Gamma(k) T^k} t^{k-1} e^{-\frac{t}{T}}\,.
\end{equation*}
In the special case $k=1$, the gamma distribution is an exponential distribution, while for $k<1$ it limits to a power-law distribution as $T\to \infty$.
The lognormal and gamma distributions were used in Refs.~\cite{Iribarren09,Vazquez07,Iribarren11} to model distributions of e-mail response times.

The center panel of each row of Fig.~\ref{figSI7} gives results for $\gamma=0$, and the right panel of each row gives results for $\gamma=0.1$. For $\gamma=0.15$, the ``good-fit'' regions have almost disappeared from these plots, so we do not show them. The left panel of each row shows the contours of the quantity
\begin{equation}
	\left\langle \tau \right\rangle = \frac{\sum_{\tau=1}^{168} \tau \, W(\tau)}{\sum_{\tau=1}^{168} W(\tau)}\,, \label{meanT}
\end{equation}
which is related to the goodness-of-fit of the recent-activity ($\gamma=0$) models. Observe that the light-coloured regions of the center panels align closely with the contours showing $\left\langle \tau \right\rangle$ values between 30 and 50 hours. Note that $\left\langle \tau \right\rangle$ is not identical to the mean response time of the distribution $P(t)$ because of the cutoff at $168$ hours in the sums of Eq.~(\ref{meanT}).
This cutoff reflects the fact that the history window of 168 hours defines the $\tau$ range upon which the recent-activity model operates for an app that was launched recently. It seems that a memory weighting that corresponds to roughly 2 days (i.e., 48 hours) of recent activity is sufficient in all of these cases to fit the  model to the data. A 2-day window was also identified as significant in the temporal clustering of adoption decisions among online friends in Ref.\cite{aral2009}.



\section*{SI5: Recent-Activity Model as a Random-Copying-with-Memory Process }

In this section, we show that one can interpret the recent-activity model ($\gamma = 0$) described in the main text as a random-copying process that is similar to those studied by Bentley et al. \cite{Bentley04,Bentley11}. We also describe these models in terms of branching processes and discuss the circumstances under which one obtains critical branching processes. In this context, a critical branching process is one in which each parent has, on average, one child over its lifetime \cite{Harrisbook}.

\begin{figure}
\centering
\epsfig{figure=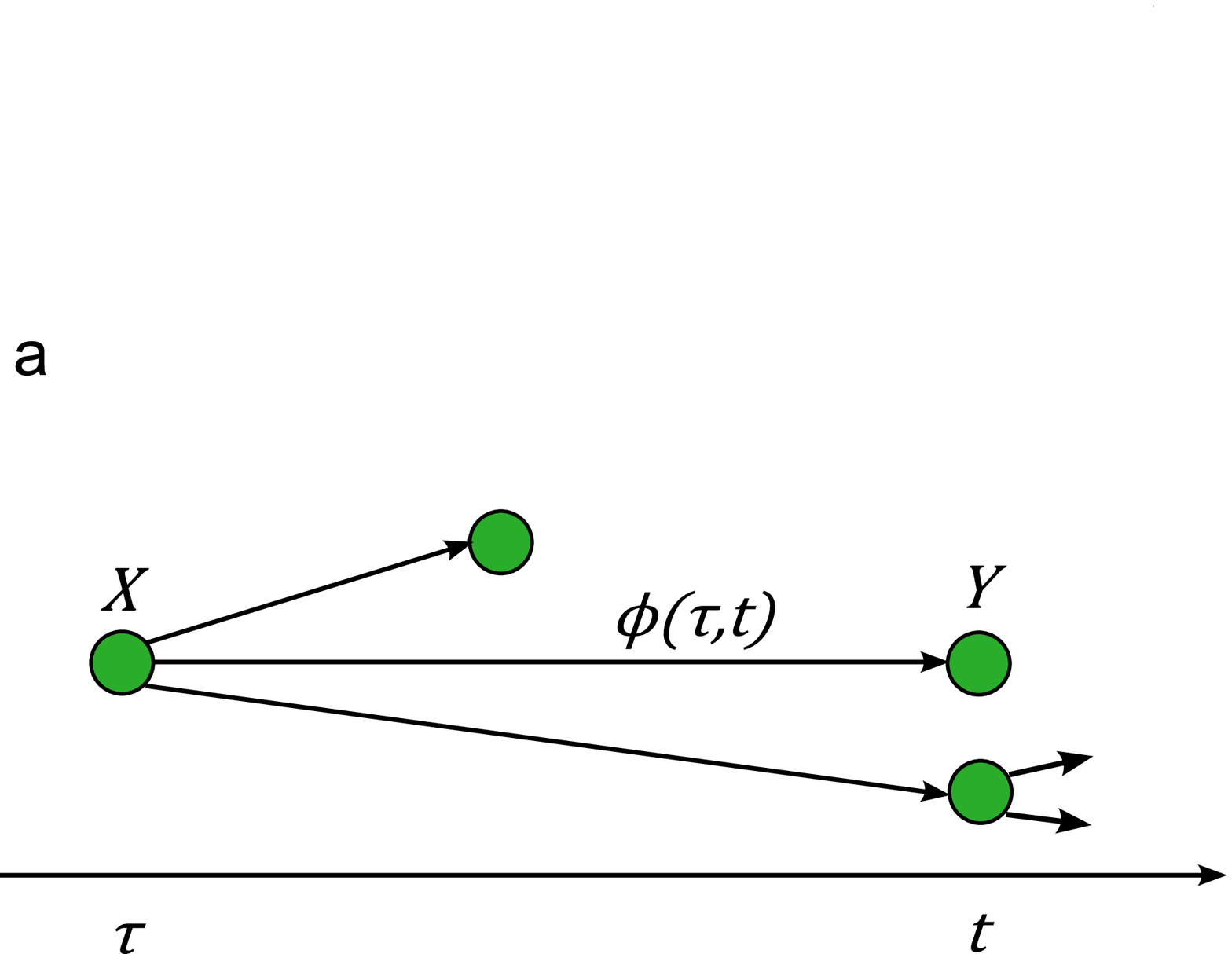,width=7.3cm}
\hspace{0.5cm}
\epsfig{figure=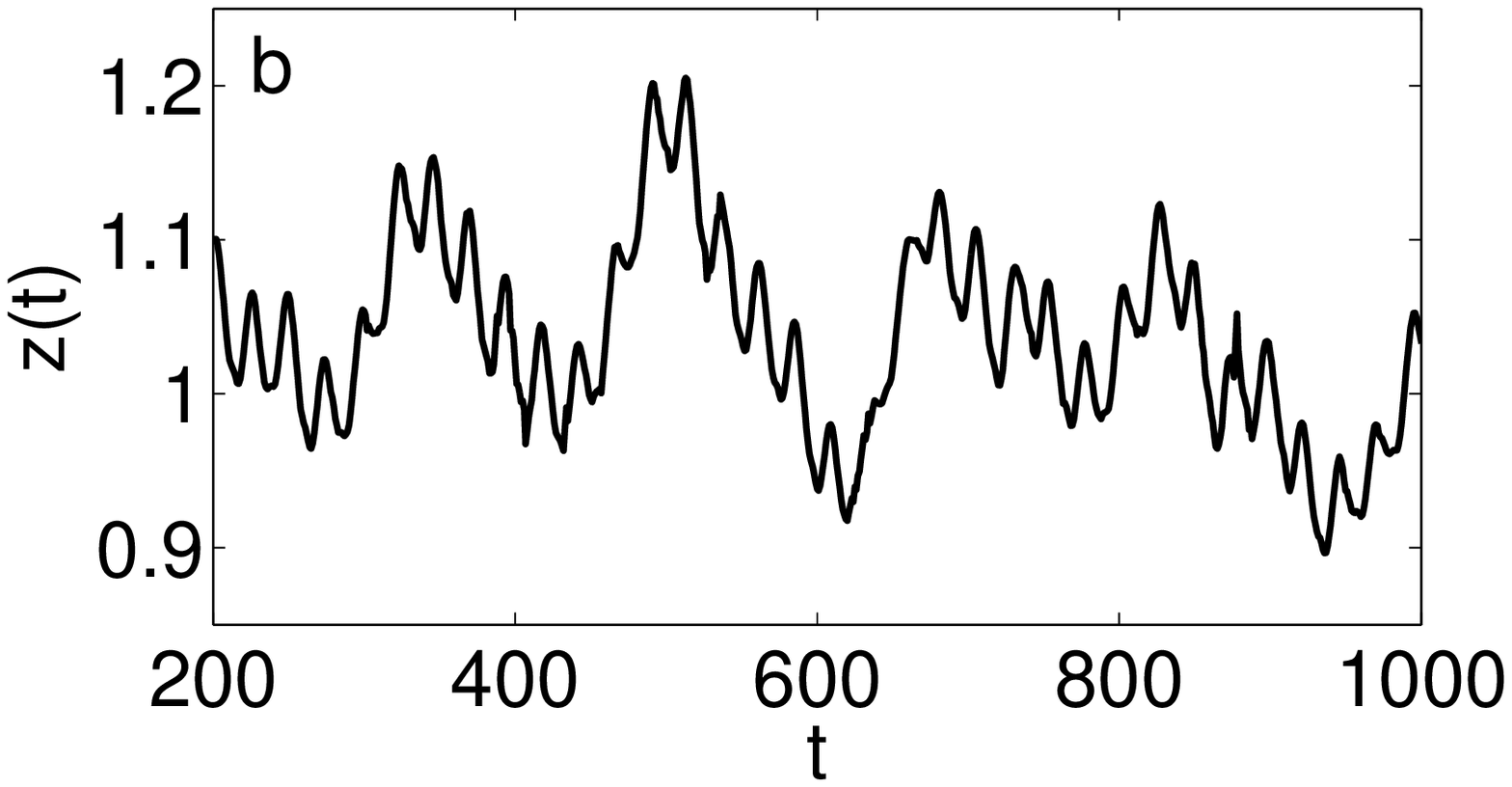,width=7.5cm}
\caption{(a) Tree schematic for the model in Section SI5. Nodes indicate the installation of an app at the time indicated on the horizontal axis. (b) Effective branching number $z(t)$ for the data from Eq.~(\ref{z}).
} \label{figSI4}
\end{figure}

We consider a random-copying model in which each individual (an agent in our simulation) at time $t$ copies the action (i.e., the choice of app to install) of an agent from a previous time step. In the schematic of Fig.~\ref{figSI4}a, we denote the copying action with an arrow from the earlier installation event to the later installation event (i.e., arrows point \emph{from} the target of the copying \emph{to} the copier). This generates a tree structure in time in which each node represents a single installation action and each arrow links a ``parent'' (target) node to some number of ``child'' (copier) nodes. Each child node has exactly one parent---this represents the installation action that was copied---but the number of children assigned to any given parent depends on the details of the random-copying process. As noted in the main text, we do not have any information on network topology, so we make the
assumption that all agents can copy the action of any earlier agent, unrestricted by network connectivity.


There are $F(t)$ agents who install an app at time $t$, and they all act independently of each other. Consider one such agent $Y$, who must choose an earlier installation to copy. Let $\Phi(X,Y)$ denote the probability that $Y$ copies the past action of a selected node $X$ (see Fig.~\ref{figSI4}a). Normalization implies that $\sum_{X} \Phi(X,Y) = 1$, where the sum is over all possible target nodes $X$ such that $X$ takes an action before $Y$. We assume that the selection probability depends only on the time $\tau$ of the target node $X$ and the time $t$ of the action $Y$, so we write $\Phi(X,Y) = \Phi\left(X(\tau),Y(t)\right) \equiv \phi(\tau,t)$. This implies that all installations at time $\tau$ are equally likely to be copied by $Y$. Moreover, we assume that the dependence on $\tau$ appears only through the time $t_e := t-\tau$ elapsed since the target event, so $\phi(\tau,t) \propto W(t-\tau)$, where $W$ is the memory function (see  SI4). Because there are $F(\tau)$ installing agents (i.e., nodes) at time $\tau$, the correctly normalized copying probability must obey $\sum_{\tau<t}\phi(\tau,t) F(\tau) = 1$.  This yields
\begin{equation}
	\phi(\tau,t) = \frac{W(t-\tau)}{\sum_{\tau^\prime=-\infty}^{t-1} W(t-\tau^\prime) F(\tau^\prime)}\,. \label{SI1}
\end{equation}
Note we are allowing a potentially infinite history, which might be appropriate for very heavy-tailed memory-functions \cite{Malmgren08,Karsai12}.

Using this random-copying model, we want to compute the probability that user $Y$ installs a given app $i$ at time $t$. There are $f_i(\tau)$ agents who install app $i$ at each time $\tau$ with $\tau<t$. (Installer $X$ in Fig.~\ref{figSI4}a is just one example of many.) Agent $Y$ can copy each of these agents with probability $\phi(\tau,t)$.
Summing over all earlier times implies that the total probability that $Y$ installs app $i$ is
\begin{equation}\label{summ}
	\sum_{\tau=-\infty}^{t-1} \phi(\tau,t) f_i(\tau)\,.
\end{equation}
Using the definition of $\phi(\tau,t)$, Eq.~(\ref{summ}) can be rewritten as
\begin{equation}
	\frac{\sum_{\tau=-\infty}^{t-1} W(t-\tau) f_i(\tau)}{\sum_{\tau^\prime=-\infty}^{t-1} W(t-\tau^\prime) F(\tau^\prime)}\,,
\end{equation}
which is precisely $p_i^r$ in the main text, where we note that $f_i(\tau)=0$ for $\tau<0$ in Eq.~(4) from the main text because data is available only from $t=0$ onwards. The normalization constant $L$ in Eq.~(4) can be written as
\begin{align*}
	L &= \left( \sum_i \sum_{\tau^\prime} W(t-\tau^\prime)  f_i(\tau^\prime) \right)^{-1} \\
		&= \left( \sum_{\tau^\prime} W(t-\tau^\prime) \sum_i  f_i(\tau^\prime) \right)^{-1} \\
		&= \left( \sum_{\tau^\prime} W(t-\tau^\prime) F(\tau^\prime) \right)^{-1}
\end{align*}
by reordering the summations and using Eq.~(1) from the main text.

Returning to the branching-process interpretation of Fig.~\ref{figSI4}a, we calculate the expected number of children for each parent in the tree. Consider node $X$, which can be copied by any one of the $F(t)$ installing agents at time $t$. Each of these agents chooses to copy $X$ with probability $\phi(\tau,t)$. Summing over $t$ gives the expected number of children of node $X$ (and indeed of any user at time $\tau$) over all future times:
\begin{equation}
	z(\tau) = \sum_{t=\tau+1}^\infty  \phi(\tau,t) F(t) =  \sum_{t=\tau+1}^\infty \frac{W(t-\tau) F(t)}{\sum_{\tau^\prime=-\infty}^{t-1} W(t-\tau^\prime) F(\tau^\prime)}\,. \label{z}
\end{equation}
This \emph{effective branching number} $z(\tau)$ depends on the time label $\tau$ of the parent node (i.e., the time at which user $X$ installed the app)
because the interaction of the variable level of installation activity $F(t)$ with the memory function $W(\tau)$ implies that installations from some times are more likely to be copied in the future than installations from other times.


If $F(t)$ is constant, it follows that $z(\tau)=1$ for all $\tau$. (In this case, letting $s=t-\tau$ and $s^\prime=t-\tau^\prime$ gives $z(t)=\sum_{s=1}^\infty W(s)/\sum_{s^\prime=1}^\infty W(s^\prime) = 1$.) Because each individual installation then has, on average, exactly one offspring, we obtain a critical branching process \cite{Harrisbook}, for which one expects to obtain power-law distributions of popularity (with exponents $\alpha \in [3/2,2)$) in the mean-field limit \cite{Zapperi95,Adami02,Goh03}.
Consequently, the competition among apps for the finite number of installer slots leads
to a critical branching process that is reminiscent of the self-organization mechanism in self-organized-criticality models \cite{Bakbook,Zapperi95,Gleeson13}. Bentley et al.~\cite{Bentley11} used numerical simulations to examine this case of constant $F(t)$, though they did not give a branching-process interpretation. Note additionally that we concentrate on the accumulated popularity over time $n_i(t)$. In contrast, rewiring models such as those in Refs.~\cite{Evans07,Beguerisse09} focus instead on the distribution of short-time increments [similar to our $f_i(t)$].

As we show in Fig.~\ref{figSI1}, the total installation activity $F(t)$ exhibits substantial variation over time due to daily human activity patterns and to the aggregate growth in popularity of Facebook applications. In Fig.~\ref{figSI4}b, we show the effective branching number $z(t)$ calculated from Eq.~(\ref{z}) using the $F(t)$ function taken from the data and the long-memory weighting function (i.e., an exponential response-time distribution with $T=50$ hours) used in Figs.~1j,k,l of the main text. Despite the growth and fluctuations in $F(t)$ that are evident in Fig.~\ref{figSI1}, the values of $z(t)$ remain close to the critical value of $1$ throughout the period of the study. This occurs because the memory of the weighting function $W$ achieves a balance: it is sufficiently long so that it dampens the impact of daily oscillations on $z(t)$, but it is sufficiently short so that it also ameliorates the effect of the slow growth in $F(t)$ on $z(t)$. The resulting branching process is therefore near-critical \cite{Adami02}, with an effective branching number between 0.9 and 1.2. This might help to explain the heavy-tailed popularity distributions that have been observed not only in this data set \cite{Onnela10} but also in many other empirical data sets \cite{Bentleybook}. Recent models for the popularity of memes on Twitter have also been shown to be poised near criticality \cite{Weng12,Gleeson13}.




\section*{SI6: Ranking Model}
\begin{figure}
\centering
\epsfig{figure=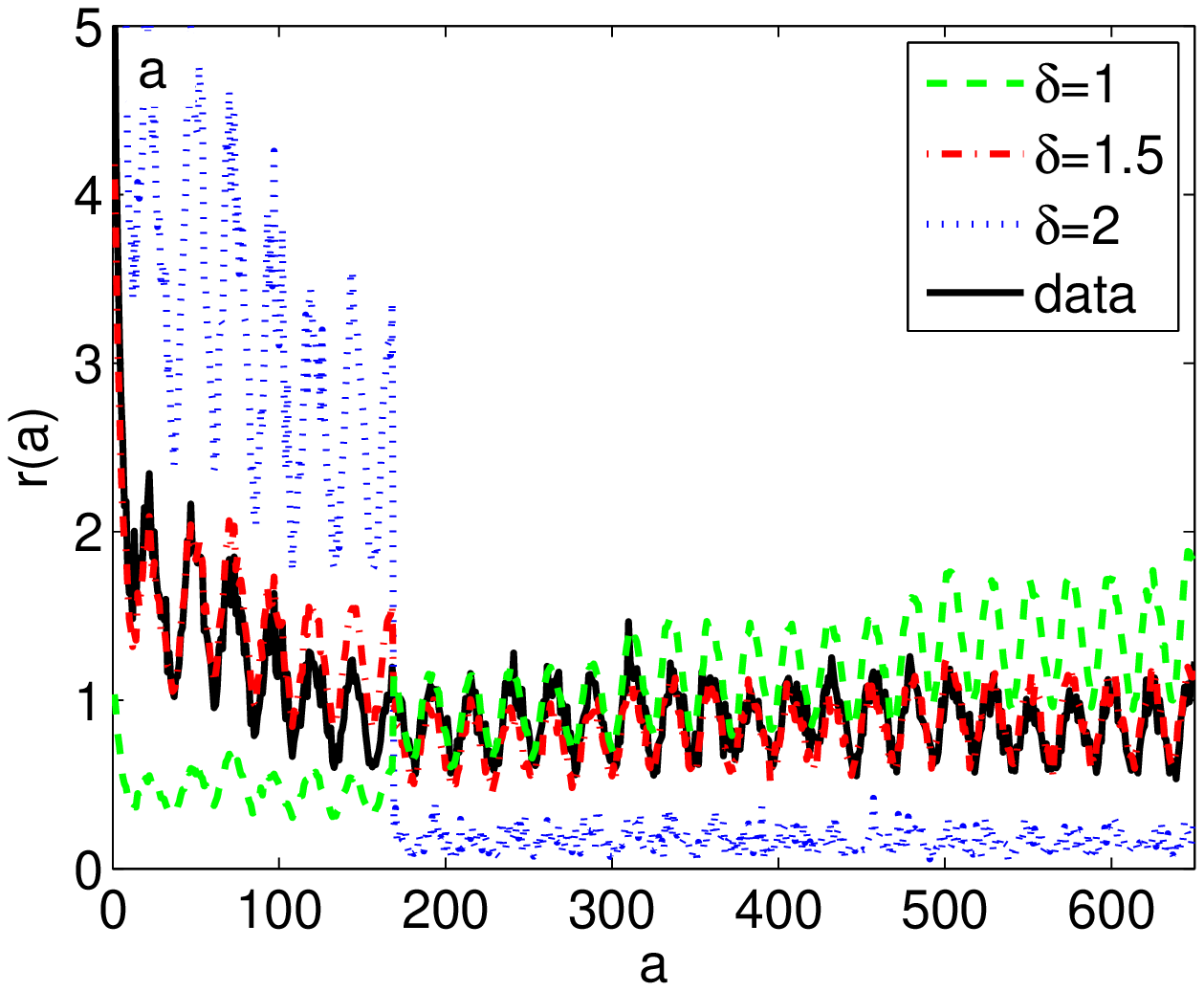,width=8cm}
\epsfig{figure=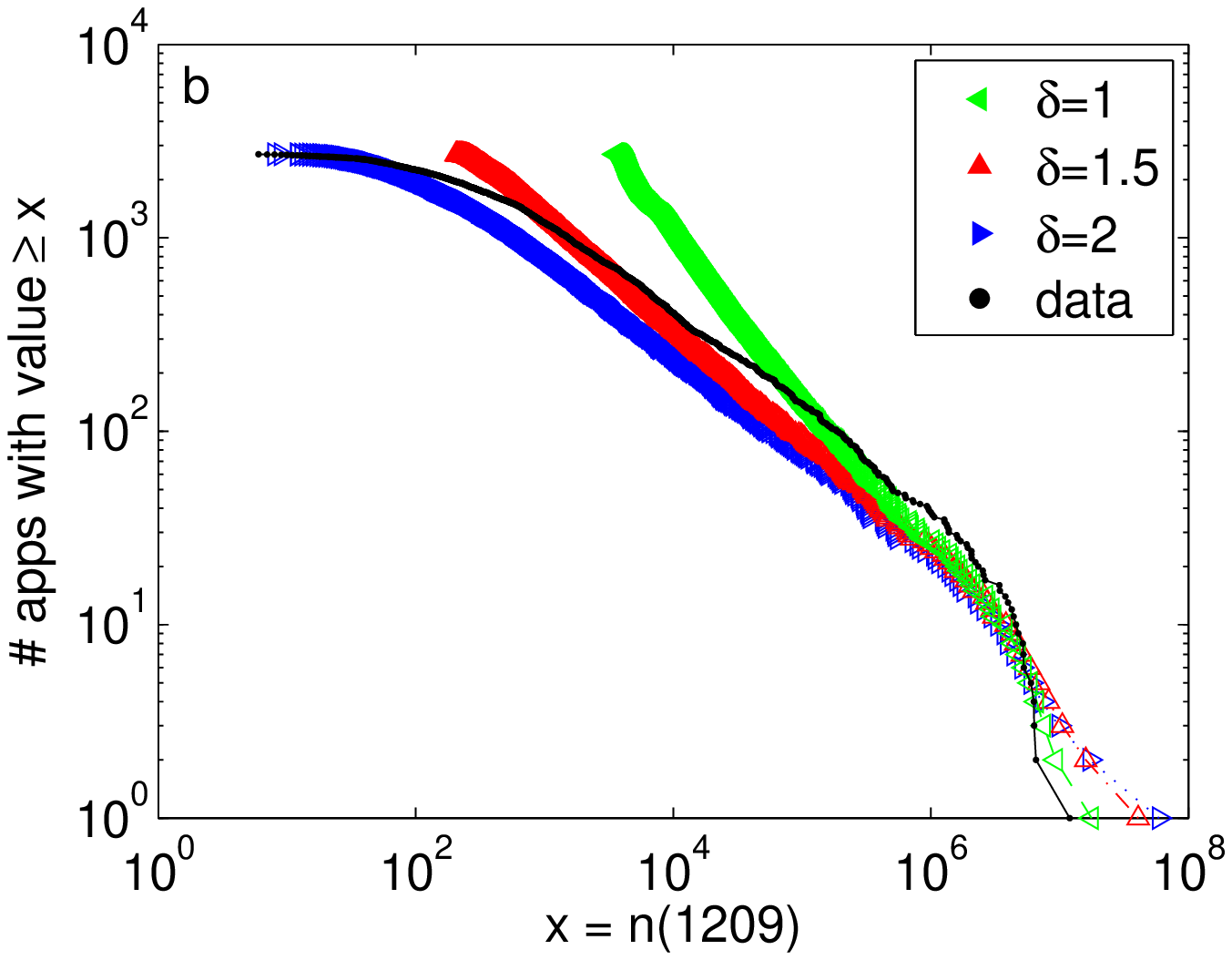,width=8cm}
\caption{As in Fig.~1a,b of the main text, but now we use the ranking model described in Section SI6.
} \label{figSI11}
\end{figure}

The ranking model introduced in Refs.~\cite{Fortunato06,Ratkiewicz10} suggests an alternative rule for how Facebook users might choose an app to install if they base their decisions only on a global listing of all apps according to their popularity. If an agent focuses only on the rank order of apps within the list and ignores the popularities (i.e., the numbers of installations) of the apps, then it is plausible that the probability of choosing app $i$ at time $t$ depends only on its ranking at time $t-1$. In the ranking model, this probability is
\begin{equation}
	p_i^r(t) = \frac{r_i^{-\delta}}{\sum_{j} r_j^{-\delta}}   \,, \label{rankmodel}
\end{equation}
where $r_i$ is the rank of app $i$ at time $t-1$ and the quantity $\delta$ is a tunable parameter. For example, the second-ranked app ($r_i=2$) is  $2^{\delta}$ times less likely to be chosen than the top-ranked app ($r_i=1$). Such rich-get-richer dynamics is different to the linear preferential attachment mechanism of Eq.~(3) of the main text, although it can also  lead to power-law distributions of popularity \cite{Fortunato06,Ratkiewicz10}.

In Fig.~\ref{figSI11}, we show the results of replacing the cumulative rule of Eq.~(3) from the main text with the ranking model rule (\ref{rankmodel}) while neglecting all recent information (i.e., putting $\gamma=1$). For $\delta=2$, the ranking model results are qualitatively similar to those of the linear preferential attachment case of Figs.~1d,e,f of the main text. Both models underpredict installations of LES apps, so the $r(a)$ curve is too low at large ages. For $\delta=1$, however, installations of (less-popular) LES apps are overpredicted by the ranking model, so the $r(a)$ curve in Fig.~\ref{figSI11}a is higher than the data curve at large $a$. In all cases---even $\delta=1.5$, for which the fit to $r(a)$ is reasonably good---the distributions of app popularities differ dramatically from the data (see Fig.~\ref{figSI11}b). We conclude that the ranking model, like the cumulative-information model that we considered in the main text, cannot provide a good fit to the data on Facebook apps.


\section*{{SI7: Cumulative-Rule Models Requiring Parameter-Fitting}}

{
In this section, we examine three extensions of the basic cumulative rule [see Eq.~(3) of the main text]. Unlike the parsimonious models that we studied in the main text, each of the extensions that we now consider includes multiple parameters---one or more for each app---that need to be fitted from the available data. In order to make a fair comparison with the results that we presented in Fig.~1 of the main text, we use a history window of $H=168$ hours to fit the parameters for each app, and we then implement a stochastic simulation using the appropriate version of the cumulative rule (with $\gamma=1$ in all cases).  In Fig.~\ref{figcumextns}, we present our results for mean scaled age-shifted growth rates, distributions of app popularity, and turnover plots. They should be compared with Fig.~1 of the main text, as that figure shows the corresponding results for the models that we described in the main text.

In each of the three models that we describe below, we define the probability $p_i(t)$ that app $i$ will be chosen by one of the $F(t)$ agents who install an app at time $t$.  To allow the models to be fitted to the history-window data of each app, we need to make an important assumption. We assume that the actual number $f_i(t)$ of installers of app $i$ at time $t$ is equal to its expected value:
\begin{equation}
	\hspace{-5 cm}\text{Assumption 1:} \quad\quad f_i(t) = p_i(t)\, F(t)\,.\label{Assumpt1}
\end{equation}
This assumption is likely to be good when the mean number of installers $p_i(t) F(t)$ is large, but it can be  inaccurate for unpopular apps that have small numbers of installations at time $t$.

To test the effect of Assumption~1, we calculate the exact installation probabilities $p_i(t)=f_i(t)/F(t)$ from the full data set and then insert these probabilities into a stochastic simulation (using a history window of $H=168$). We show the results of this calculation in Fig.~\ref{figcheck}, from which it is clear that Assumption 1 does not cause the simulation results to differ appreciably from the data (see Fig.~1a,b,c). This test also provides an important check on our stochastic simulations: when the probabilities $p_i(t)$ are set correctly, it is evident that the data can indeed be accurately reproduced by our simulations.

We proceed to consider several models that are based on extensions of the cumulative rule. Each model is motivated by an example from the extensive literature on modelling heavy-tailed distributions \cite{Simkin11,Simkin07,Borgs07,Wu07,Wang13,Shen14}.
In Sections SI7.1--SI7.3, we use $H=168$ data points for each app to fit the parameters of the model. This gives a fair comparison with the situations that we considered in the main text. In Section SI7.4, we check whether we can obtain better results if we use \emph{all} available data for model fitting. We conclude that the parsimonious recent-activity model of the main text gives superior performance to the alternatives that we consider in this section, as it can produce accurate results based on a history window of only 168 data points.

\begin{figure}
\centering
\epsfig{figure=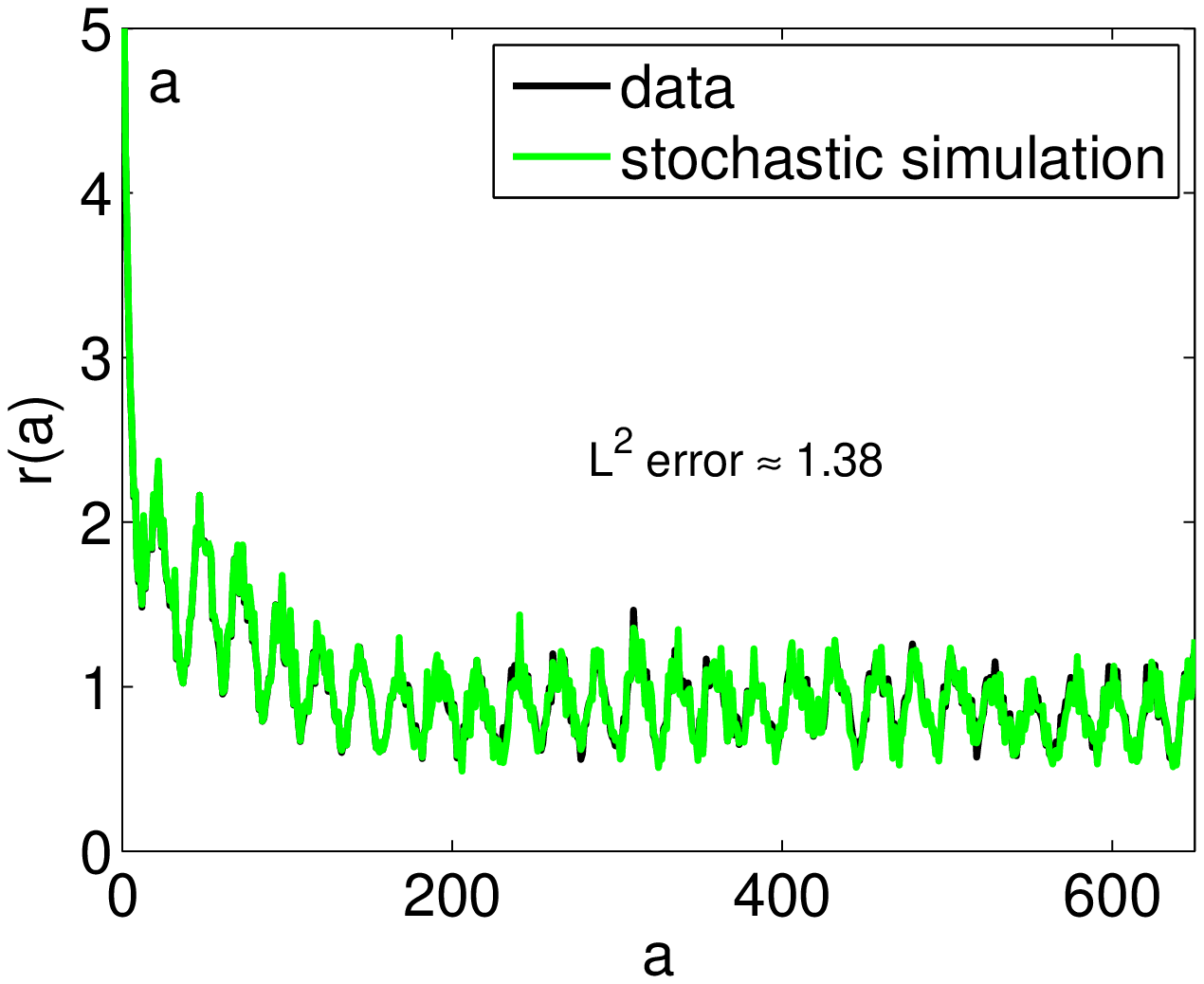,width=5.25cm} 
\epsfig{figure=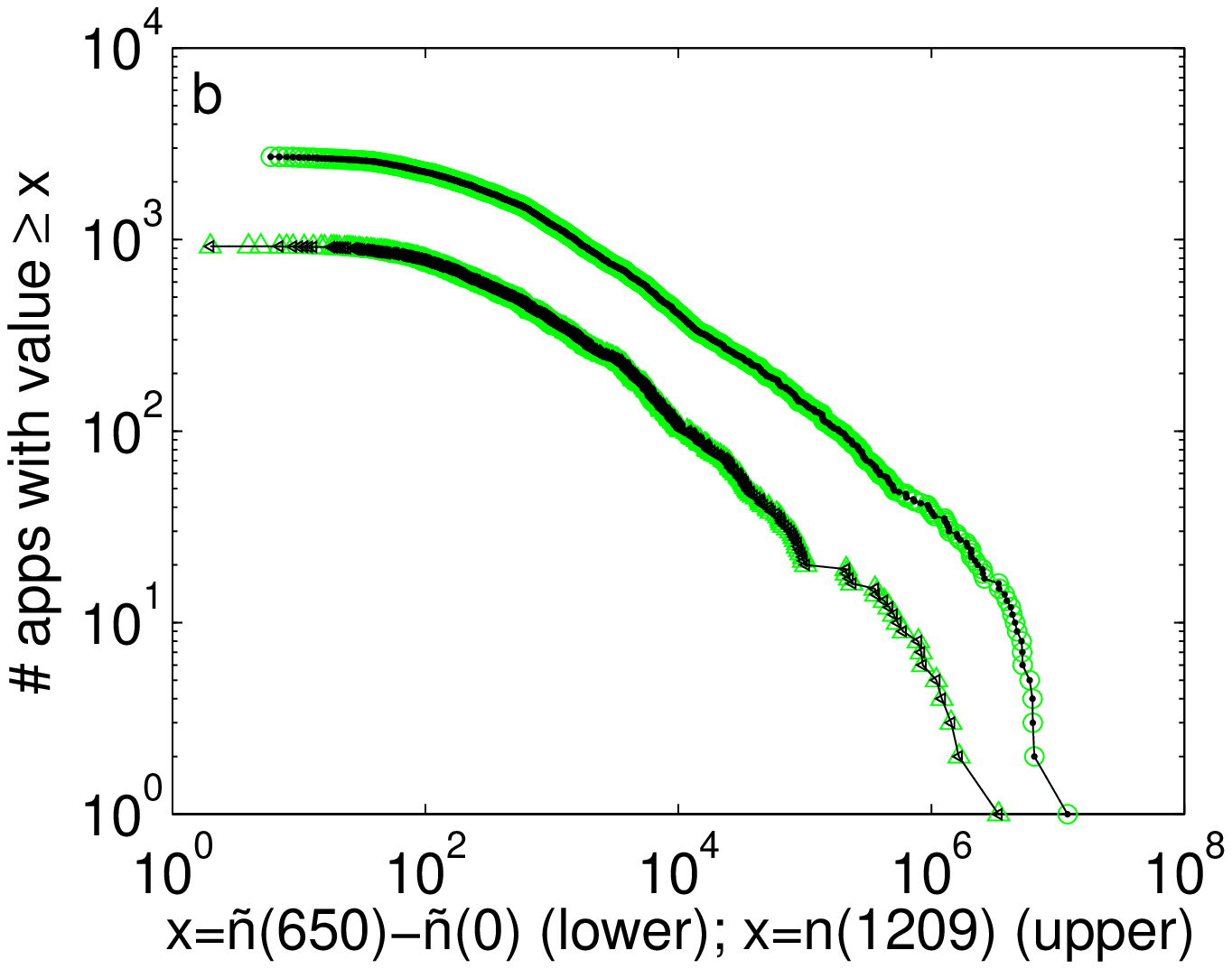,width=5.25cm}
\epsfig{figure=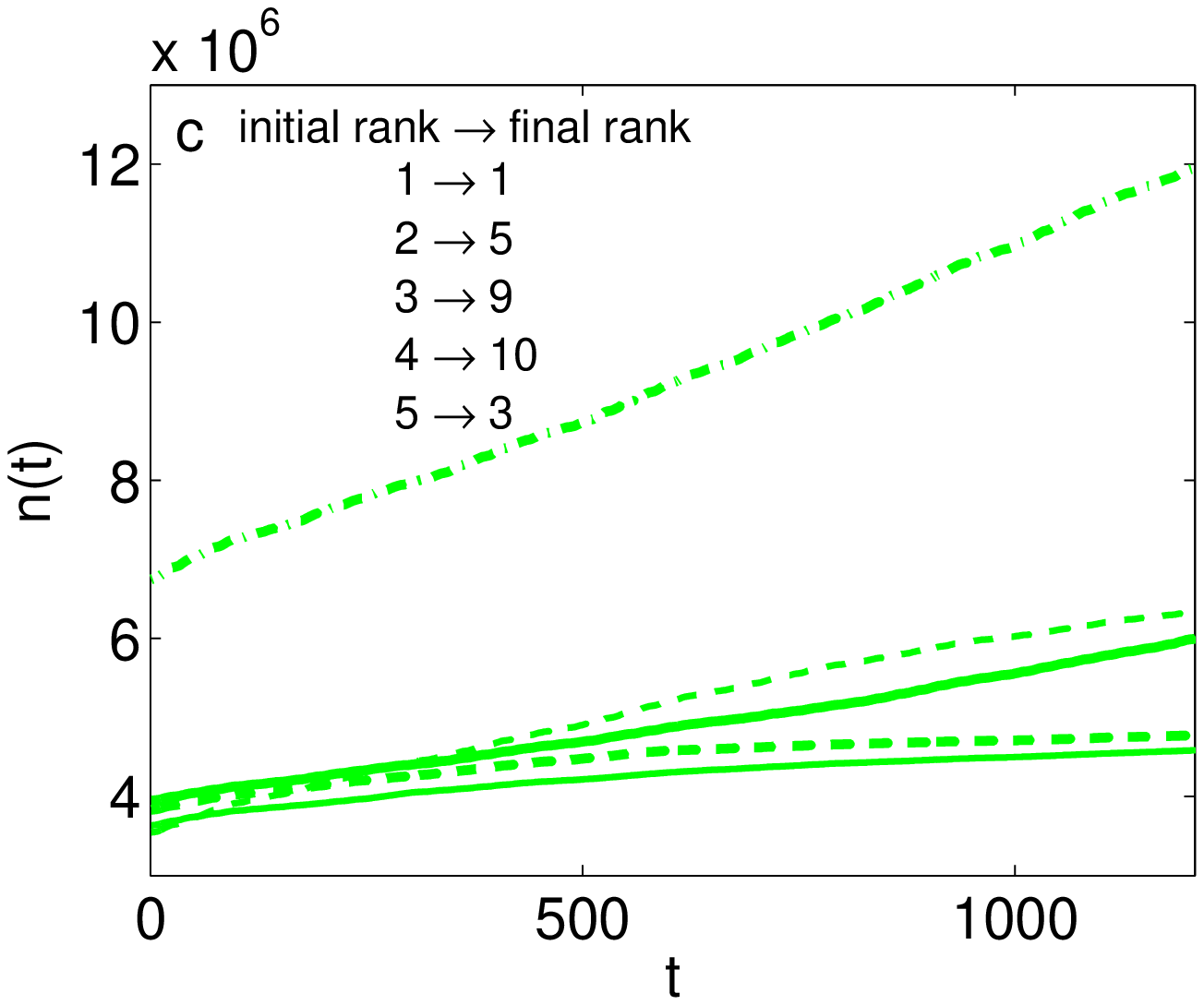,width=5.25cm} 
\caption{{ Results of using the exact installation probabilities $p_i(t)=f_i(t)/F(t)$ in stochastic simulators. The excellent match to data (compare these results to Figs.~1a,b,c of the main text) implies that any violations of Assumption 1 do not cause appreciable errors in the simulation results.
}
}
\label{figcheck}
\end{figure}

\begin{figure}
\centering
\epsfig{figure=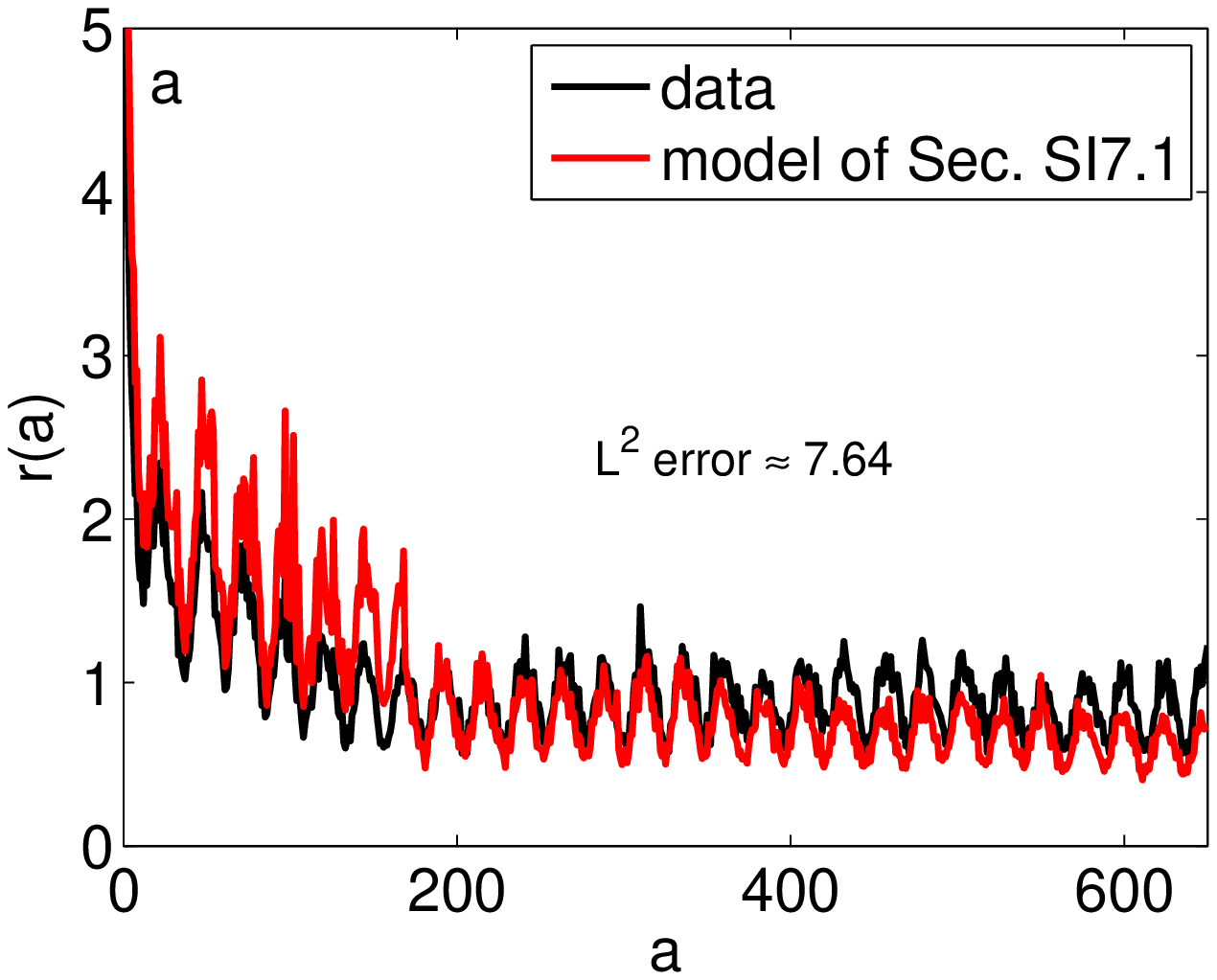,width=5.25cm} 
\epsfig{figure=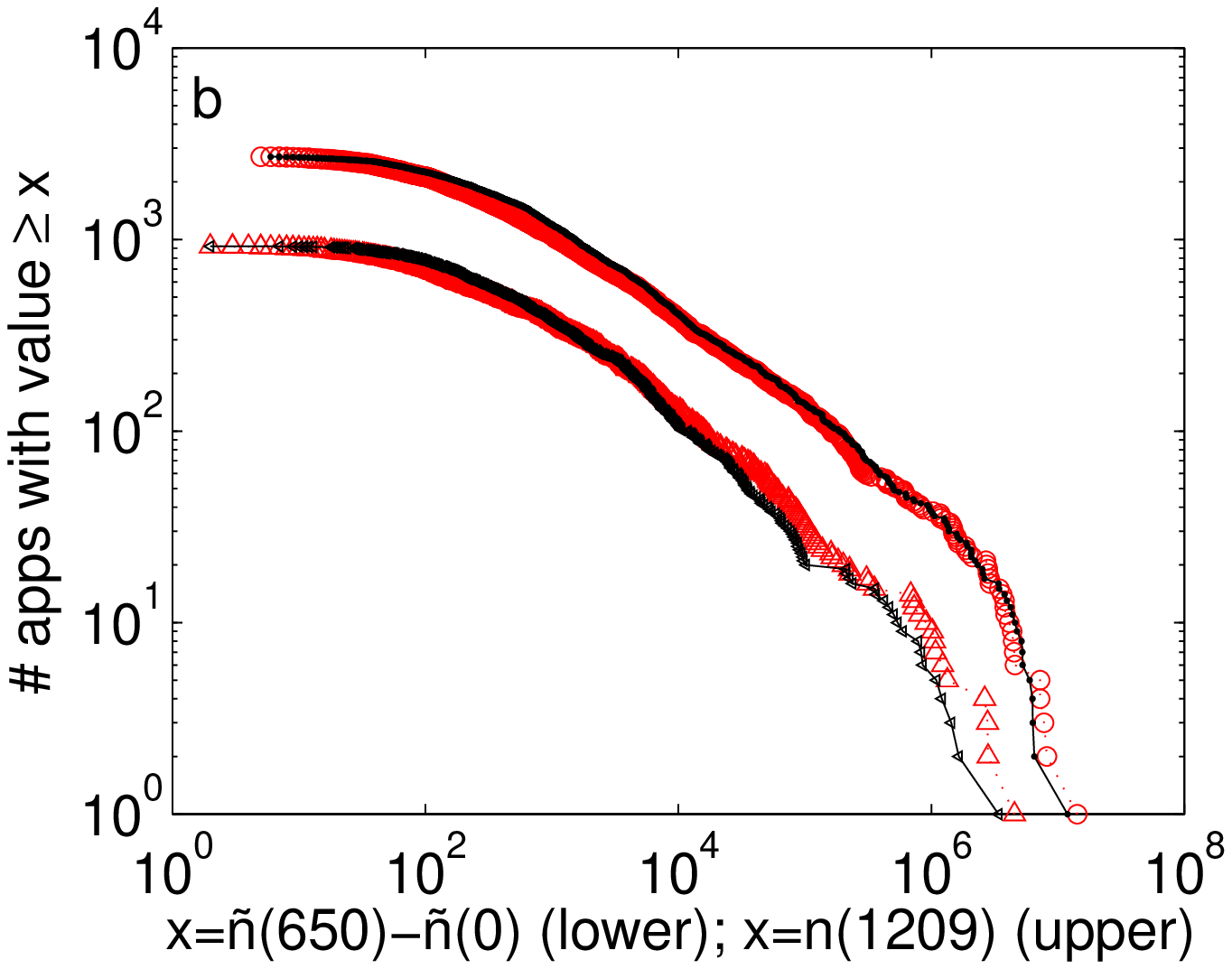,width=5.25cm}
\epsfig{figure=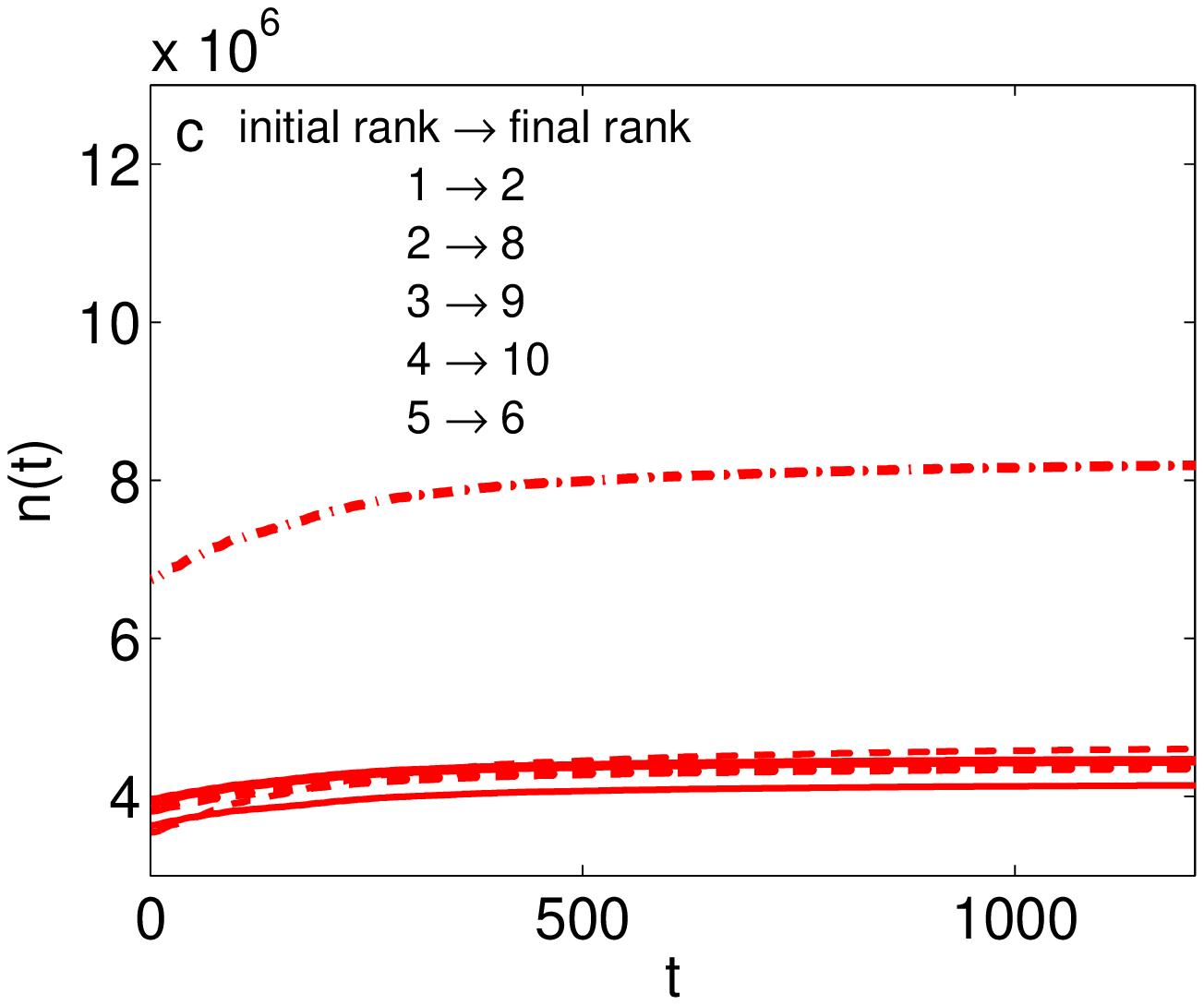,width=5.25cm} 
\epsfig{figure=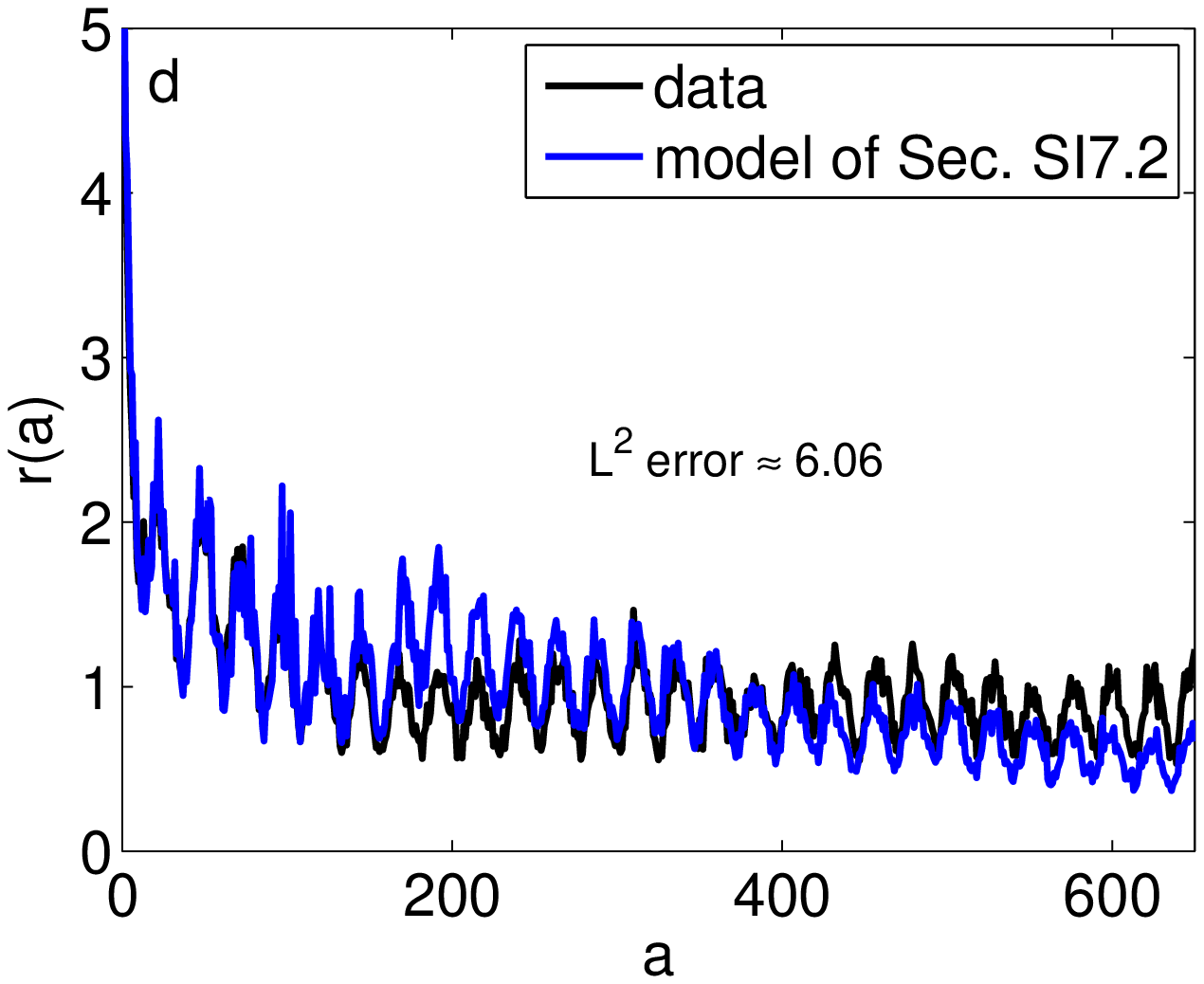,width=5.25cm}
\epsfig{figure=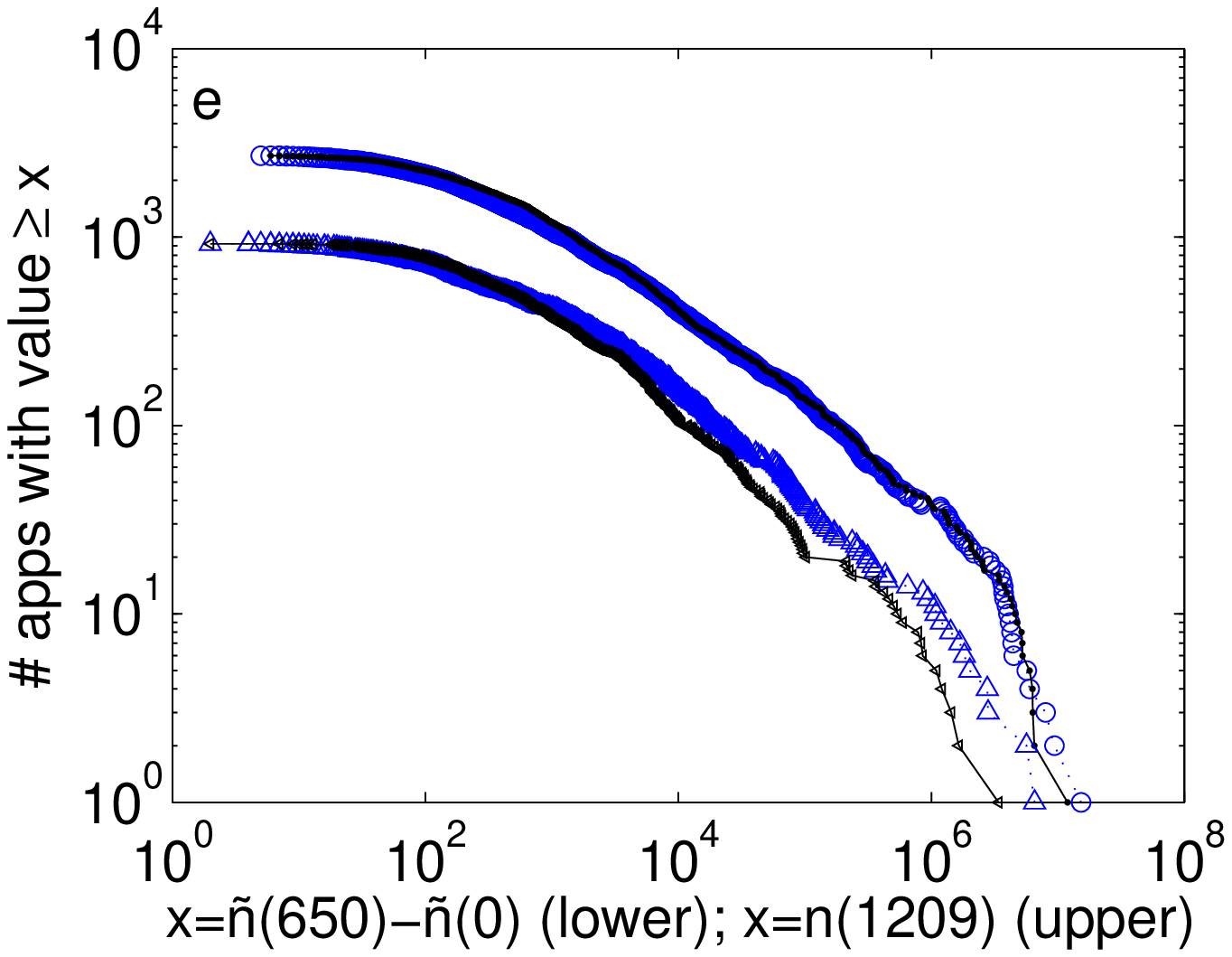,width=5.25cm}
\epsfig{figure=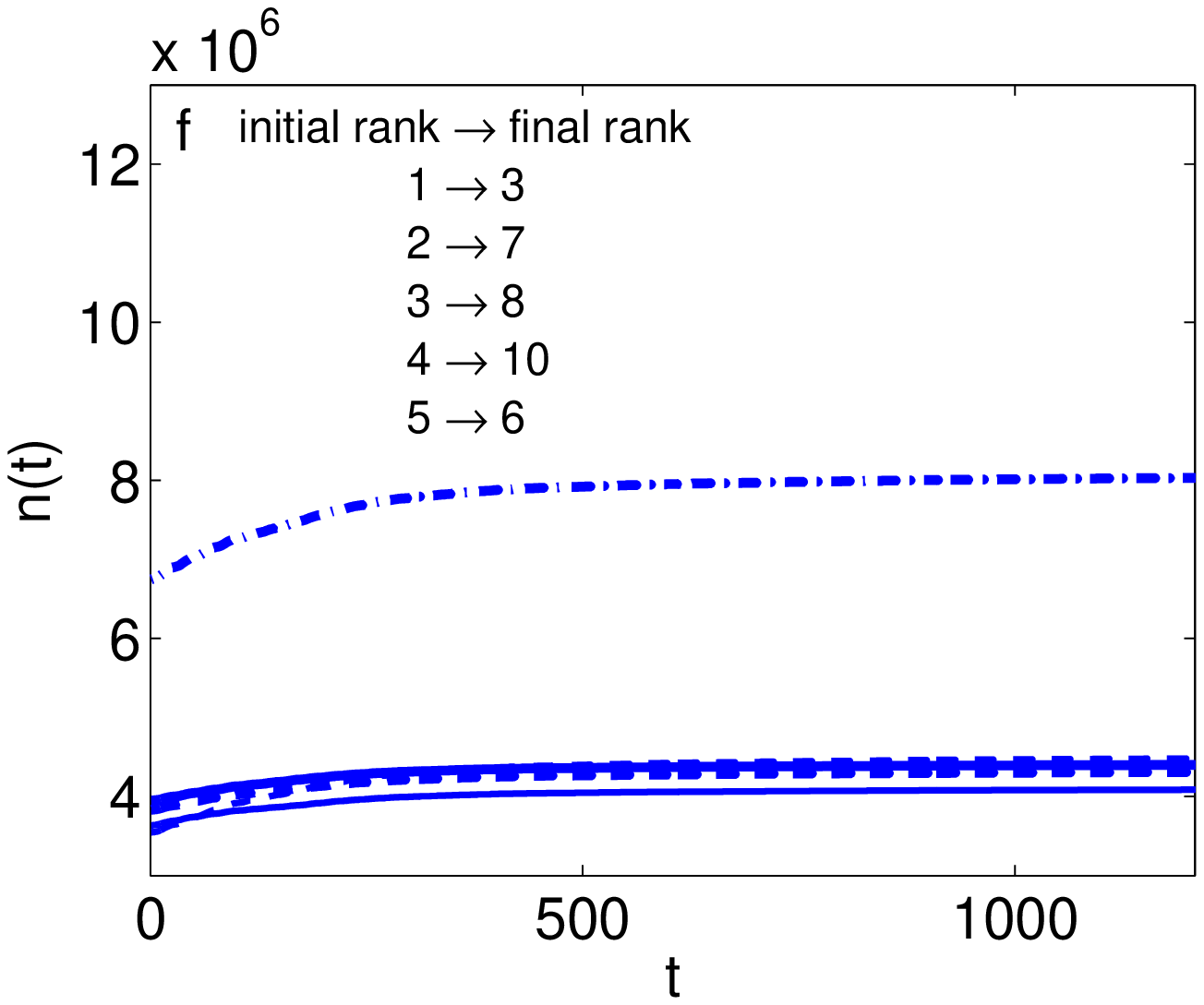,width=5.25cm} 
\epsfig{figure=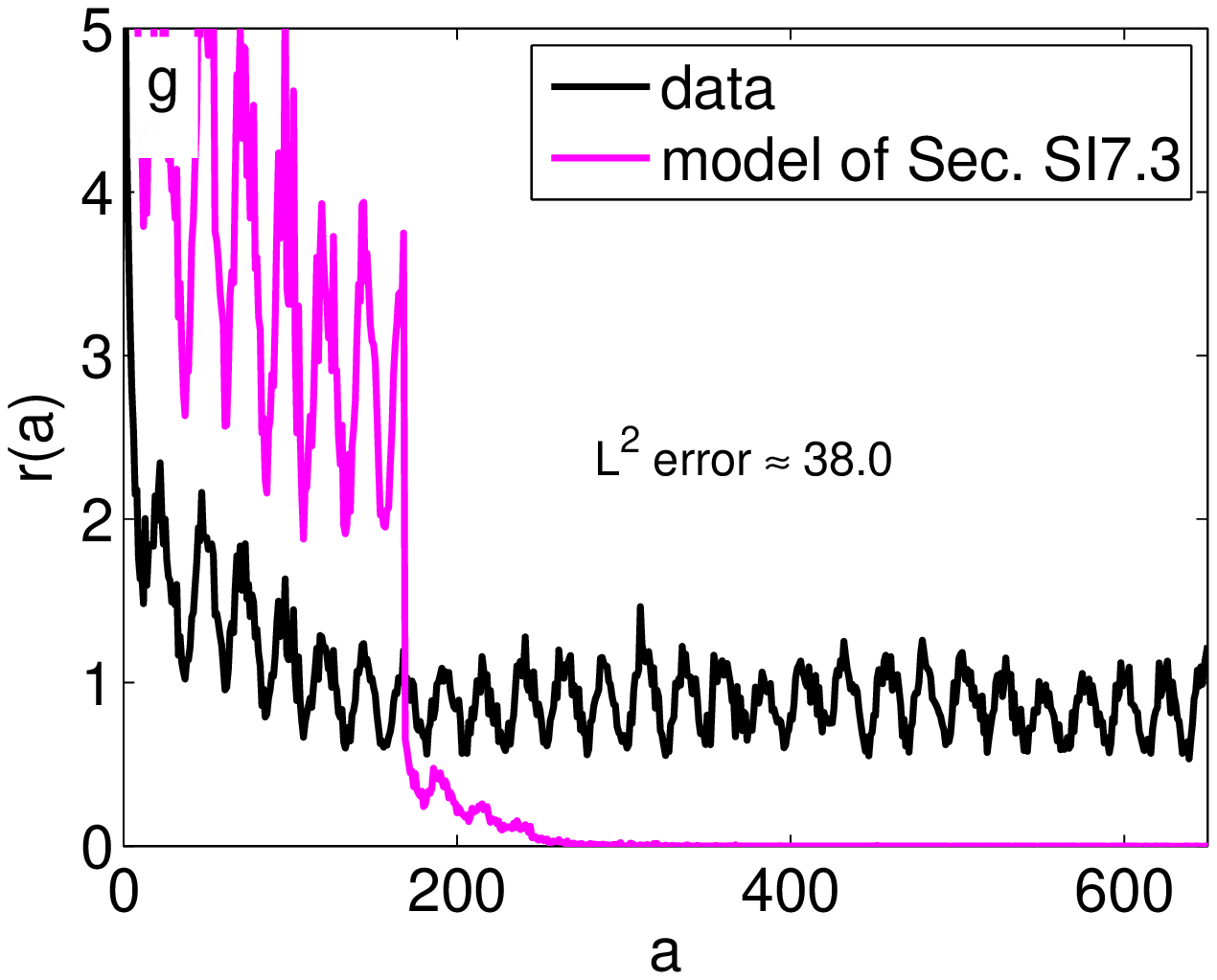,width=5.25cm}
\epsfig{figure=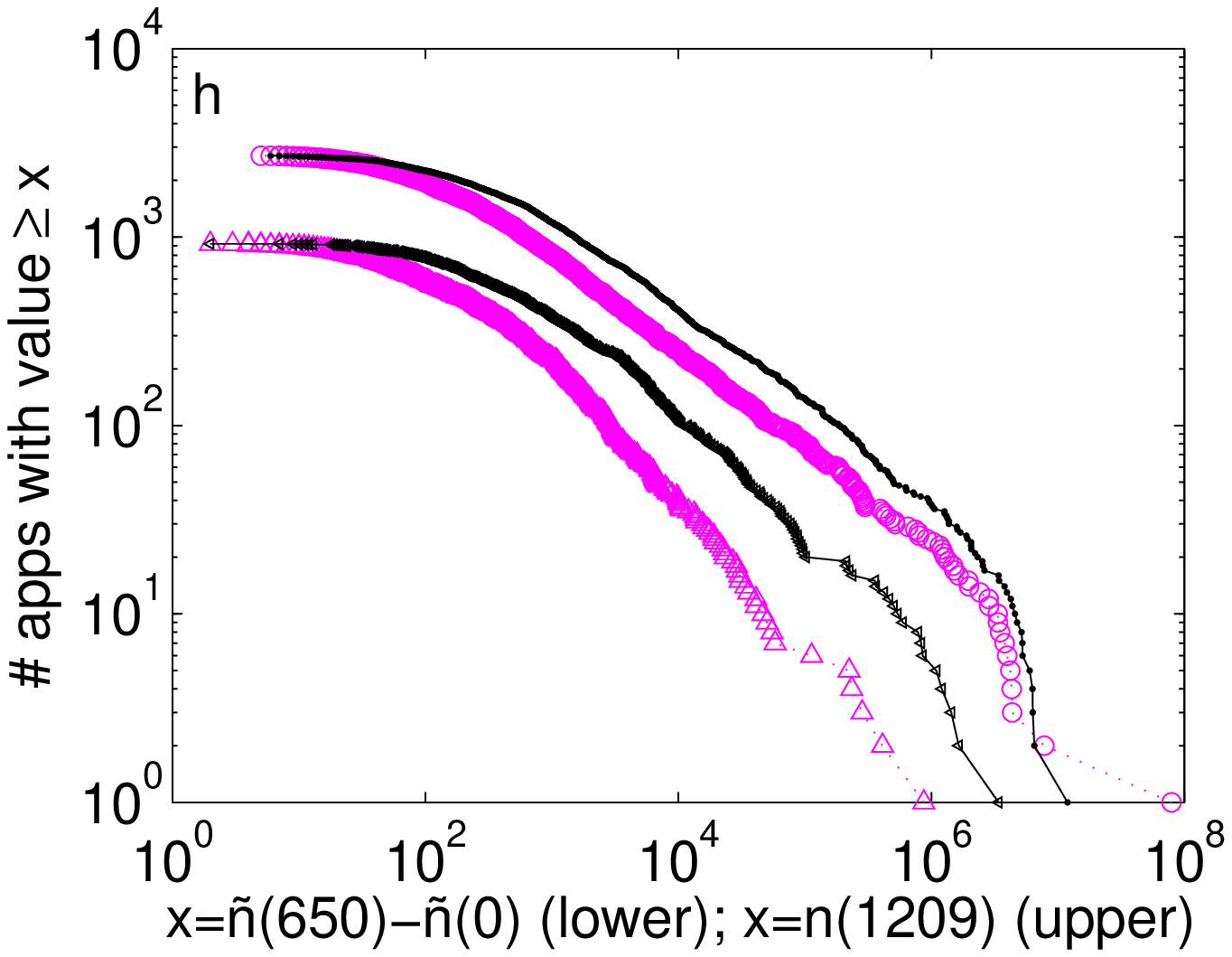,width=5.25cm}
\epsfig{figure=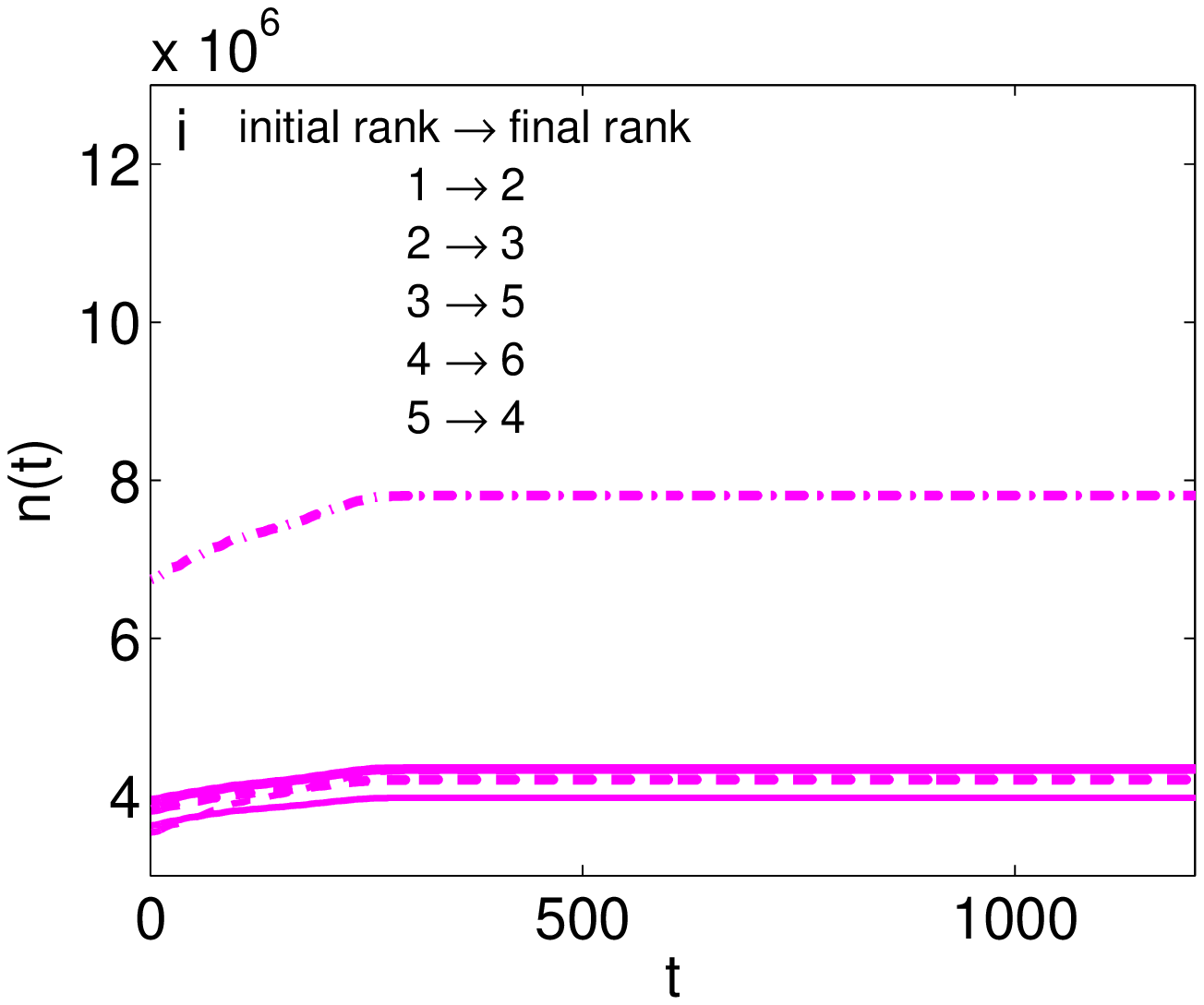,width=5.25cm} 
\epsfig{figure=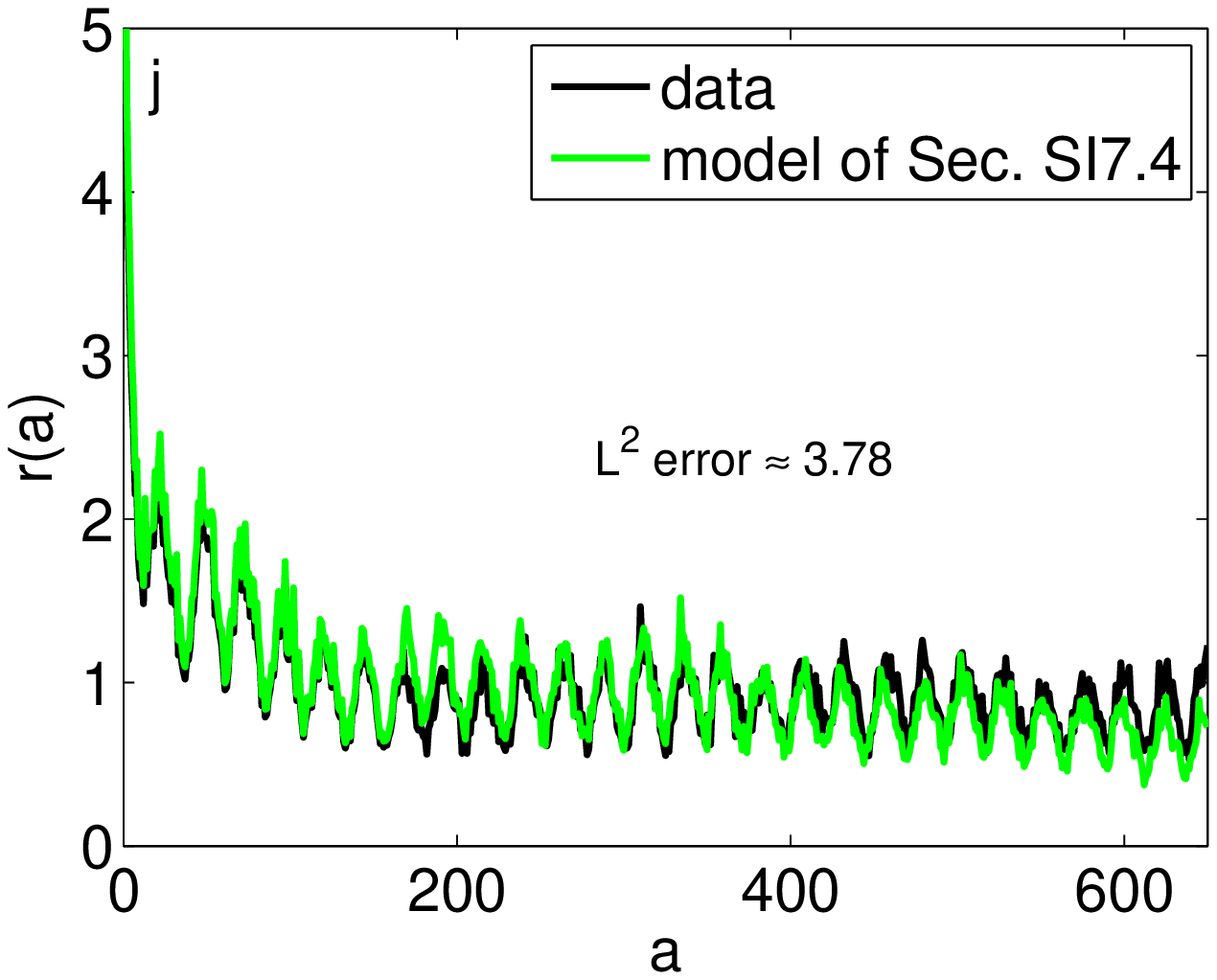,width=5.25cm}
\epsfig{figure=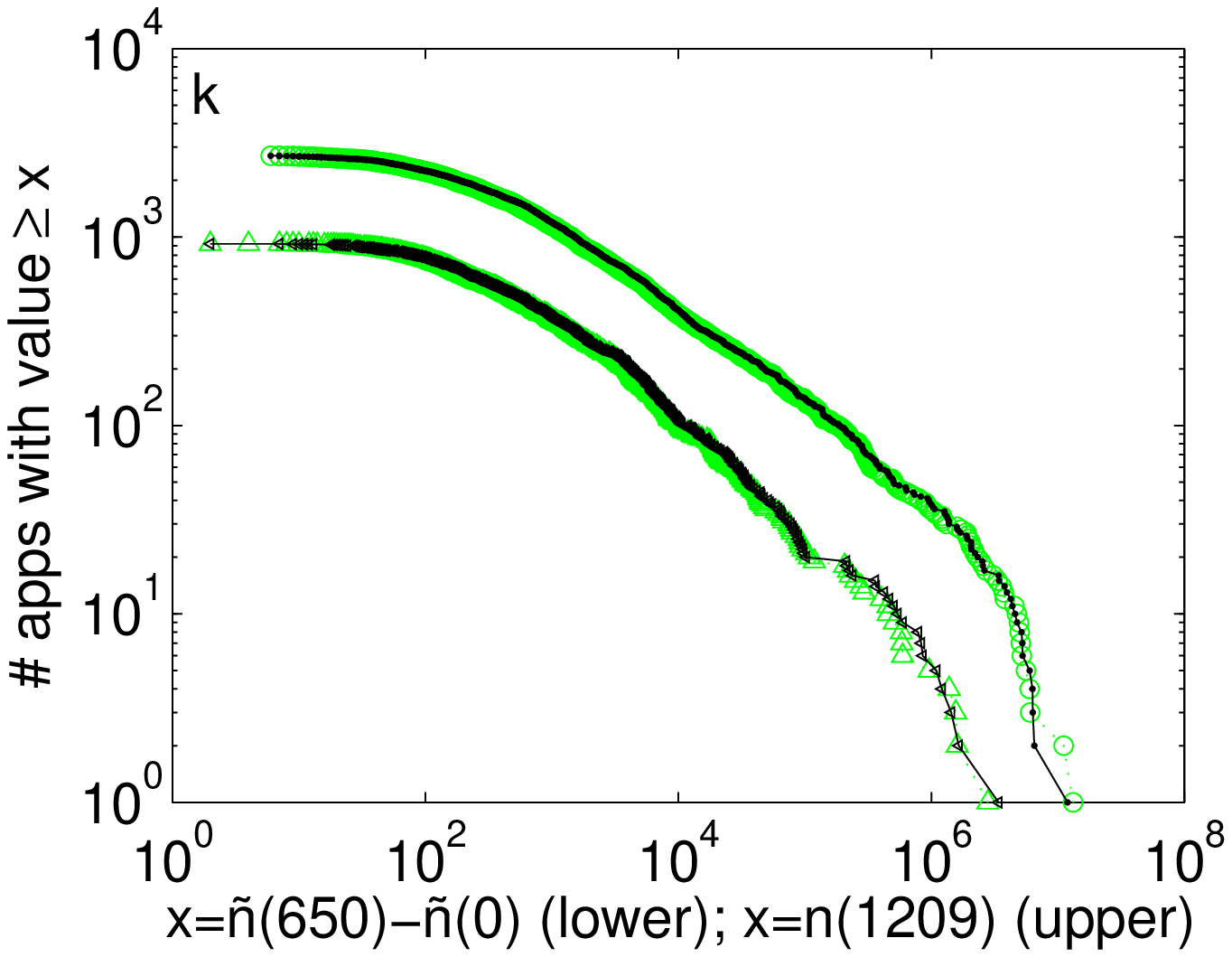,width=5.25cm}
\epsfig{figure=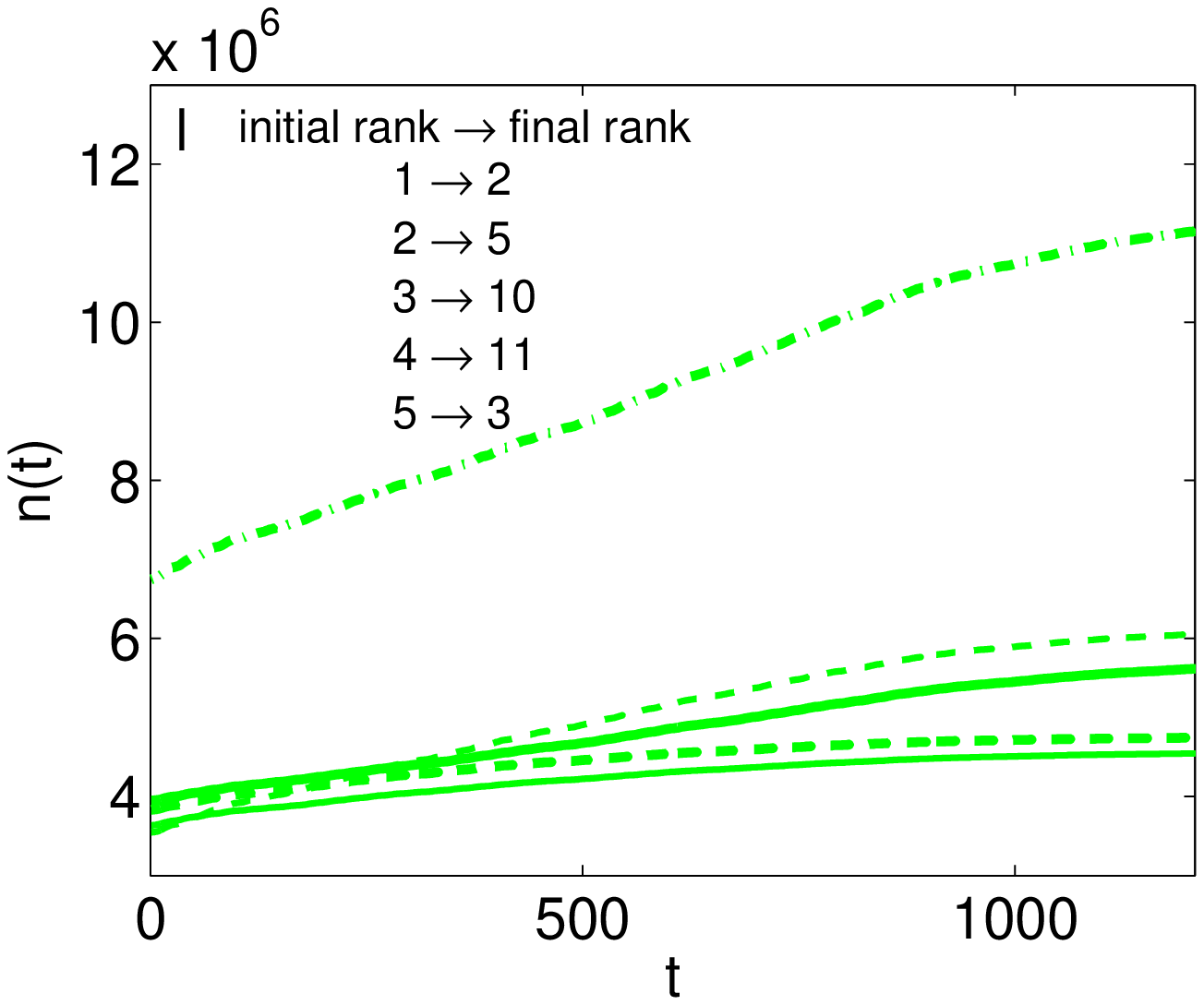,width=5.25cm} 
\caption{{ As in Fig.~1 of the main text (with $H=168$), but now we show results for the extensions to the cumulative-information model that we examine in Sections SI7.1--SI7.4. {(a,b,c)} Results for the cumulative-information-with-fitness model of Section SI7.1. {(d,e,f)} Results for the cumulative rule with fitness and novelty decay (see Section SI7.2). {(g,h,i)} Results for the model with app-specific novelty decay (see Section SI7.3). Our parameter estimation for panels (a)--(i) uses 168 data points for each app. {(j,k,l)} Results for the model with app-specific novelty decay, but with parameters estimated using all available data (see Section SI7.4).
}
}
\label{figcumextns}
\end{figure}


\subsection*{SI7.1: Cumulative Advantage with Heterogeneous Fitnesses}\label{secSI71}

The first extension of the basic cumulative rule (see Eq.~(3) of the main text) is based on the idea that a cumulative model supplemented with fitnesses can allow new entrants a head start \cite{Borgs07}.
To implement this idea, we replace the original cumulative rule by a refined version:
\begin{equation}
	p_i^c(t) = \lambda_i\, K\, n_i(t-1)\,, \label{cumfit}
\end{equation}
where $\lambda_i$ is the \emph{fitness} of app $i$ (cf.~Section SI8) and the constant $K$ is determined by the usual normalization: $\sum_i p_i^c(t)=1$ [so $K=1/\sum_i \lambda_i n_i(t-1)$]. As noted above, we use the history window (with $H=168$) of data for each app to infer the values of the $\lambda_i$ parameters and then run stochastic simulations based on the rule (\ref{cumfit}).

To estimate the  $\lambda_i$ values for this model, we begin with the full data set (i.e., the exact values of $f_i(t)$ and $n_i(t)$ for all $i$ and all $t$). If the rule (\ref{cumfit}) were exact, then Assumption 1 would imply that
\begin{align}
f_i(t) & = p_i^c(t)\,F(t)\nonumber\\
        &= \lambda_i\, K\, n_i(t-1)\, F(t) \nonumber\\
        & = \frac{\lambda_i\, n_i(t-1)}{\sum_j \lambda_j\, n_j(t-1)}F(t) \label{fit0}
\end{align}
for all times $t$ and all apps $i$. We can thus write the unknown $\lambda_i$ values in terms of known quantities:
\begin{equation}
	\frac{\lambda_i\, n_i(t-1)}{\sum_j \lambda_j\, n_j(t-1)} = \frac{f_i(t)}{F(t)}\,. \label{fit1}
\end{equation}
Recalling from Eq.~(1) in the main text that $F(t)=\sum_j f_j(t)$, we obtain a solution of Eq.~(\ref{fit1}) by setting $\lambda_i \, n_i(t-1)$ equal to $f_i(t)/F(t)$ for each app $i$. Solving for the fitnesses then yields
\begin{equation}
	\lambda_i = \frac{f_i(t)}{F(t)\, n_i(t-1)} \,. \label{fit2}
\end{equation}

Because the model is not exact, the right-hand side of Eq.~(\ref{fit2}) is not constant.
 To estimate the parameters in a manner consistent with the models that we study below, we sum both sides of Eq.~(\ref{fit2}) over the history window of app $i$ to obtain the relation
\begin{equation}
	\lambda_i (T-t_i) = \sum_{t=t_i+1}^T \frac{f_i(t)}{F(t) n_i(t-1)} \quad \text{ for } T \in \{t_i+1,\ldots,t_i+H\}\,. \label{fitnew1}
\end{equation}
We calculate the values of the right-hand side of this relation from the history-window data, and then estimate the parameter $\lambda_i$ using least-squares fitting on the $H$ data points.

In Figs.~\ref{figcumextns}a,b,c, we show the results of using the rule (\ref{cumfit}) with the fitness values inferred in the way that we just described. The turnover plot in Fig.~\ref{figcumextns}c highlights the shortcoming of this model: the app that was initially most popular (and that continues to grow linearly in time in the real data, as illustrated in Fig.~\ref{figcheck}c and in Fig.~1c of the main text) has a sudden decrease in installation rate as soon as it exits its history window.
 By comparing with the benchmark case of Fig.~\ref{figcheck}c, we identify the reason for this loss of popularity: the inferred fitnesses of many other apps give installation probabilities of $p_i^c(t)$ at time $t=H+1$ that substantially exceed their true probabilities from Eq.~(\ref{Assumpt1}). Because the total number of installing agents at each time step is restricted to be exactly $F(t)$, there is competition between the apps for the limited resource of agent attention. (Such competition has been examined in several data sets of online social networks \cite{lerman2013attent,lerman2013attent2,Weng12}.)
 Therefore, when many apps have installation probabilities that are too high, some other apps must suffer the consequence of fewer installations. In this case, the initially most-popular apps become victims of the intense competition. Apps that were initially less popular but have high fitnesses rise to take the top ranking by $t=t_\text{max}$.  It is clear that this model---despite having 2705 fitted parameters---does not do as well in reproducing the temporal behaviour of the data as the recent-activity model of the main text (see, e.g., Fig.~1l of the main text).


\subsection*{SI7.2: Cumulative Rule with Fitness and Novelty Decay}\label{secSI72}

Wu and Huberman \cite{Wu07} examined data from the news web site \url{digg.com} and proposed a model that includes an age-dependent decay in the novelty value of stories. In our notation, their basic idea is a further refined version of the cumulative rule of Eq.~(\ref{cumfit}):
\begin{equation}
	p_i^c(t) = \lambda_i\, K\, n_i(t-1)\, d(t-t_i)\,,\label{fit4}
\end{equation}
where $d = d(a)$ is a decaying function of its argument (recall $a=t-t_i$ is the age of app $i$ at time $t$, because $t_i$ is its launch time) that models the loss of attractiveness due to novelty decay over time. As before, $K$ is a normalization constant.

We begin by considering how to estimate the unknown parameters in Eq.~(\ref{fit4}) using only the data for each app within its history window (with $H=168$). Following the same steps as those leading from Eq.~(\ref{fit0}) to Eq.~(\ref{fit2}) yields the relation
\begin{equation}
	\lambda_i \, d(t-t_i) = \frac{f_i(t)}{F(t)n_i(t-1)} \label{fit5}
\end{equation}
for each app $i$ and for all times
$t>t_i$. Because the novelty-decay function $d(a)$ is assumed to be the same for all apps (we will relax this assumption in Section SI7.3), it can be computed explicitly, up to a scaling factor, by averaging Eq.~(\ref{fit5}) over all apps $i$ in the LES subset $\mathcal{I}$ (see Eq.~(2) of the main text).  We thereby obtain
\begin{equation}
	d(a) \propto \left< \frac{\tilde{f}_i(a)}{F(t_i+a) \tilde{n}_i(a-1)}\right>_\mathcal{I} \quad\text{ for }a \in \{1,\ldots,H\}\,.
\end{equation}

\begin{figure}
\centering
\epsfig{figure=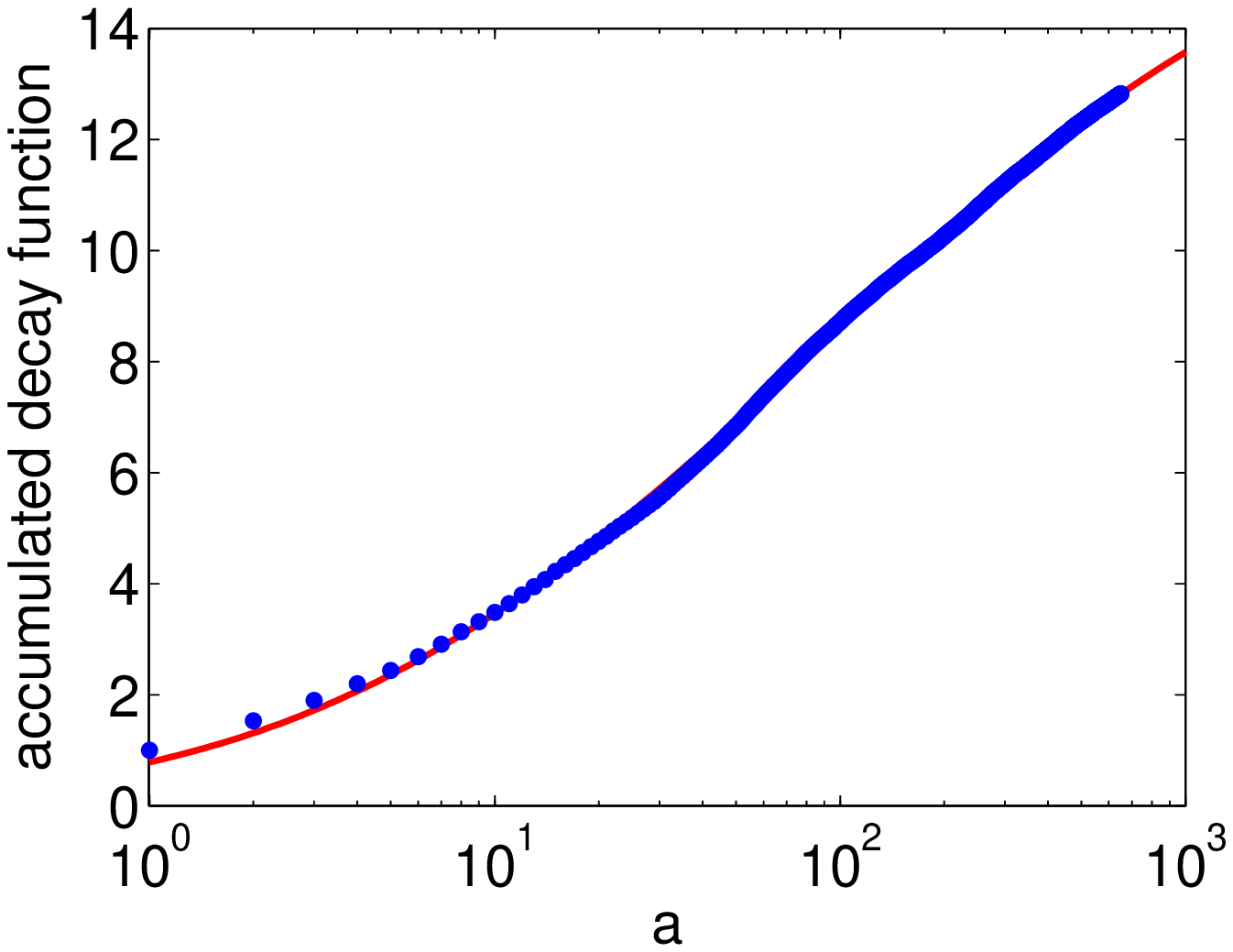,width=8cm}
\caption{{ Accumulated novelty-decay function---i.e., the left-hand side of Eq.~(\ref{fit6})---as a function of age $a$ fitted by $\lambda \Phi\left((\log(a)-\mu)/\sigma\right)$ with parameters $\lambda \approx 16.58$, $\mu \approx 4.47$, and $\sigma \approx 2.68$.
}
} \label{figd}
\end{figure}

We find that one can fit the novelty-decay function by a lognormal function of age. Specifically, Fig.~\ref{figd} illustrates a successful fit to the accumulated decay function
\begin{equation}
	\sum_{s=1}^a d(s) \propto \Phi\left(\frac{\log(a)-\mu}{\sigma}\right) \label{fit6}\,,
\end{equation}
where $\Phi(x)= (2\pi)^{-1/2} \int_{-\infty}^x e^{-y^2/2} dy$ is the cumulative normal distribution and the parameters of the lognormal decay function are $\mu \approx 4.47$ and $\sigma \approx 2.68$. This form of novelty decay contrasts to the stretched exponential function fitted to data from \url{digg.com} in \cite{Wu07}, but a lognormal decay function was successfully used in \cite{Wang13} to model the likelihood of a paper being cited at a time $a$ after its publication (see Section SI7.3).


Now that we have estimated the novelty-decay function using the LES apps, we determine the fitness parameter $\lambda_i$ for each LES app $i$ by summing both sides of Eq.~(\ref{fit5}) over the app's history window and using Eq.~(\ref{fit6}):
\begin{align}
	\lambda_i \sum_{t=t_i+1}^T d(t-t_i) &= \sum_{t=t_i+1}^T \frac{f_i(t)}{F(t) n_i(t-1)} \quad \text{ for } T \in \{t_i+1,\ldots,t_i+H\}\,, \nonumber\\
\Rightarrow \lambda_i \,\Phi\left(\frac{\log(T-t_i)-\mu}{\sigma}\right) &= \sum_{t=t_i+1}^T \frac{f_i(t)}{F(t) n_i(t-1)}\quad \text{ for } T \in \{t_i+1,\ldots,t_i+H \}\,. \label{fit7}
\end{align}
Because we know the right-hand side of Eq.~(\ref{fit7}) from the data, we can use least-squares fitting to the determine the best fit parameter $\lambda_i$ for each app. Recall that for those apps that are not launched in the study window, we set $t_i=0$; we examine the effects of this approximation in Section SI7.4.

Now that we have used the history window for each app to estimate the parameters for this model, we run stochastic simulations using rule (\ref{fit4}). We show our results in Figs.~\ref{figcumextns}d,e,f. As we also saw for the model of Section SI7.1, we observe that the competition between the apps quickly causes the growth of the largest apps to deviate from their exact trajectories, leading to a turnover plot (see Fig.~\ref{figcumextns}f) that is very different to that in the data (see Fig.~\ref{figcheck}c).


\subsection*{SI7.3: App-Specific Novelty Decay}\label{secSI73}

Wang, Song, and Barab\'{a}si recently proposed a cumulative-advantage model for the the number of citations that scientific papers garner over time \cite{Wang13}. In our notation, their model can be expressed in a manner similar to the Wu and Huberman model (see Section SI7.2), but with app-specific novelty-decay functions $d_i = d_i(a)$ replacing the universal decay function $d(a)$ of Eq.~(\ref{fit4}):
\begin{equation}
	p_i^c(t) = \lambda_i\, K\, n_i(t-1)\, d_i(t-t_i)\,.\label{fit9}
\end{equation}
Wang et al.~used a lognormal function to describe the novelty decay observed in their data, and our analysis of LES apps in Section SI7.2 supports a similar choice for our study. We therefore assume that the $d_i(a)$ are lognormal functions with app-specific parameters $\mu_i$ and $\sigma_i$. All of the derivations of Section SI7.2 also hold for this model, with the consequence that we estimate the values of $\lambda_i$, $\mu_i$, and $\sigma_i$ from the data by least-squares fitting of the relation [compare to Eq.~(\ref{fit7})]
\begin{equation}
	\lambda_i \,\Phi\left(\frac{\log(T-t_i)-\mu_i}{\sigma_i}\right) = \sum_{t=t_i+1}^T \frac{f_i(t)}{F(t) n_i(t-1)}\quad \text{ for } T \in \{t_i+1,\ldots,t_i+H \}\,. \label{fit10}
\end{equation}
As in Section SI7.2, we set $t_i=0$ for those apps that are not launched within the study window (see Section SI7.4).

As an aside, we note that one can make the connection to the model in Ref.~\cite{Wang13} explicit by taking the continuous-time approximation
\begin{equation}
	\tilde{f}_i(a) \approx \frac{d}{d a}\tilde{n}_i(a)\,,
\end{equation}
setting $F(t)$ to be constant, and replacing sums by integrals. Equation (\ref{fit10}) then becomes
\begin{equation}
	\lambda_i \,\Phi\left(\frac{\log(a)-\mu_i}{\sigma_i}\right) = \int_0^a \frac{1}{\tilde{n}_i(a')} \frac{d}{d a'} \tilde{n}_i(a')\, da'\,,
\end{equation}
and its solution
\begin{equation}\label{sosol}
	\tilde{n}_i(a) \propto \exp\left[ \lambda_i \, \Phi\left(\frac{\log(a)-\mu_i}{\sigma_i}\right)\right]
\end{equation}
gives the popularity of app $i$ at age $a$.  Equation (\ref{sosol}) reproduces, up to an additive constant, Eq.~(3) of Ref.~\cite{Wang13}.

Returning to the least-squares fitting of Eq.~(\ref{fit10}), we estimate
the $3\times 2705$ parameters for this model, and then use our stochastic-simulation framework to make predictions. We show our results in Figs.~\ref{figcumextns}g,h,i. As we have seen for the other extensions of the cumulative-information model, competition between apps (which is not considered in Refs.~\cite{Wu07,Wang13}) amplifies any error in the fitting functions of the model. We conclude that none of these adaptations of cumulative-advantage models provide a generative mechanism that describes the Facebook apps data as well as the recent-activity model that we described in the main text.


\subsection*{SI7.4: App-Specific Novelty Decay Using All Data}\label{secSI74}

As we noted in Sections SI7.2 and SI7.3, there are 980 apps that were launched prior to the study window and thus have unknown launch times. Throughout our work, we assume that $t_i=0$ for these apps. It is possible that this assumption might adversely affect the fitting of the models that rely on age-dependent novelty decay, as the ages of some apps will be misrepresented.
To check the impact of this assumption, we therefore recalibrate the model of Wang et al.~\cite{Wang13} that we described in Section SI7.3 by using \emph{all} available data for every app to estimate parameters rather than just the 168 hours used in Section SI7.3 as a priori information. To do this, we replace the set of $T$ values in Eq.~(\ref{fit10}) by $T \in \{t_i+1, \ldots, t_\text{max}\}$.
The extra data is helpful for the model, as it enables it to perform much better in stochastic simulations---see Figs.~\ref{figcumextns}j,k,l---although it is still not quite as accurate as the recent-activity model of Figs.~1j,k,l of the main text. The improvement in accuracy from using extra data implies that the inaccurate launch times do not prevent this model from fitting reasonably well to the data. However, the quantity of data required to estimate the parameters is
much larger than the history window of 168 hours that suffices to produce good results for the recent-activity model of Figs.~1j,k,l.  The model of Wang et al.~also has many more fitting parameters than the model that we presented in the main text.
}


\section*{SI8: Recent-Activity Model With Heterogeneous Fitnesses}

We now consider replacing the recent-activity rule [see Eq.~(4) of the main text] with an alternative that includes a fitness parameter $\lambda_i$ for app $i$. The refined recent-activity rule is
\begin{equation}
	p_i^r(t) = \lambda_i L  \sum_{\tau=0}^{t-1} W(t-\tau) f_i(\tau)\,, \label{4new}
\end{equation}
which is normalized so that $\sum p_i^r(t) =1$.  All else being equal, apps with higher fitnesses are more likely to be selected for installation than apps with lower fitnesses. Thus far for the recent-activity model, we have focused on the so-called \emph{neutral-model} \cite{Bentleybook,Pinto11} scenario, in which all fitnesses are equal (with $\lambda_i=1$ for all $i$). Noting from Fig.~1k of the main text that some of the largest LES app popularities are underpredicted by the otherwise successful recent-activity, long-memory model with homogeneous fitnesses (e.g., for $n \in [10^5,10^6]$), it is natural to ask whether heterogeneous fitnesses might lead to a better fit to the data.

\begin{figure}
\centering
\epsfig{figure=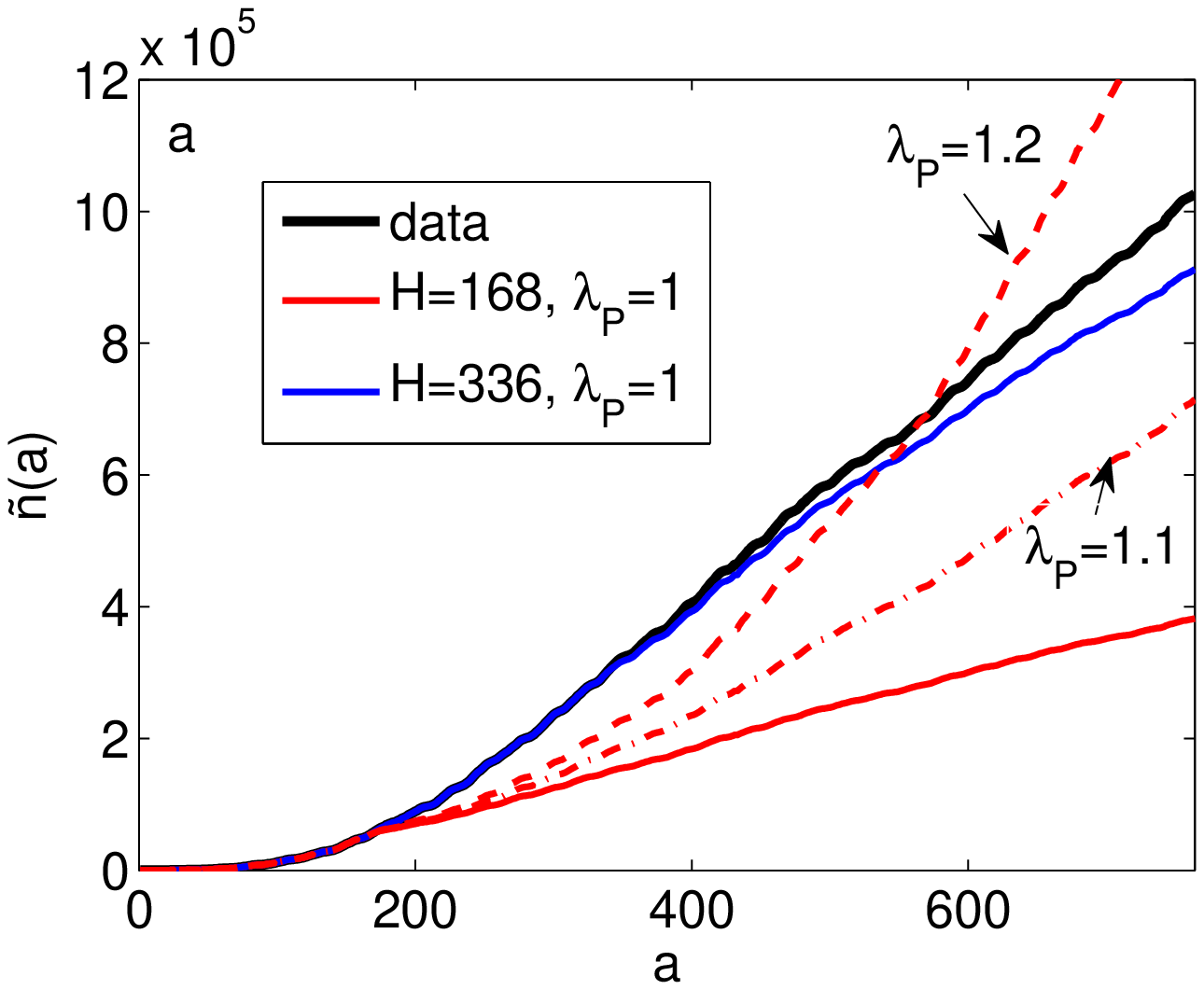,width=8cm}
\epsfig{figure=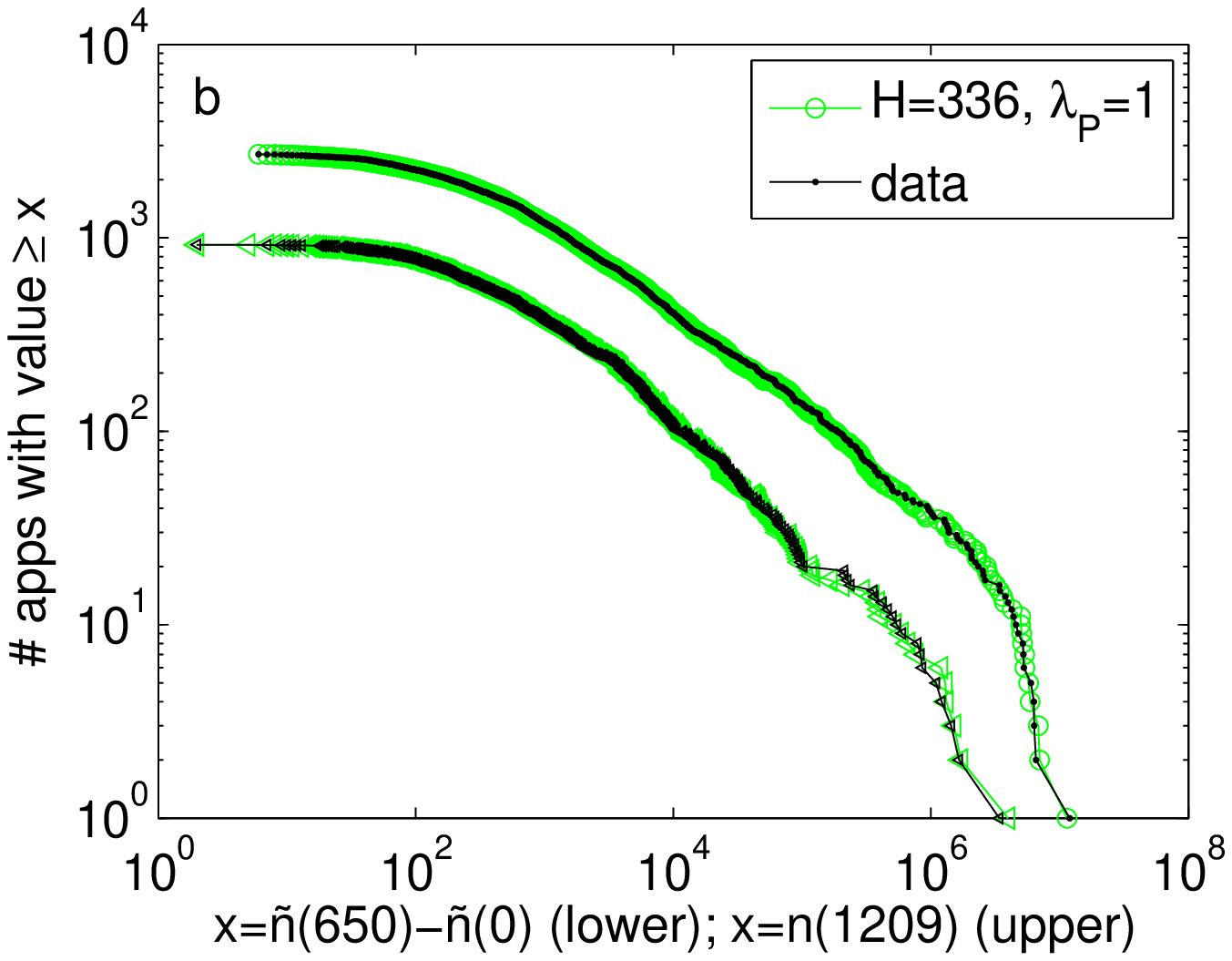,width=8cm}
\caption{(a) Growth trajectory for ``Pirates vs.~Ninjas'' from data (black) and from models (red and blue). (b) Popularity distributions as in Figs.~1e,h,k but for $H=336$ hours (i.e., 2 weeks).
} \label{figSI8}
\end{figure}

In Fig.~\ref{figSI8}, we show the growth of ``Pirates vs.~Ninjas'', the 7th most popular (at age $t = t_\text{LES}$) LES app (see panel 7 of Fig.~\ref{figSI2}). This is one of the apps in which the recent-activity, long-memory model of the main text with a 1-week history window gives an inaccurate prediction (solid red curve).
This leads to notable differences between the popularity distributions of LES apps in Fig.~1h of the main text near $n=10^6$.  We thus consider changing the fitness of this particular app to a value $\lambda_P > 1$, while maintaining $\lambda_i = 1$ for all other apps.  In Fig.~\ref{figSI8}a, we show the results of typical simulations using the dashed red curves. Although it is clearly possible to increase the popularity of this app by changing its fitness, we note that the $\lambda_P>1$ trajectories exhibit increasing curvature, and the growth
is super-linear in time rather than linear in time.
For comparison, we also show results of an equal-fitness simulation in which we use a larger history window of 2 weeks (i.e., $H=336$ hours) for all apps. In this case, the model's linear growth is much closer to the data, because the history window now includes the transition from novelty to post-novelty regimes (see Table~\ref{tabSI1} and the heuristic fit of Fig.~\ref{figSI2}) at about 236 hours (i.e., about 1.4 weeks). The plots in Fig.~\ref{figSI8}b confirm that using this longer history window leads to a much closer match between model and data.

We conclude that there does not appear to be strong evidence for heterogeneous fitnesses [as defined in our model through Eq.~(\ref{4new})] among the apps, at least in the post-novelty regime. This conclusion is consistent with the findings of Bentley et al.~regarding the applicability of the neutral model to other instances of choice among multiple alternatives \cite{Bentley04} as well as with the experimental results of Salganik et al.~\cite{Salganik06}, who showed that attractiveness of downloaded music is influenced more heavily by the actions of other downloaders than by the inherent quality of the music itself.

\section*{SI9: Fluctuation-Scaling Relations}

\begin{figure}
\centering
\epsfig{figure=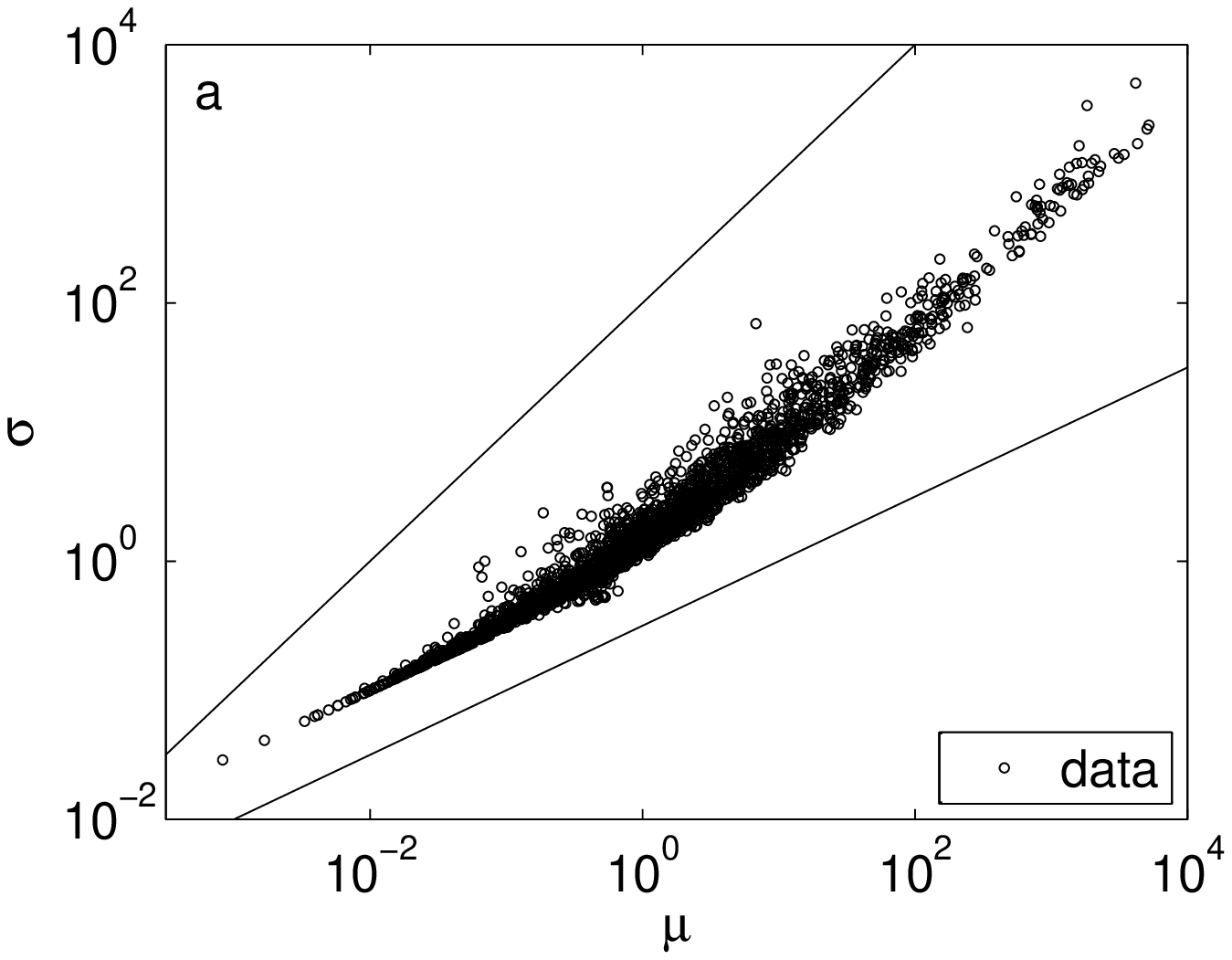,width=8cm}
\epsfig{figure=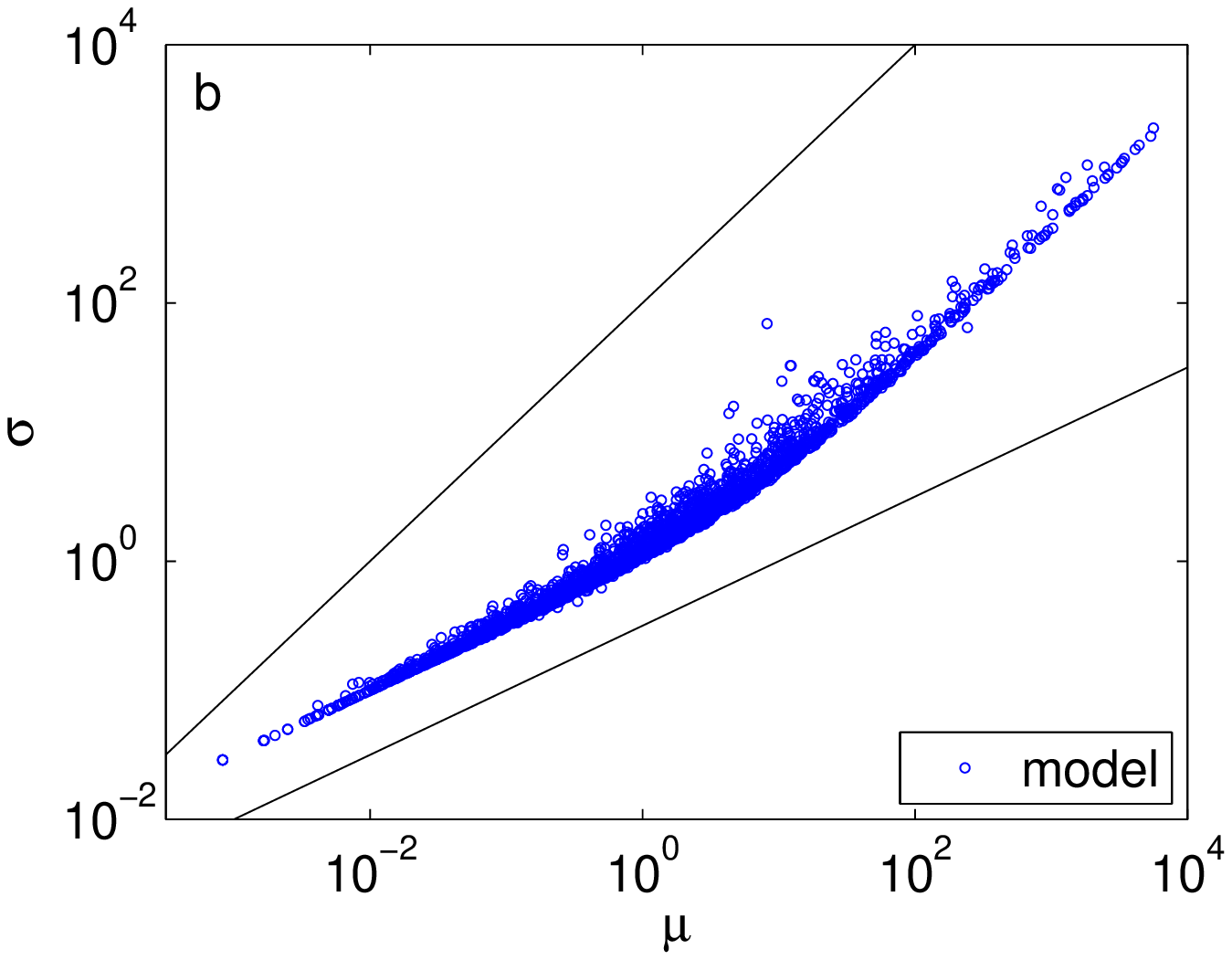,width=8cm}
\caption{Fluctuation-scaling plots for (a) data and (b) the recent-activity, long-memory model described in the main text. The straight lines correspond to scaling exponents of $\beta=1/2$ (lower line) and $\beta=1$ (upper line).}
 \label{fig9}
\end{figure}

In Fig.~\ref{fig9}a, we show a fluctuation-scaling (FS) plot of the Facebook apps data. As in Ref.~\cite{Onnela10}, we calculate for each app $i$ the mean $\mu_i$ and standard deviation $\sigma_i$ of the increments $f_i(t)$ over times $t$ from launch time $t_i$ (recall we set $t_i=0$ if the launch time is unknown) to the end of the data (i.e., $t=t_\text{max}$). We then plot $\mu_i$ versus $\sigma_i$ for all $i$ to generate Fig.~\ref{fig9}a. Reference~\cite{Onnela10} highlighted the existence of two
FS regimes: the relation $\sigma_i\sim \mu_i^\beta$ with $\beta\approx 1/2$ is evident for small-$\mu_i$ apps, whereas a larger $\beta$ value ($\beta\approx 0.85$) occurs for large-$\mu_i$ apps. In Fig.~\ref{fig9}b, we show the corresponding FS plot for the simulated results from the recent-activity, long-memory model of Figs.~1j,k,l in the main text. Clearly, the plot is qualitatively similar to that of the data. In particular, it has scaling regimes with FS exponents of $\beta\approx 1/2$ for low-$\mu_i$ (i.e., low popularity) apps and $\beta\approx 1$ for high-$\mu_i$ (i.e., high popularity) apps. We now use our model to further analyze these two regimes.  (The possible nature of the transition between these regimes is discussed in Ref.~\cite{Onnela10}.)


As we discuss below, our model reveals that the $\beta \approx 1$ scaling of the large-$\mu_i$ apps is related intimately to the large diurnal oscillations in Facebook user activity. Recall that we represent such oscillations at the population level using the function $F(t)$. In simulations using non-oscillatory versions of $F(t)$, we find that the $\beta=1/2$ regime extends to much larger values of $\mu_i$, which suggests that the $\beta\approx 1$ regime in Fig.~\ref{fig9} appears because very popular apps exhibit coherent diurnal oscillations in their levels of installation activity.  By contrast, small-$\mu_i$ apps receive a mean of fewer than 2 installations per hour, and their $f_i(t)$ time series appear similar to shot noise, for which one expects an FS exponent of $\beta=1/2$.


\begin{figure}
\centering
\epsfig{figure=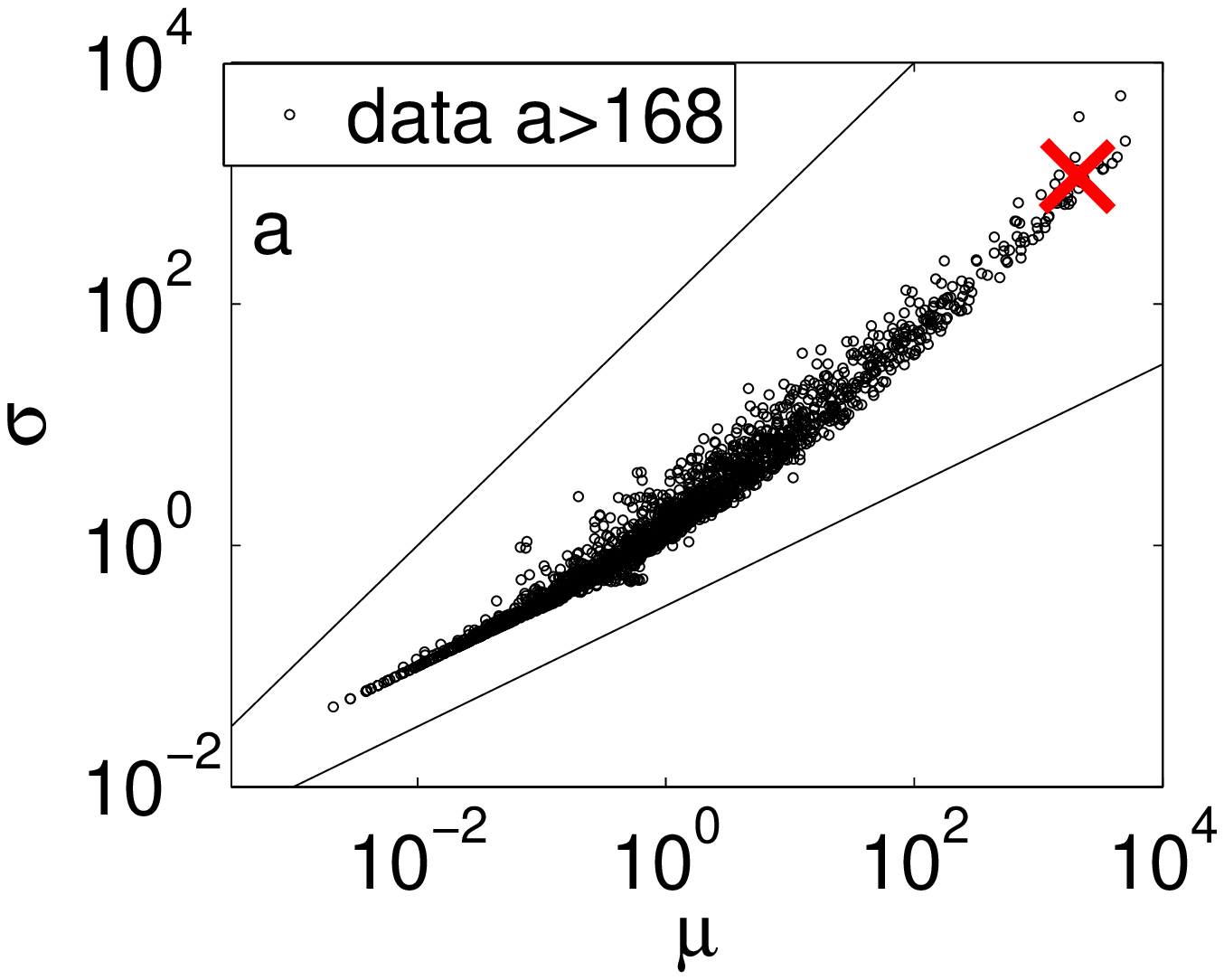,width=5.2cm}
\epsfig{figure=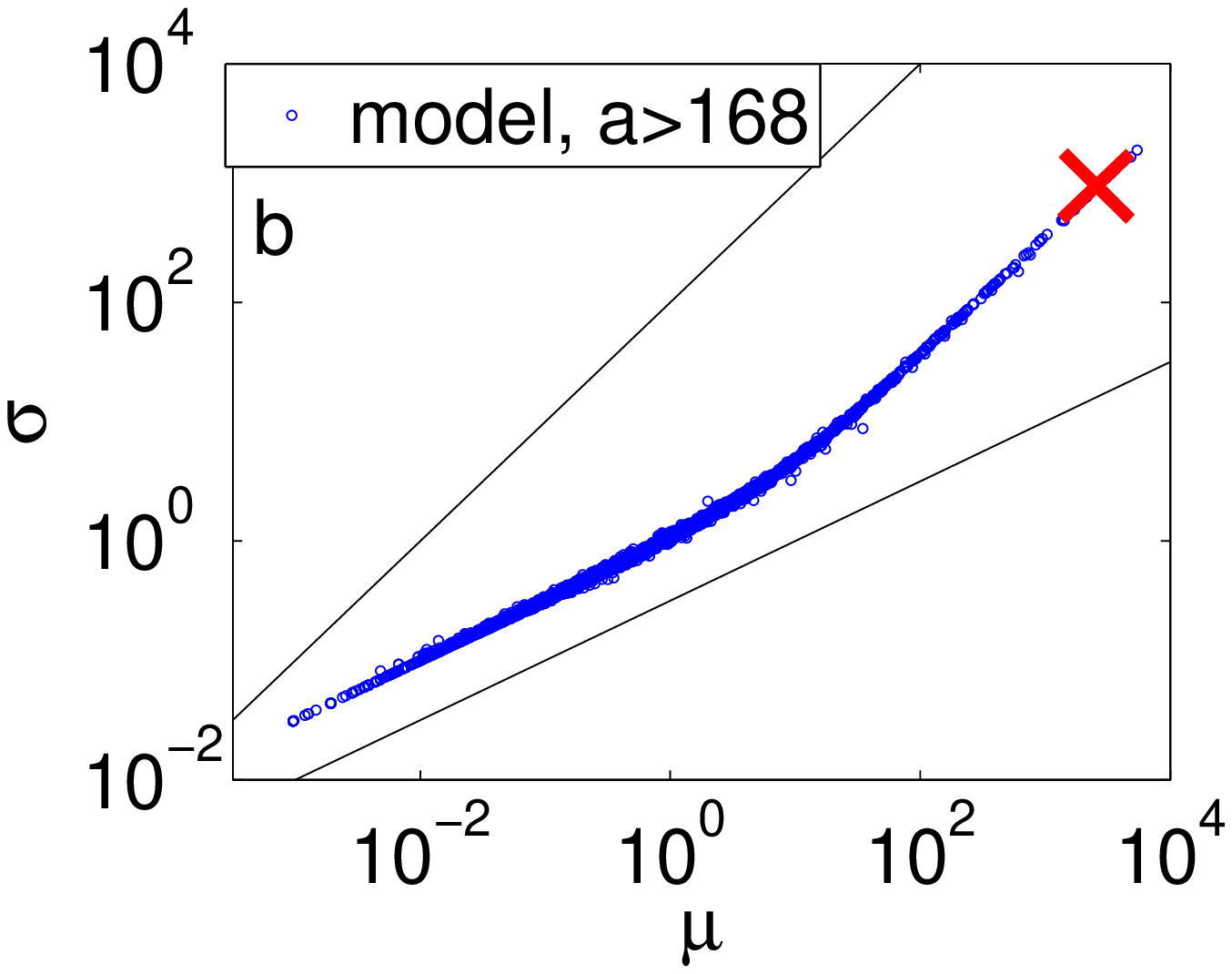,width=5.2cm}
\epsfig{figure=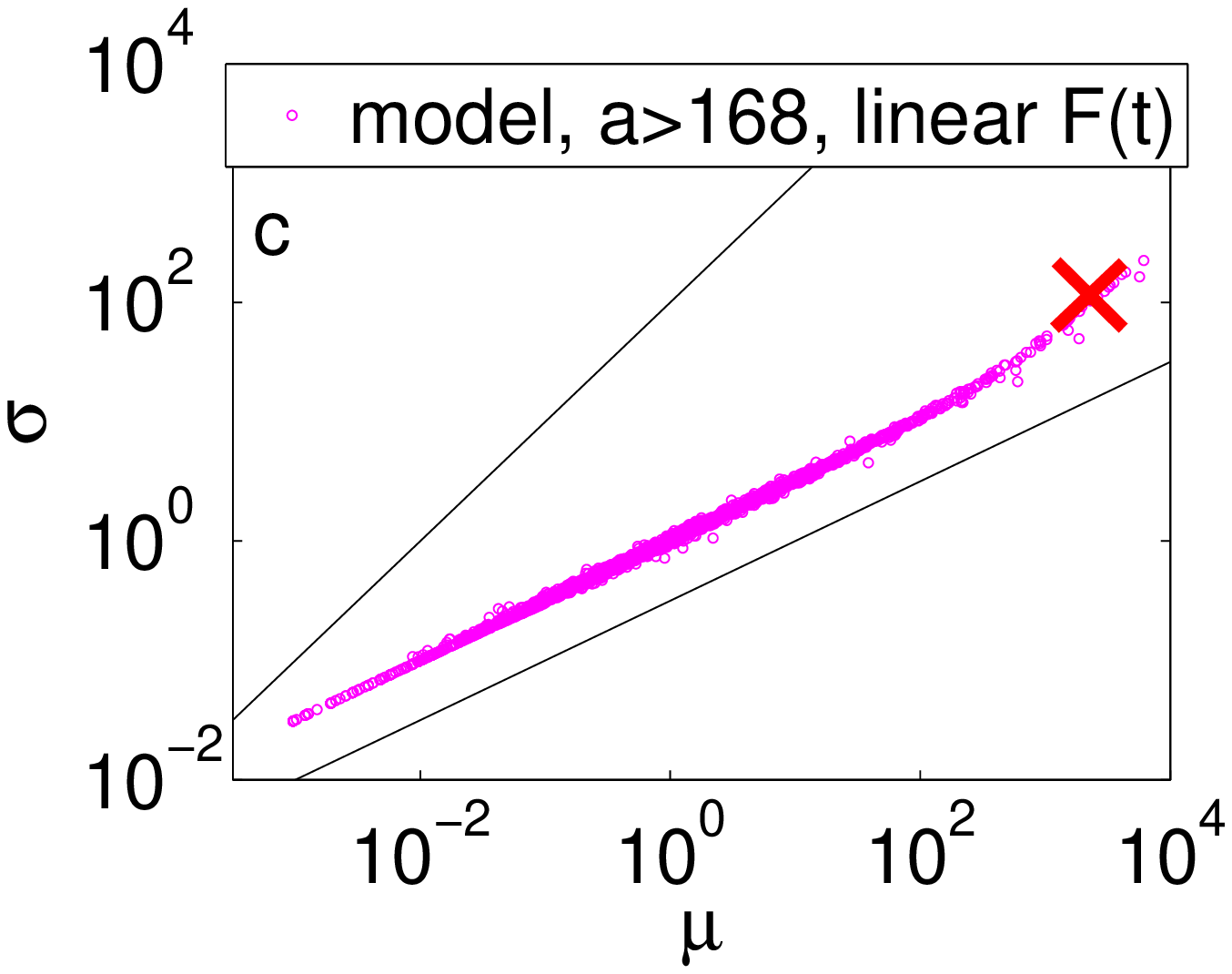,width=5.2cm}
\epsfig{figure=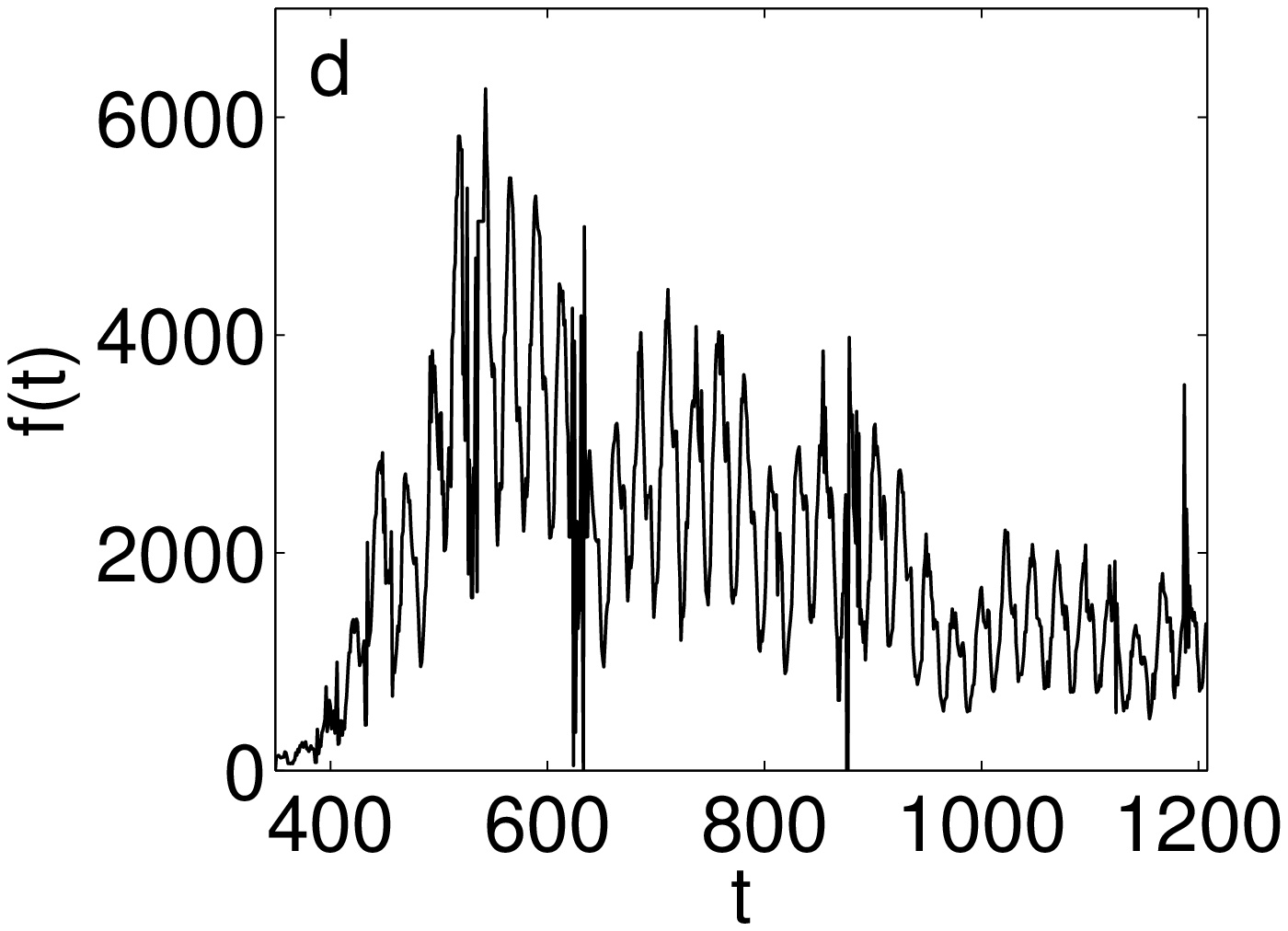,width=5.2cm}
\epsfig{figure=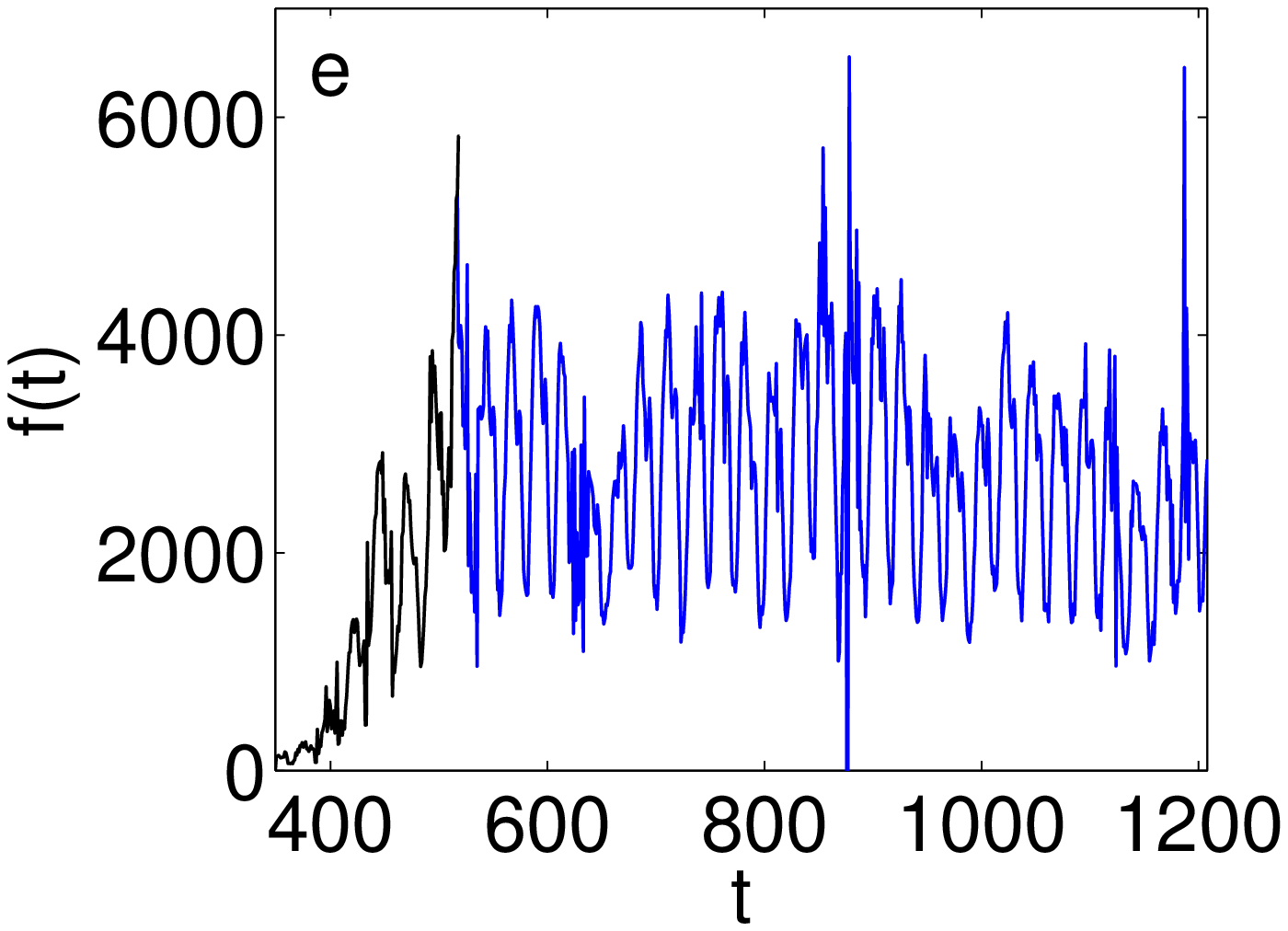,width=5.2cm}
\epsfig{figure=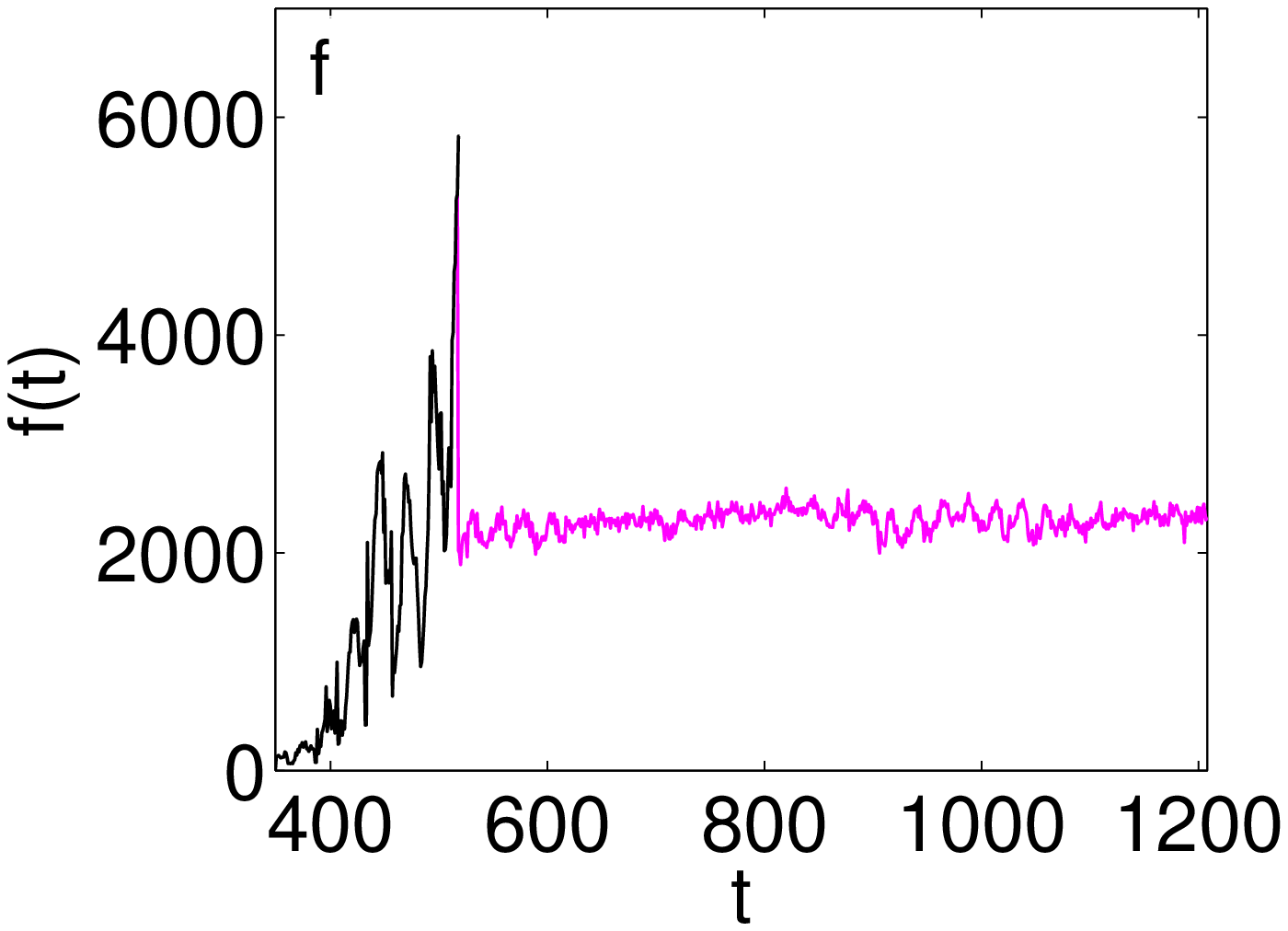,width=5.2cm}
\caption{(Top row) Fluctuation-scaling plots using only values of $a$ such that $a>168$.  The red crosses in the top row and the $f_i(t)$ time series in the bottom row correspond to the app ``What's your stripper name?''.
} \label{figSI9}
\end{figure}

In the top row of Fig.~\ref{figSI9}, we show FS plots for the data and the model with a slightly different way of calculating $\mu_i$ and $\sigma_i$ from the one that discussed above.  (Recall that we calculated $\mu_i$ as the temporal mean of the increments $f_i(t)$ from  $t=t_i$ to the final time $t=t_\text{max}$; we calculated the standard deviation $\sigma_i$ similarly.) In Fig.~\ref{figSI9}, however, we instead begin the temporal averaging at $t=t_i+168$. (If $t_i+168>t_\text{max}$, then we drop this point from the plot.) This implies that we calculate the means and standard deviations only over ages from 1 week onwards, so we neglect the novelty regimes for most apps. Comparing Fig.~\ref{figSI9}a with Fig.~\ref{fig9}a, we see that this change in definition of $\mu_i$ and $\sigma_i$ does not strongly affect the FS plot of the data. However, as one can see by comparing  Fig.~\ref{figSI9}b to Fig.~\ref{fig9}b, the model results clearly are impacted by ignoring the novelty regime in calculating $\mu_i$ and $\sigma_i$.  This arises from the relatively small fluctuations in the model for very popular apps.   For example, the panels in the bottom row of Fig.~\ref{figSI9} show the $f_i(t)$ time series for the app ``What's your stripper name?'' (see panel 3 of Fig.~\ref{figSI2}). In the data (Fig.~\ref{figSI9}d), the $f_i(t)$ time series decays slowly with the age of the app. However, the model does not reproduce this decay (see Fig.~\ref{figSI9}e), as it instead has a
mostly unchanging envelope of $f_i(t)$ values in the post-novelty regime, and the fluctuations are due mainly to the aggregate activity level $F(t)$ that is input into the model.   These fluctuations clearly give the main contribution to the standard deviation $\sigma_i$ in our revised calculation. Indeed, the diurnal variations are inherited directly from the $F(t)$ function, and these fluctuations have the same order of magnitude as the mean. See, in particular, the function $\psi(t)$ in Fig.~\ref{figSI1}. This implies that $\sigma_i \sim \mu_i$ in this case, and it thereby yields an FS scaling exponent of $\beta=1$.

We generate the third panel in each row of Fig.~\ref{figSI9} using a further modification of our model: we replace the total activity function $F(t)$ that we input into the model with the linear growth function $A(t)$ from Fig.~\ref{figSI1}. This revised model has a total installation activity that grows linearly in time, but it does not experience the system-wide diurnal variations of the data. We still copy the $H=168$ hour history window for newly-launched apps directly from the data (see the black curve in Fig.~\ref{figSI9}f). This introduces some residual 24-hour variation, but it is much less prominent than in the model before modififcation. The resulting post-novelty standard deviations $\sigma_i$ for popular apps are much smaller than in the other cases considered.
Moreover, the scaling $\sigma_i\sim \mu_i^{1/2}$ holds for a much larger range of $\mu_i$ values. (Compare Fig.~\ref{figSI9}c to Fig.~\ref{figSI9}b.) We conclude that the high-$\mu_i$ scaling of $\beta\approx 1$ is connected intimately with diurnal variations in the activity levels of Facebook users.


\bibliographystyle{unsrt}

\bibliography{FB_bib2}

\end{document}